%% file: main.tex
\documentclass[12pt,a4paper]{scrartcl}
\pdfoutput=1

\usepackage[utf8]{inputenc}
\usepackage[T1]{fontenc}
\usepackage{amsmath,amssymb,mathtools}
\usepackage[separate-uncertainty=true,retain-unity-mantissa=false]{siunitx}
\usepackage{slashed}
\usepackage{graphicx}
\usepackage{booktabs,multicol,multirow}
\usepackage[shortlabels]{enumitem}
\usepackage{hyperref}
\usepackage[capitalize]{cleveref}
\usepackage{xspace}
\usepackage[noblocks]{authblk}
\usepackage[dvipsnames]{xcolor}
\usepackage{soul}
\usepackage[textsize=tiny]{todonotes}
\usepackage{subcaption}
\usepackage{ulem}
\usepackage[sorting=none,%
  citestyle=numeric-comp,%
  bibstyle=mynumeric,%
  giveninits=true]{biblatex}
\input{abbr.tex}

\addtokomafont{disposition}{\rmfamily}

\title{\Large External leg corrections as an origin\\[.3em]
of large logarithms}
\author[1,2]{Henning Bahl\footnote{\href{mailto:hbahl@uchicago.edu}{hbahl@uchicago.edu}}}
\author[2]{Johannes Braathen\footnote{\href{mailto:johannes.braathen@desy.de}{johannes.braathen@desy.de}}}
\author[2,3]{Georg Weiglein\footnote{\href{mailto:georg.weiglein@desy.de}{georg.weiglein@desy.de}}}
\affil[1]{University of Chicago, Department of Physics, 5720 South Ellis Avenue, Chicago, IL~60637~USA}
\affil[2]{Deutsches Elektronen-Synchrotron DESY, Notkestr.~85, 22607 Hamburg, Germany}
\affil[3]{II. Institut für Theoretische Physik, Universität Hamburg, Luruper Chaussee 149, 22761 Hamburg, Germany}
\date{}
\titlehead{\hfill \parbox{3cm}{\vspace{-2cm}\begin{flushright} \texttt{DESY 21-231} \\ \texttt{EFI 12-10} \end{flushright}}}

\begin{document}
\maketitle

\begin{abstract}\noindent
    \input{sec_abstract.tex}
\end{abstract}
\setcounter{footnote}{0}

\newpage

\tableofcontents

\newpage

%%%%%%%%%%%%%%%%%%%%%%%%%%%%%%%%%%%%%%%%%%%%%%%%%%%%%%%%%%%%%%%%%%%%%%%%%%%%%%
%%%%%%%%%%%%%%%%%%%%%%%%%%%%%%%%%%%%%%%%%%%%%%%%%%%%%%%%%%%%%%%%%%%%%%%%%%%%%%

\section{Introduction}
\label{sec:intro}

\input{sec_intro.tex}

%%%%%%%%%%%%%%%%%%%%%%%%%%%%%%%%%%%%%%%%%%%%%%%%%%%%%%%%%%%%%%%%%%%%%%%%%%%%%%
%%%%%%%%%%%%%%%%%%%%%%%%%%%%%%%%%%%%%%%%%%%%%%%%%%%%%%%%%%%%%%%%%%%%%%%%%%%%%%

\section{Toy model}
\label{sec:toy_model}

\input{sec_toy.tex}

%%%%%%%%%%%%%%%%%%%%%%%%%%%%%%%%%%%%%%%%%%%%%%%%%%%%%%%%%%%%%%%%%%%%%%%%%%%%%%
%%%%%%%%%%%%%%%%%%%%%%%%%%%%%%%%%%%%%%%%%%%%%%%%%%%%%%%%%%%%%%%%%%%%%%%%%%%%%%

\section{Scalar external leg corrections as an origin of large logarithms}
\label{sec:wfr_ll}

\input{sec_wfr_ll.tex}

%%%%%%%%%%%%%%%%%%%%%%%%%%%%%%%%%%%%%%%%%%%%%%%%%%%%%%%%%%%%%%%%%%%%%%%%%%%%%%
%%%%%%%%%%%%%%%%%%%%%%%%%%%%%%%%%%%%%%%%%%%%%%%%%%%%%%%%%%%%%%%%%%%%%%%%%%%%%%

\section{Applications}
\label{sec:applications}

\input{sec_applications.tex}

%%%%%%%%%%%%%%%%%%%%%%%%%%%%%%%%%%%%%%%%%%%%%%%%%%%%%%%%%%%%%%%%%%%%%%%%%%%%%%
%%%%%%%%%%%%%%%%%%%%%%%%%%%%%%%%%%%%%%%%%%%%%%%%%%%%%%%%%%%%%%%%%%%%%%%%%%%%%%

\section{Conclusions}
\label{sec:conclusions}

\input{sec_conclusions.tex}

%%%%%%%%%%%%%%%%%%%%%%%%%%%%%%%%%%%%%%%%%%%%%%%%%%%%%%%%%%%%%%%%%%%%%%%%%%%%%%
%%%%%%%%%%%%%%%%%%%%%%%%%%%%%%%%%%%%%%%%%%%%%%%%%%%%%%%%%%%%%%%%%%%%%%%%%%%%%%

\section*{Acknowledgements}
\sloppy{We thank I.~Sobolev for collaboration in the early stages of this project. We thank W.~Hollik, D.~Meuser, and P.~Slavich for useful discussions as well as S.~Paasch for help with the N2HDM model file. We acknowledge support by the Deutsche Forschungsgemeinschaft (DFG, German Research Foundation) under Germany‘s Excellence Strategy -- EXC 2121 ``Quantum Universe'' – 390833306. H.B.\ acknowledges support by the Alexander von Humboldt foundation. }

%%%%%%%%%%%%%%%%%%%%%%%%%%%%%%%%%%%%%%%%%%%%%%%%%%%%%%%%%%%%%%%%%%%%%%%%%%%%%%
%%%%%%%%%%%%%%%%%%%%%%%%%%%%%%%%%%%%%%%%%%%%%%%%%%%%%%%%%%%%%%%%%%%%%%%%%%%%%%
%%%%%%%%%%%%%%%%%%%%%%%%%%%%%%%%%%%%%%%%%%%%%%%%%%%%%%%%%%%%%%%%%%%%%%%%%%%%%%

\appendix
\section{Definition of loop functions}
\label{app:def_loopfn}

\input{app_loopfndef}

\section{Derivatives of two-loop functions}
\label{app:deriv_loopfn}

\input{app_derivativesloopfunctions.tex}
%%%%%%%%%%%%%%%%%%%%%%%%%%%%%%%%%%%%%%%%%%%%%%%%%%%%%%%%%%%%%%%%%%%%%%%%%%%%%%
%%%%%%%%%%%%%%%%%%%%%%%%%%%%%%%%%%%%%%%%%%%%%%%%%%%%%%%%%%%%%%%%%%%%%%%%%%%%%%
%%%%%%%%%%%%%%%%%%%%%%%%%%%%%%%%%%%%%%%%%%%%%%%%%%%%%%%%%%%%%%%%%%%%%%%%%%%%%%

\clearpage
\printbibliography

\end{document}

%% file: abbr.tex
\renewcommand{\Re}{\text{Re}}
\renewcommand{\Im}{\text{Im}}
\newcommand{\cp}{\ensuremath{{\cal CP}}}

\newcommand{\DR}{{\ensuremath{\overline{\text{DR}}}}\xspace}

\newcommand{\gev}{\;\text{GeV}\xspace}
\newcommand{\tev}{\;\text{TeV}\xspace}

\newcommand{\SUSY}{{\text{SUSY}}}

\newcommand{\BE}{\begin{equation}}
\newcommand{\EE}{\end{equation}}

\newcommand{\msusy}{\ensuremath{M_\SUSY}\xspace}

\newcommand{\mir}{m_{\text{IR}}}

\newcommand{\xt}{\ensuremath{\widehat X_t}\xspace}
\newcommand{\yt}{\ensuremath{\widehat Y_t}\xspace}

\newcommand{\lln}{\overline{\ln}}
\newcommand{\nn}{\nonumber}
\newcommand{\msbar}{\ensuremath{\overline{\text{MS}}}~}

\NewBibliographyString{refname}
\NewBibliographyString{refsname}
\DefineBibliographyStrings{english}{%
  refname = {Ref\adddot},
  refsname = {Refs\adddot}
}
\DeclareCiteCommand{\ccite}
  {%
  \ifnum\thecitetotal=1
    \bibstring{refname}%
  \else%
    \bibstring{refsname}%
  \fi%
  \addnbspace\bibopenbracket%
  \usebibmacro{cite:init}%
   \usebibmacro{prenote}}
  {\usebibmacro{citeindex}%
   \usebibmacro{cite:comp}}
  {}
  {\usebibmacro{cite:dump}%
   \usebibmacro{postnote}%
   \bibclosebracket}
\newrobustcmd*{\Ccite}{\bibsentence\ccite}

%% file: sec_abstract.tex
The appearance of large logarithmic corrections is a well-known phenomenon in the presence of widely separated mass scales. In this work, we point out the existence of large Sudakov-like logarithmic contributions related to external-leg corrections of heavy scalar particles which cannot be resummed straightforwardly using renormalisation group equations. Based on a toy model, we discuss in detail how these corrections appear in theories containing at least one light and one heavy particle that couple to each other with a potentially large trilinear coupling. We show how the occurrence of the large logarithms is related to infrared singularities. In addition to a discussion at the one-loop level, we also explicitly derive the two-loop corrections containing the large logarithms. We point out in this context the importance of choosing an on-shell-like renormalisation scheme. As exemplary applications, we present results for the two-loop external-leg corrections for the decay of a gluino into a scalar top quark and a top quark in the Minimal Supersymmetric extension of the Standard Model as well as for a heavy Higgs boson decay into two tau leptons in the singlet-extended Two-Higgs-Doublet Model.

%% file: sec_intro.tex
Until now, only one scalar particle without a known substructure has been found: the Higgs boson discovered at the Large Hadron Collider (LHC) in 2012~\cite{ATLAS:2012yve,CMS:2012qbp}. In the Standard Model (SM), the Higgs boson arises as part of an $SU(2)_L$ doublet alongside with the neutral and charged ``would-be'' Goldstone fields. For suitable parameter choices, the presence of the Higgs potential triggers the spontaneous breakdown of the $SU(2)_L\times U(1)_Y$ gauge symmetry. As a direct consequence, the ``would-be'' Goldstone fields become the longitudinal degrees freedom of the gauge bosons $W^\pm$ and $Z$ which in this way acquire mass.

While the SM provides a phenomenological description of electroweak symmetry breaking, an understanding of the underlying dynamics is lacking so far. Furthermore, a number of experimental observations --- e.g., the presence of Dark Matter or the existence of non-zero neutrino masses --- and theoretical issues --- e.g., the hierarchy problem --- cannot be explained within the framework of the SM and are, therefore, hints for the existence of physics beyond the SM (BSM). Indeed, BSM models with an extended Higgs sector, featuring additional Higgs bosons, provide a description of the available data that is at least comparable or even better than for the case of the SM, see e.g.\ \ccite{Biekotter:2021qbc} for a recent discussion of LHC results and \ccite{Bagnaschi:2017tru,GAMBIT:2017zdo} for global fits.

So far no direct evidence for BSM particles has been found via the searches at the LHC, which instead have placed increasingly strong lower bounds on the masses of potential BSM particles. While this does not rule out the possibility of light BSM particles with relatively small couplings to SM particles, in many BSM scenarios that are currently investigated it implies a rather large hierarchy between the electroweak (EW) scale of the SM and the scale of at least some of the BSM particles.

For the description of BSM scenarios where the masses of the new particles are so heavy that they are beyond the reach of present and future accelerators, effective field theory (EFT) frameworks --- e.g., SMEFT~\cite{Buchmuller:1985jz,Grzadkowski:2010es} --- offer a conceptually clean way to parameterise the effects of heavy BSM particles on low-scale physics. Moreover, they provide a way to resum large logarithmic corrections involving the EW and the BSM scales appearing in the theoretical predictions for physics at the EW scale, such as the prediction for the mass of the SM-like Higgs boson in supersymmetric theories, see e.g.\ \ccite{Slavich:2020zjv} for a recent review.

By construction, these EFT frameworks, where the heavy particles are integrated out, do not allow the description of the dynamics of heavy BSM particles. Such a description, however, is of interest since the LHC high-luminosity run and also future colliders like an $e^+e^-$ linear collider or the FCC offer a significant potential for discovering new particles.

Precise theoretical predictions for the production and decay of new heavy particles are important for determining the viable parameter space of the considered BSM scenarios and for assessing the sensitivity for discoveries and for discriminating between different possible realisations of BSM physics. In those predictions, which have been obtained in the literature for a variety of models of BSM physics, the appearance of large logarithmic corrections potentially spoiling the reliability of the perturbative expansion is a re-occurring issue. For describing the dynamics of heavy BSM particles, it is, however, obviously not possible to resum these logarithmic corrections by integrating out the heavy particles.

A well-known example in this context are the fermionic decays of a heavy BSM Higgs boson. In order to avoid large QCD corrections, the fermion mass entering via the Yukawa coupling of the Higgs boson to a pair of fermions should be evolved to the heavy Higgs mass scale~\cite{Braaten:1980yq,Drees:1990dq}. Large logarithms develop, however, not only in this kind of QCD corrections, but also in the form of electroweak Sudakov logarithms (see e.g.\ \ccite{Domingo:2018uim,Domingo:2019vit}) which are related to the exchange of $W$, $Z$ or light Higgs bosons. In principle, these electroweak Sudakov logarithms can be resummed using soft-collinear effective theory (SCET)~\cite{Chiu:2007yn,Chiu:2007dg,Chiu:2008vv}. A specific SCET approach for resumming large logarithms in the decay of BSM particles to SM particles has been developed in \ccite{Alte:2018nbn} and applied to various example models in \ccite{Alte:2019iug,Heiles:2020plj,Mecaj:2020opd}.

In the present paper, we carry out a detailed investigation of the occurrence of large logarithmic corrections in processes with external scalar BSM particles. We point out that large logarithms can arise via external leg corrections involving an interaction between two heavy scalar particles and one light scalar particle. The size of these corrections can be further enhanced\footnote{As a consequence of the chiral symmetry in the case of fermions and of the gauge symmetry in the case of gauge bosons, analogous enhanced logarithmic corrections do not appear for BSM fermions or BSM gauge bosons --- hence our focus on scalar corrections in this paper.} by large trilinear couplings between the involved scalars. Focusing first on a simple toy model, we explicitly show the origin of these logarithms at the one-loop level. Moreover, we point out how these logarithms are related to infrared divergencies which can be cured by light scalar radiation or by resumming the mass corrections of the light scalar. While in principle one would expect that it should be possible to resum these Sudakov-like logarithms with the methods of e.g.~\ccite{Alte:2018nbn} (the corresponding EFT realisation to the best of our knowledge has not been obtained in the literature so far), we follow a more direct approach in the present paper by explicitly calculating the two-loop corrections involving the large logarithms. In this context, we compare different renormalisation schemes for the involved parameters. As a result, we stress the importance of choosing on-shell-like schemes in order to avoid corrections that are enhanced by powers of the heavy BSM scale over the EW scale. In addition to these conceptual studies, we discuss various exemplary applications. First, we discuss the decay of a gluino into a top and a scalar top quark in the Minimal Supersymmetric extension of the Standard Model (MSSM). As a second example, we study heavy Higgs decays to two tau leptons in the next-to-minmal Two-Higgs-Doublet Model (N2HDM). For each of these examples, we explicitly derive the quantum corrections involving the large
logarithms at the one-loop and the two-loop order.

Our paper is organised as follows. In \cref{sec:toy_model}, we introduce our toy model. The occurrence of large logarithms for this toy model is then discussed in detail in \cref{sec:wfr_ll}. The various exemplary applications are presented in \cref{sec:applications}. \cref{sec:conclusions} contains the conclusions. Details on the two-loop calculations carried out in this paper are given in \cref{app:def_loopfn,app:deriv_loopfn}.

%% file: sec_toy.tex
In order to illustrate our discussion as clearly as possible, we start by focussing on a toy model containing three real singlet scalars $\phi_{1,2,3}$ that are coupled to a Dirac fermion $\chi$. This model offers a simple setting for discussing the issue of large trilinear-coupling-enhanced logarithms arising via external scalar leg corrections, and the different mass configurations in which they can appear.  Moreover, it can be directly mapped to many BSM models (as done in \cref{sec:applications} for some examples).

We endow this model with a $\mathbb{Z}_2$ symmetry under which the scalars and fermions transform as
\begin{align}
\phi_1 \rightarrow - \phi_1, \hspace{.4cm} \phi_2\rightarrow - \phi_2, \hspace{.4cm} \phi_3\rightarrow \phi_3, \hspace{.4cm} \chi\rightarrow \chi\,.
\end{align}
We assume that the Lagrangian parameters are chosen such that this $\mathbb{Z}_2$ symmetry is not broken spontaneously. Therefore, only $\phi_3$ acquires a vacuum expectation value (VEV), which we denote as $v_3$. We can, however, redefine the couplings in the Lagrangian in order to absorb this VEV $v_3$ (and thus work with a field $\phi_3$ having $\langle\phi_3\rangle=0$) and we can furthermore assume without loss of generality that the scalar mass matrix is diagonal. The Lagrangian of this toy model can then be written as
\begin{align}
\label{EQ:toymodel_lagr}
\mathcal{L} ={}& \frac{1}{2}\sum_{i=3}^3(\partial_\mu\phi_i\partial^\mu\phi_i - m_i^2\phi_i^2) \nonumber\\
&  - \frac12A_{113} \phi_1^2\phi_3 - A_{123} \phi_1\phi_2\phi_3 - \frac{1}{2}A_{223} \phi_2^2\phi_3 - \frac{1}{6}A_{333}\phi_3^3 \nonumber\\
&-\frac{1}{24}\lambda_{1111}\phi_1^4-\frac{1}{6}\lambda_{1112}\phi_1^3\phi_2-\frac{1}{4}\lambda_{1122}\phi_1^2\phi_2^2-\frac{1}{6}\lambda_{1222}\phi_1\phi_2^3-\frac{1}{24}\lambda_{2222}\phi_2^4\nonumber\\%
&-\frac{1}{4}\lambda_{1133}\phi_1^2\phi_3^2-\frac{1}{2}\lambda_{1233}\phi_1\phi_2\phi_3^2-\frac{1}{4}\lambda_{2233}\phi_2^2\phi_3^2-\frac{1}{24}\lambda_{3333}\phi_3^4\nonumber\\%
& + \bar\chi(i\slashed{\partial} - m_\chi )\chi + y_3 \phi_3 \bar\chi\chi\,,
\end{align}
where the trilinear couplings $A_{ijk}$ will be of special interest in the discussion below. Note that since the fermions are not charged under the $\mathbb{Z}_2$ symmetry, only $\phi_3$ can couple to them. Furthermore, because the three scalars are real, the trilinear, quartic, and Yukawa couplings in the Lagrangian are also real.

We are mostly interested in the situation in which $\phi_1$ is much lighter than $\phi_2$ and $\phi_3$, while $\phi_2$ and $\phi_3$ are approximately mass-degenerate. Accordingly, we will consider in the following the double limit in which $m_1 \ll m_2\sim m_3$ and concurrently $m_2\to m_3$. From a phenomenological point of view, $\phi_2$ and $\phi_3$ can be regarded as heavy BSM fields whereas $\phi_1$ represents a SM field with a mass around the electroweak scale.

%% file: sec_wfr_ll.tex
One important ingredient in obtaining predictions for physical observables is to ensure that the external particles have the correct on-shell properties as required by the LSZ formalism~\cite{Lehmann:1954rq}. For unstable particles that can mix with each other this can be achieved by employing a (in general non-unitary) matrix, denoted by $\mathbf{\hat Z}$ in the following (see e.g.\ \ccite{Frank:2006yh,Fuchs:2016swt,Fuchs:2017wkq}), which relates the (renormalised) one-particle irreducible vertex functions involving the external loop-corrected mass eigenstate fields $\phi_a^\text{physical}$ to the (renormalised) vertex functions involving the internal lowest-order fields $\phi_j$,
\begin{align}
\hat\Gamma_{\phi_a^\text{physical}} =
\sum_j\mathbf{\hat Z}_{aj} \hat\Gamma_{\phi_j} .
\end{align}
The $(aj)$ element of the
$\mathbf{\hat Z}$ matrix can be written as a combination of the usual LSZ factor for the case with non-vanishing mixing and a term accounting for the mixing between the states $i$ and $j$,
\begin{align}
\mathbf{\hat Z}_{aj} = \sqrt{\hat Z^a_i} {\hat Z}^a_{ij},
\end{align}
where no summation over repeated indices is implied. The terms on the right-hand side are given by
\begin{align}
\label{EQ:generalLSZfactor}
\hat Z_i^a = \frac{1}{1 + \hat\Sigma^{\mathrm{eff} \, \prime}_{ii}(p^2 = \mathcal{M}_a^2)} \, , \qquad
\hat Z_{ij}^a = \left.\frac{\Delta_{ij}(p^2)}{\Delta_{ii}(p^2)}\right\vert_{p^2=\mathcal{M}_{a}^2} .
\end{align}
The superscript ``$^\prime$'' denotes a derivative with respect to the external momentum squared. The elements of the propagator matrix, $\Delta_{ii}(p^2)$, $\Delta_{ij}(p^2)$, \ldots , are obtained from inverting the matrix involving the renormalised self-energies of the fields $\phi_i$, $\phi_j$, \ldots , denoted as $\hat\Sigma_{ii}(p^2)$, $\hat\Sigma_{ij}(p^2)$, \ldots , where the diagonal and off-diagonal entries read $p^2 - m_i^2 + \hat\Sigma_{ii}(p^2)$ and $\hat\Sigma_{ij}(p^2)$, respectively. Here $m_i$ is the tree-level mass of $\phi_i$. The effective self-energy $\hat\Sigma^{\mathrm{eff}}_{ii}(p^2)$ is composed of the usual self-energy $\hat\Sigma_{ii}(p^2)$ and mixing contributions. For the example of the case where the three particles $i, j, k$
can mix with each other it reads
\begin{align}
    \hat\Sigma^{\mathrm{eff}}_{ii}(p^2) = \hat\Sigma_{ii}(p^2) +
    \frac{\Delta_{ij}(p^2)}{\Delta_{ii}(p^2)} \hat\Sigma_{jj}(p^2) +
    \frac{\Delta_{ik}(p^2)}{\Delta_{ii}(p^2)} \hat\Sigma_{kk}(p^2) .
\end{align}
The quantities in \cref{EQ:generalLSZfactor} are evaluated at the complex propagator pole, $p^2 = \mathcal{M}_a^2$, which is associated with $\phi_i$.

We note in this context that the elements of the $\mathbf{\hat Z}$ matrix given above contain higher-order contributions that are crucial for the description of the resonant mixing of two or more unstable particles that are nearly mass-degenerate~\cite{Fuchs:2016swt,Fuchs:2017wkq}. In the following discussion focusing on large logarithmic contributions arising from external-leg corrections of heavy scalar particles we will always perform a strict perturbative expansion, keeping only the terms contributing to the order --- one- or two-loop --- at which we are working. In this way a mixing of orders in perturbation theory is avoided, so that there is no associated residual dependence on unphysical gauge or field-renormalisation contributions (as discussed, e.g., in Refs.~\cite{Domingo:2020wiy,Domingo:2021kud}). Thus, we do not provide here a treatment of the resonance-type behaviour of the nearly mass-degenerate heavy fields $\phi_2$ and $\phi_3$ that would in general be expected to occur (this refers in particular to parameter regions where their mass difference is smaller than the sum of their total widths). We note that in the considered toy model the mixing between the fields $\phi_2$ and $\phi_3$ is forbidden by the imposed $\mathbb{Z}_2$ symmetry, but we will study a model with non-vanishing scalar mixing in \cref{sec:MSSM_Xt_case2}. The treatment of the resonance-type behaviour, as outlined in \cite{Fuchs:2016swt,Fuchs:2017wkq}, can be carried out in addition to the analysis of large logarithmic contributions that will be presented in the following.

%%%%%%%%%%%

\subsection{Large trilinear-coupling-enhanced logarithms appearing in scalar external leg corrections}
\label{sec:large_logs}

The external leg corrections can become a source of large logarithmic contributions. We discuss this here in the context of the toy model introduced in \cref{sec:toy_model}. The discussion is, however, straightforwardly transferable to other models (as we will demonstrate in \cref{sec:applications}).

%%%%%%%%%%% figure %%%%%%%%%%%
\begin{figure}
\centering
\includegraphics[scale=1]{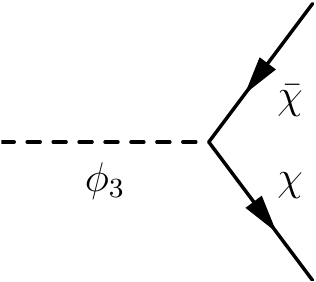}
\caption{$\phi_3\rightarrow\bar\chi\chi$ decay process at tree level.}
\label{fig:chichiphi3_tree}
\end{figure}
%%%%%%%%%%% figure %%%%%%%%%%%

As an example process, we investigate the $\phi_3\rightarrow\bar\chi\chi$ decay process, which is shown at tree level in \cref{fig:chichiphi3_tree}. In more realistic models, this process could for example correspond to the decay of a heavy Higgs boson to two SM fermions (see also \cref{sec:N2HDM}). The corresponding process for the other heavy field, $\phi_2\rightarrow\bar\chi\chi$, is forbidden as a consequence of the imposed $\mathbb{Z}_2$ symmetry (which also eliminates interference effects in the production and decay of $\phi_2$ and $\phi_3$). We note, however, that both the decay processes of $\phi_2$ and $\phi_3$ need to be taken into account in order to ensure the cancellation of infrared divergences in the limit where $m_2 = m_3$ and $m_1 = 0$, see the discussion below.

%%%%%%%%%%% figure %%%%%%%%%%%
\begin{figure}
    \centering
    \includegraphics[scale=1]{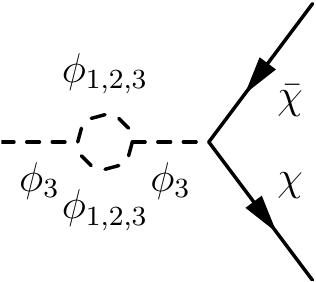}\hspace{.3cm}
    \includegraphics[scale=1]{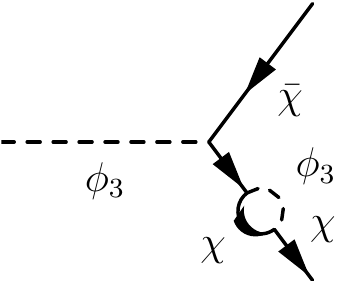}\hspace{.3cm}
    \includegraphics[scale=1]{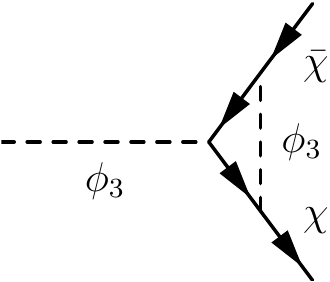}
    \caption{One-loop virtual corrections to the $\phi_3\to\bar\chi\chi$ decay process.}
    \label{fig:chichiphi3_virt}
\end{figure}
%%%%%%%%%%% figure %%%%%%%%%%%

At tree-level, the decay width for this process reads
\begin{align}
    \Gamma^{(0)}(\phi_3\to\chi\bar\chi)=\frac{1}{8\pi}m_3\left(1-\frac{4m_\chi^2}{m_3^2}\right)^{3/2}y_3^2\,.
\end{align}
Once radiative corrections are taken into account, one must compute additional contributions to this process, as illustrated in \cref{fig:chichiphi3_virt}. We denote the relative one- and two-loop corrections to the decay width by $\Delta\hat\Gamma^{(1)}_{\phi_3\to\chi\bar\chi}$ and $\Delta\hat\Gamma^{(2)}_{\phi_3\to\chi\bar\chi}$ respectively, so that
\begin{align}
    \hat\Gamma(\phi_3\to\chi\bar\chi)=\Gamma^{(0)}(\phi_3\to\chi\bar\chi)\big[1+\Delta\hat\Gamma^{(1)}_{\phi_3\to\chi\bar\chi}+\Delta\hat\Gamma^{(2)}_{\phi_3\to\chi\bar\chi}\big]\,.
\end{align}
Among the radiative corrections, one must take into account one-particle-reducible diagrams corresponding to field renormalisation contributions of the external particles --- or equivalently, to the LSZ factors on the external legs (see above).

Here, we focus on large logarithmic corrections that are proportional\footnote{Logarithmic terms can also arise via the external leg corrections of the fermion $\chi$. They can, however, easily be avoided by choosing the renormalisation scale $\sim m_3$. } to powers of the trilinear couplings $A_{ijk}$. These arise only via the LSZ factor of the external $\phi_3$ leg. Taking into account only terms proportional to at least two powers of trilinear couplings, we obtain for the one-loop corrections to the $\phi_3\to\bar\chi\chi$ decay process\footnote{For calculating the loop amplitudes at the one- and two-loop level, we have employed \texttt{FeynArts}~\cite{Hahn:2000kx} and \texttt{FormCalc}~\cite{Hahn:1998yk}.}
\begin{align}
\label{eq:toy_IRdiv}
 \Delta\hat\Gamma_{\phi_3\to\bar\chi\chi}^{(1)}\supset -\frac{1}{2}k y_3
 \Re\bigg[&(A_{113})^2\frac{d}{dp^2}B_0(p^2,m_1^2,m_1^2)+2(A_{123})^2\frac{d}{dp^2}B_0(p^2,m_1^2,m_2^2) \nn\\
 & + (A_{223})^2\frac{d}{dp^2}B_0(p^2,m_2^2,m_2^2)+(A_{333})^2\frac{d}{dp^2}B_0(p^2,m_3^2,m_3^2)\bigg]\bigg|_{p^2=m_3^2} \nn \\
 & + \cdots\,,
\end{align}
where $k\equiv (4\pi)^{-2}$ is used to indicate the loop order, $B_0$ is the usual Passarino-Veltmann function (we recall its definition in \cref{EQ:defAB}), and the ellipsis denotes terms that are not proportional to at least two powers of trilinear couplings.

As already noted in \cref{sec:toy_model}, we concentrate on the situation in which $m_1 \ll m_2 \sim m_3$, and hence the second term in \cref{eq:toy_IRdiv} is of particular interest because it is  infrared divergent in the double limit of $m_1\to0$ and $m_3\to m_2$. At the one-loop level no other terms proportional to $A_{123}^2$ appear that could cancel this divergence. For this mass hierarchy, we can distinguish two different cases,
\begin{enumerate}
  \item $\phi_2$ and $\phi_3$ are almost mass-degenerate, $\phi_1$ is light ($m_1 \rightarrow 0$, $m_2 \rightarrow m_3$),
  \begin{align}
  \label{EQ:B0p_smallm1}
   \frac{d}{d p^2}B_0(p^2,m_1^2,m_2^2)\big|_{p^2=m_3^2} = \frac{1}{m_3^2}\left(\frac{1}{2}\ln\frac{m_3^2}{m_1^2} - 1 + \mathcal{O}\left(\epsilon^{1/2}\right)\right)\,,
  \end{align}
  where $\epsilon\equiv m_3^2 - m_2^2$ and $m_1^2 \sim \epsilon$, and with the expansion in $m_1$ performed first. We note here that as long as $m_1$ is non-zero, it regulates the possible IR divergence in the derivative of the $B_0$ function in the equation above, while the limit $m_2\to m_3$ can be taken without any issue. 
  \item $\phi_2$ and $\phi_3$ are almost mass-degenerate, $\phi_1$ is massless ($m_1 = 0$, $m_2 \rightarrow m_3$),
  \begin{align}
  \label{EQ:B0p_m1zero}
   \frac{d}{d p^2}B_0(p^2,0,m_2^2)\big|_{p^2=m_3^2} = \frac{1}{m_3^2}\left(\ln\left(-\frac{m_3^2}{\epsilon}\right) - 1 + \mathcal{O}\left(\epsilon\right)\right)\,,
  \end{align}
  where $\epsilon\equiv m_3^2 - m_2^2$.
\end{enumerate}
In both cases, logarithms appear which become large in the limit $m_1\to 0$ or $\epsilon \to 0$. The suppression by $1/m_3^2$ can be compensated by the prefactor $A_{123}^2$ if $A_{123}\sim m_3$, a situation which can easily be realised in many BSM theories. If one of the scalar fields would obtain a vev, similar terms proportional to two powers of the vev times a scalar quartic coupling would appear. However, such terms are expected to be smaller than genuine BSM trilinear terms because they are of order vev times an $\mathcal{O}(1)$ number --- the size of the quartic coupling being limited by unitarity --- while the natural scale for BSM trilinear couplings is the BSM mass scale itself. Note on the other hand that the magnitude of trilinear couplings can also be constrained by unitarity (at finite energies), see e.g.\ \ccite{Goodsell:2018tti}, or by other considerations like vacuum stability, as is for instance the case with $X_t$ in the MSSM, see for instance \ccite{Hollik:2018wrr}.

The described logarithms do not involve the renormalisation scale and can, therefore, not be avoided by an appropriate scale choice. Also a straightforward EFT approach, in which $\phi_3$ and $\phi_2$ are integrated out, would not circumvent the issue, since we consider here a process in which $\phi_3$ appears as an external particle. Consequently, the appearing logarithms can not straightforwardly be resummed using renormalisation group equations.\footnote{As already mentioned in the introduction, we expect the SCET approach worked out in \ccite{Alte:2018nbn} to provide a way to resum these logarithms. While the approach seems to be suitable, the needed concrete EFT has not been worked out yet.}  We also want to remark that this type of divergence, caused by a light scalar close to or in the massless limit, is similar to the so-called ``Goldstone boson catastrophe,'' which has been discussed in the literature -- see in particular \ccite{Elias-Miro:2014pca,Martin:2014bca,Kumar:2016ltb,Braathen:2016cqe, Braathen:2017izn,Espinosa:2017aew}.

%%%%%%%%%%%

\subsection{The infrared limit}
\label{sec:infrared_limit}

As stated above, the external leg corrections to the $\phi_3\to \chi\bar\chi$ process develop infrared divergences in the limit $m_1\to 0 $ or $\epsilon\to 0$, and this obviously raises the question of how to address them. In the following, we discuss two methods to cancel out the IR divergences: firstly the inclusion of real radiation, as prescribed by the KLN theorem~\cite{Kinoshita:1962ur,Lee:1964is}, and secondly a resummation of $\phi_1$ contributions.\footnote{For completeness, we should mention another possible approach, following \ccite{Passera:1998uj}, that is to include the width $\Gamma_3$ of the $\phi_3$ scalar, and to evaluate the process at the complex pole, $\mathcal{M}_3^2=m_3^2-i\Gamma_3 m_3$. However, this approach for curing the IR divergence works only at the one-loop order if no non-zero width is taken into account for the internal particles. Moreover, in some models or scenarios, the width of $\phi_3$ can vanish.}

%%%%%%%%%%%

\subsubsection{Inclusion of real corrections}

%%%%%%%%%%% figure %%%%%%%%%%%
\begin{figure}
\centering
\includegraphics[scale=1]{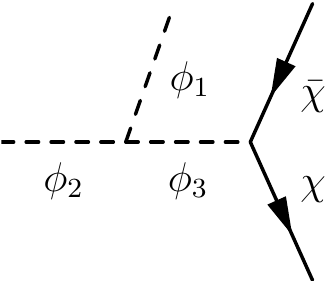}
\caption{Real $\phi_1$ emission diagram cancelling the IR divergence in the virtual correction to the $\phi_3\to\bar\chi\chi$ process.}
\label{fig:phi3phi3_se}
\end{figure}
%%%%%%%%%%% figure %%%%%%%%%%%

In the case where $m_2=m_3$ and $m_1=0$, the radiation of a real $\phi_1$ scalar becomes kinematically allowed, and the appearance of the IR divergence in the $\phi_3\to\chi\bar\chi$ decay width can be understood as being due to considering an IR-unsafe observable. Following the KLN theorem, in order to obtain an IR-safe observable, one should include also the soft part of the Bremsstrahlung process $\phi_2\to\chi\bar\chi\phi_1$, shown in \cref{fig:phi3phi3_se}, which cannot be experimentally resolved from the $\phi_3\to\chi\bar\chi$ decay. Reintroducing a mass $m_1$ as an IR regulator, we obtain
\begin{align}
\label{EQ:toymodel_softrad}
  \Gamma^{(0)}(\phi_2\to\chi\bar\chi\phi_1)\big|^\text{soft}= \Gamma^{(0)}(\phi_3\to\chi\bar\chi)\cdot k \frac{(A_{123})^2}{m_3^2}\bigg[&-\frac{E_\ell}{\sqrt{E_\ell^2+m_1^2}}-\frac{1}{2}\ln m_1^2\nn\\
  &+\ln(E_\ell+\sqrt{E_\ell^2+m_1^2})\bigg]\nn\\
  =\Gamma^{(0)}(\phi_3\to\chi\bar\chi)\cdot k \frac{(A_{123})^2}{m_3^2}\bigg[&-1-\frac{1}{2}\ln m_1^2+\ln(2E_\ell)+\mathcal{O}(m_1)\bigg]\,,
\end{align}
where $E_\ell$ denotes the energy resolution of a detector that would be used to study this process. We can observe that the second term in the brackets in the lower equation exactly cancels the divergence caused by $\frac{d}{dp^2} B(p^2,m_1^2,m_3^2)|_{p^2=m_3^2}$ in the limit $m_1\to0$ --- c.f. \cref{EQ:B0p_smallm1} --- and we find
\begin{align}
\label{EQ:toymodel_totwsoftrad}
    \hat\Gamma(\phi_3\to\chi\bar\chi)+\Gamma^{(0)}(\phi_2\to\chi\bar\chi\phi_1)\big|^\text{soft}=&\  \Gamma^{(0)}(\phi_3\to\chi\bar\chi)\cdot\bigg[1+k\frac{(A_{123})^2}{m_3^2}\ln\frac{2E_\ell}{m_3}\bigg] \nn \\
    & +\cdots\,,
\end{align}
where the ellipsis denotes IR-safe terms, not involving $A_{123}$. We note that the inclusion of the soft real radiation not only cancels the IR-divergent logarithm, but also changes the finite part of the result (c.f. the $-1$ piece in \cref{EQ:toymodel_softrad}).

If we instead set $m_1=0$ and regulate the IR divergence by $\epsilon$ $(=m_3^2-m_2^2)$, the real radiation diagram in \cref{fig:phi3phi3_se} evaluates to
\begin{align}
  \Gamma^{(0)}(\phi_2\to\chi\bar\chi\phi_1)\big|^\text{soft}= \Gamma^{(0)}(\phi_3\to\chi\bar\chi)\cdot k \frac{(A_{123})^2}{m_3^2}\bigg[&-1 + \ln 2 + \ln \frac{m_3 E_\ell}{\epsilon} \bigg]\,, \label{eq:toy_model_real_rad_m1zero}
\end{align}
cancelling the IR divergent term in \cref{EQ:B0p_m1zero}, and giving for the terms of order $(A_{123})^2$ in the sum $\hat\Gamma(\phi_3\to\chi\bar\chi)+\Gamma^{(0)}(\phi_2\to\chi\bar\chi\phi_1)\big|^\text{soft}$ the same expression as in \cref{EQ:toymodel_totwsoftrad}.

Finally, we note that while including only the soft part of the real $\phi_1$ radiation process introduces, as mentioned already, a dependence on a new parameter $E_\ell$, this can be avoided by computing also the hard part of the real radiation process. We will
do this in the following in our numerical analysis, where we perform the necessary three-body phase space integration numerically.

%%%%%%%%%%%

\subsubsection{Resummation of \texorpdfstring{$\phi_1$}{phi1} contributions}

Inspired by one of the proposed solutions to the ``Goldstone boson catastrophe,'' a second possibility to address these problematic IR-divergent terms is to resum the contributions from $\phi_1$, following \ccite{Elias-Miro:2014pca,Braathen:2016cqe,Espinosa:2017aew}. While the resummation approach was devised for the IR divergences from Goldstone bosons in the effective potential (and its derivatives), it can also be applied for a more general situation, such as the one here. The reason for this is that scenarios with large mass hierarchies lead to significant corrections to the light-scalar mass(es). This means that the perturbative expansion is not adequately organised in this case, because all diagrams with (subloop) self-energy insertions yield large contributions, and a resummation is necessary to capture the relevant contributions of higher-order diagrams with self-energy insertions.

We note that the approach of setting the $\phi_1$ (the Goldstone-like state in our toy model) on shell, as done in \ccite{Braathen:2016cqe}, would not work here as there is no tree-level contribution involving $\phi_1$ that would produce a one-loop shift in the calculation of the $\phi_3\to\bar\chi\chi$ process in such a way as to cancel the IR divergences.

%%%%%%%%%%% figure %%%%%%%%%%%
\begin{figure}
    \centering
    \includegraphics[scale=1]{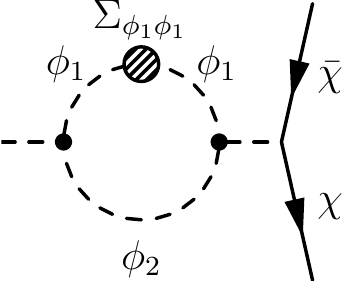}\hspace{.5cm}
    \includegraphics[scale=1]{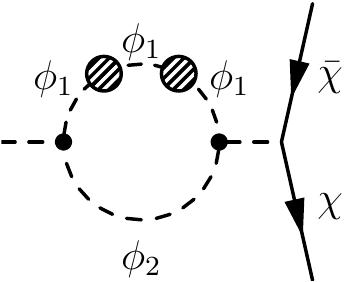}\hspace{.5cm}
    \includegraphics[scale=1]{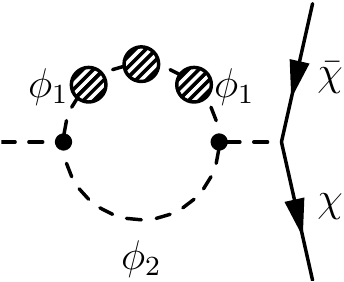}
    \caption{Examplary Feynman diagrams appearing in the resummation of $\phi_1$ contributions. The hashed dots denote the one-loop $\phi_1$ self-energy $\hat\Sigma^{(1)}_{11}$.}
    \label{fig:chichiph3_phi1resum}
\end{figure}
%%%%%%%%%%% figure %%%%%%%%%%%

We show in \cref{fig:chichiph3_phi1resum} some examples of the diagrams that are resummed by applying this procedure. At the cost of mixing orders in perturbation theory, this resummation generates a finite mass for $\phi_1$ and therefore avoids divergences in the derivatives of the $B_0$ function in \cref{eq:toy_IRdiv}. For $m_1=0$, the contributions to the $\phi_1$ mass read
\begin{align}
\label{EQ:delta_m1}
 \Delta m_{\phi_1}^2=\hat\Sigma^{(1)}_{11}(p^2=0)%\big|_\text{hard}
 =-k \bigg[& \frac12\lambda_{1122}A_0(m_2^2)+\frac12\lambda_{1133}A_0(m_3^2)\nn\\
 &+(A_{113})^2B_0(0,0,m_3^2)+(A_{123})^2B_0(0,m_2^2,m_3^2)\bigg]\,,
\end{align}
where $\hat\Sigma^{(1)}_{11}(p^2)$ denotes the one-loop $\phi_1$ self-energy at external squared momentum $p^2$, and the loop function $A_0(m^2)$ is defined in \cref{EQ:defAB} below. We recall also that the quartic couplings $\lambda_{1122}$ and $\lambda_{1133}$ are defined in the Lagrangian in \cref{EQ:toymodel_lagr}. The $(A_{123})^2$ terms in the one-loop decay width are then obtained as
\begin{align}
    \hat\Gamma(\phi_3\to\chi\bar\chi)\supset\Gamma^{(0)}(\phi_3\to\chi\bar\chi)\cdot \bigg[1+k\frac{(A_{123})^2}{m_3^2}\left(\frac12\ln\frac{\Delta m_1^2}{m_3^2}+1\right)\bigg]\,,
\end{align}
which is free of IR divergences. We should point out here that employing this resummation means that one interprets the occurrence of the IR divergence in the one-loop corrections to the $\phi_3$ decay as being due to the breakdown of the perturbative expansion explained above, rather than a lack of inclusiveness of the observable we are computing. As we will see in the following section, at two loops, there are additional IR-divergent contributions that are not contained in the class of diagrams that are incorporated by the resummation as indicated in \cref{fig:chichiph3_phi1resum}, and the resummation approach therefore would have to be extended for applications beyond one-loop order to include further classes of diagrams. It should furthermore be noted that this approach of curing IR divergences would have to be modified or may not be realised if a symmetry enforces the considered particle to be strictly massless (like e.g.\ in the case of the photon).

%%%%%%%%%%%

\subsection{Scalar external leg corrections at the two-loop level}
\label{sec:wfr_ll_2L}

We now investigate the external leg corrections to the $\phi_3\to\chi\bar\chi$ decay process at the two-loop order. For this discussion, we set $m_2^2=m_3^2=m^2$ and $m_1^2=\epsilon$.

Following the discussion at the beginning of \cref{sec:wfr_ll}, the external-leg corrections to the $\phi_3\to\chi\bar\chi$ decay width can be expanded up to two loops as 
\begin{align}
\label{EQ:gen_phi3decay}
    \hat\Gamma(\phi_3\to\chi\bar\chi)=\Gamma^{(0)}(\phi_3\to\chi\bar\chi)\bigg\{&1-\mathrm{Re}\hat\Sigma_{33}^{(1)\prime}(m^2)-\mathrm{Re}\hat\Sigma_{33}^{(2)\prime}(m^2)\nn\\
            &+\big(\mathrm{Re}\hat\Sigma_{33}^{(1)\prime}(m^2)\big)^2-\frac12\big(\mathrm{Im}\hat\Sigma_{33}^{(1)\prime}(m^2)\big)^2\nn\\
            &+\mathrm{Im}\hat\Sigma_{33}^{(1)}(m^2)\cdot\mathrm{Im}\hat\Sigma_{33}^{(1)\prime\prime}(m^2)+\mathcal{O}(k^3)\bigg\}\,,
\end{align}
where $\hat\Sigma_{33}$ is the renormalised $\phi_3$ self energy.
In particular, the two-loop contributions are given by
\begin{align}
 \Delta\hat\Gamma^{(2)}_{\phi_3\to\chi\bar\chi}=\Gamma^{(0)}(\phi_3\to\chi\bar\chi)\bigg[%\left|\hat\Sigma_{33}^{(1)\prime}(m^2)\right|^2
 &-\Re\hat\Sigma_{33}^{(2)\prime}(m^2)+\big(\mathrm{Re}\hat\Sigma_{33}^{(1)\prime}(m^2)\big)^2\nn\\
 &-\frac12\big(\mathrm{Im}\hat\Sigma_{33}^{(1)\prime}(m^2)\big)^2+\mathrm{Im}\hat\Sigma_{33}^{(1)}(m^2)\cdot\mathrm{Im}\hat\Sigma_{33}^{(1)\prime\prime}(m^2)\bigg]\,.\label{eq:toy_2L_matrix_element}
\end{align}

%%%%%%%%%%% figure %%%%%%%%%%%
\begin{figure}
\centering
%%%
\begin{subfigure}[c]{.3\linewidth}\centering
\includegraphics[scale=1]{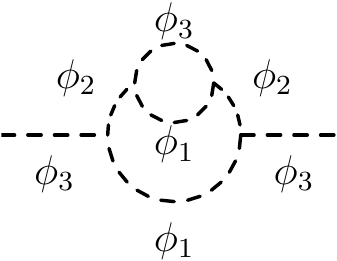}
\end{subfigure}
%%%
\begin{subfigure}[c]{.3\linewidth}\centering
\includegraphics[scale=1]{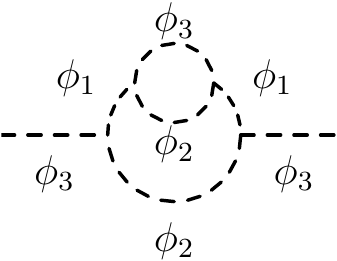}
\end{subfigure}
%%%
\begin{subfigure}[c]{.3\linewidth}\centering
\includegraphics[scale=1]{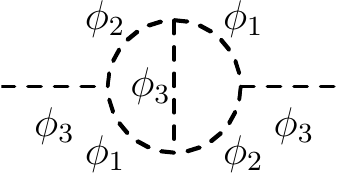}
\end{subfigure}
%%%
\caption{$\phi_3$ self energy at the two-loop level.}
\label{fig:phi3phi3_se_2L}
\end{figure}
%%%%%%%%%%% figure %%%%%%%%%%%

To leading powers in the trilinear coupling $A_{123}$, the one- and genuine two-loop contributions to the $\phi_3$ self-energy (see \cref{fig:phi3phi3_se_2L}) read in terms of $\overline{\text{MS}}$-renormalised parameters
\begin{align}
\hat\Sigma_{33}^{(1)}(p^2)=&\ k (A_{123})^2B_0(p^2,\epsilon,m^2)\,,\nn\\
\hat\Sigma_{33}^{(2\text{, genuine})}(p^2)=&\ k^2 (A_{123})^4\big[T_{11234}(p^2,m^2,m^2,\epsilon,m^2,\epsilon)+T_{11234}(p^2,\epsilon,\epsilon,m^2,m^2,m^2)\nn\\
&\hspace{2.0cm}+T_{12345}(p^2,m^2,\epsilon,m^2,\epsilon,m^2)\big].
\end{align}
The superscript ``genuine'' for the two-loop self-energy indicates that we consider therein only pure two-loop contributions. Definitions\footnote{Note that our definitions of the $T$ integrals differ slightly from the usual ones (see e.g.\ \ccite{Weiglein:1993hd}), because we already subtract pieces corresponding to subloop renormalisation (following \ccite{Martin:2003it,Martin:2003qz}).} of the two-loop $T_{11234}$ and $T_{12345}$ loop integrals as well as expressions for their derivatives (obtained in the \msbar scheme) can be found in \cref{app:def_loopfn,app:deriv_loopfn}. The two-loop integrals contain
potentially dangerous terms of the form $1/\epsilon$, $1/\sqrt{\epsilon}$, $\ln^2\epsilon/m^2$, and $\ln\epsilon/m^2$, which in turn give rise to potentially even more problematic terms of the form $1/\epsilon$ and $1/\sqrt{\epsilon}$ (as well as $\ln^2\epsilon/m^2$, $\ln\epsilon/m^2$) in the first derivatives that enter the external leg corrections. Indeed, we have up to order $\mathcal{O}(\epsilon^0)$
\begin{align}
    \hat\Sigma_{33}^{(1)\prime}(m^2)=&\ \frac{k (A_{123})^2}{m^2}\left[\frac12\ln\frac{m^2}{\epsilon}-1\right]\,,\nn\\
    \hat\Sigma_{33}^{(2\text{, genuine})\prime}(m^2)=&\ k^2 (A_{123})^4\bigg[-\frac{\lln m^2}{2 m^2 \epsilon}+\frac{\pi (4+\lln m^2)}{8\sqrt{\epsilon}m^3}\nn\\
    &+\frac{-18 \lln\epsilon \lln m^2+9 \lln^2\epsilon+60 \lln\epsilon+9 \lln^2m^2-78 \lln m^2+9\pi^2-64}{72 m^4} \nn\\
  &-\frac{\pi^2\ln2-3/2\zeta(3)}{m^4}\bigg]\,,
\end{align}
where the results in \cref{app:deriv_loopfn} for the derivatives of the $T$ integrals have been applied, and the notation $\lln x\equiv \ln x/Q^2$ ($Q$ being the renormalisation scale) has been used.

Applying \cref{EQ:gen_phi3decay}, we obtain a pure \msbar expression for the dominant external-leg contributions in powers of $A_{123}$ to the $\phi_3\to\chi\bar\chi$ decay width, which reads
\begin{align}
    \hat\Gamma(\phi_3\to\chi\bar\chi)=\Gamma^{(0)}(\phi_3\to\chi\bar\chi)\bigg\{&1-\frac{k(A_{123})^2}{m^2}\bigg[\frac12\ln\frac{m^2}{\epsilon}-1\bigg]\nn\\
            &+\frac{k^2(A_{123})^4}{m^4}\bigg[\frac{m^2\lln m^2}{2 \epsilon} - \frac{m \pi (4 + \lln m^2)}{8 \sqrt{\epsilon}}\nn\\
            &\hspace{2.25cm}+\frac{17}{9}-\frac{\pi^2}{8}+\frac{1}{8}\ln^2\frac{m^2}{\epsilon}+\frac{1}{6}\lln\epsilon+\frac{1}{12}\lln m^2\nn\\
            &\hspace{2.25cm}+\pi^2\ln 2-\frac32 \zeta(3)\bigg]\bigg\}\,.
\end{align}
We will use this pure \msbar expression for comparison in our numerical analysis below, but otherwise for obtaining our results we
employ an on-shell renormalisation scheme --- at least for the scalar masses. This implies that
finite contributions from subloop renormalisation
need to be included, which for the parameters in our result
are of the form
\begin{align}
    \hat\Sigma_{33}^{(2\text{, subloop})}(p^2)=k(A_{123})^2\bigg[&\bigg(\frac{2\delta^{(1)}A_{123}}{A_{123}}+\delta^{(1)}Z_3\bigg)B_0(p^2,m_1^2,m_2^2) \nn\\
    & + \delta^{(1)}m_1^2\frac{\partial}{\partial m_1^2}B_0(p^2,m_1^2,m_2^2)\nn\\
    & + \delta^{(1)}m_2^2\frac{\partial}{\partial m_2^2}B_0(p^2,m_1^2,m_2^2)\bigg]\,,
\end{align}
where $\delta^{(1)}m_1^2$, $\delta^{(1)}m_2^2$, $\delta^{(1)}A_{123}$, and $\delta^{(1)}Z_3$ are the one-loop mass counterterms for $m_1^2$ and $m_2^2$, the one-loop counterterm for $A_{123}$, and the one-loop
field renormalisation counterterm for $\phi_3$, respectively.

We now derive the dominant contributions to these counterterms in powers of $A_{123}$. It should be noted that because the expressions for $\hat\Sigma_{33}^{(1)}$ and $\hat\Sigma_{33}^{(2,\text{genuine})}$ are \msbar-renormalised quantities, only the finite parts of the counterterms need to be considered here (as the UV divergences have already been cancelled).  Adopting an on-shell renormalisation scheme for the scalar masses, the mass counterterms are
\begin{align}
 \delta^{(1)}m_1^2&=k(A_{123})^2\mathrm{Re}B_0(m_1^2,m_2^2,m_3^2)\,,\nn\\
 \delta^{(1)}m_2^2&=k(A_{123})^2\mathrm{Re}B_0(m_2^2,m_1^2,m_3^2)\,.
\end{align}
Next, renormalising the $\phi_3$ field in the \msbar scheme, we simply have
\begin{align}
    \delta^{(1)}Z_3\big|^{\text{fin.}}=0.
\end{align}
It should be noted that the \msbar field renormalisation drops out in the sum of the vertex correction and the LSZ factor. Therefore the prescription that has been chosen for the field renormalisation is insignificant.

Turning finally to the trilinear coupling counterterm $\delta^{(1)}A_{123}$, there are several possible choices for the renormalisation of $A_{123}$. We consider here the following options.
\begin{enumerate}[(i)]
    \item The simplest choice is to retain an \msbar renormalisation of $A_{123}$ --- in terms of the finite part of the counterterm this simply means $\delta^{(1)}_{\msbar} A_{123}\big|^{\text{fin.}}=0$. In this case, the subloop renormalisation contributions become
    \begin{align}
    \hat\Sigma_{33}^{(2\text{, subloop})}(p^2)=\ k^2(A_{123})^4\bigg[&
     \mathrm{Re}B_0(m_1^2,m_2^2,m_3^2) \frac{\partial}{\partial m_1^2}B_0(p^2,m_1^2,m_2^2) \nn\\
    & + \mathrm{Re}B_0(m_2^2,m_1^2,m_3^2)\ \frac{\partial}{\partial m_2^2}B_0(p^2,m_1^2,m_2^2)\bigg]\,.\label{eq:toy_subloop_A123_MSbar}
    \end{align}
    Taking now the derivative of $\hat\Sigma_{33}^{(2\text{, subloop})}$ with respect to $p^2$, we can observe that the $1/\epsilon$ and $1/\sqrt{\epsilon}$ divergent terms contained in these two terms exactly cancel with the ones in the derivative of $\hat\Sigma_{33}^{(2\text{, genuine})}$. Indeed, the first term of \cref{eq:toy_subloop_A123_MSbar} is
    \begin{align}
        &B_0(m_1^2,m_2^2,m_3^2)\frac{\partial}{\partial m_1^2}B_0^\prime(m_3^2,m_1^2,m_2^2) = \nn\\
        &= \left(-\lln m^2 + \frac{\epsilon}{6m^2}+\mathcal{O}(\epsilon^2)\right)\left(-\frac{1}{2 \epsilon m^2}+\frac{3 \pi}{8 \sqrt{\epsilon} m^3}+\frac{-1 + 2 \ln\frac{\epsilon}{m^2}}{4 m^4}+\mathcal{O}(\epsilon^{1/2})\right) = \nn\\
        &= \frac{\lln m^2}{2 \epsilon m^2} - \frac{3 \pi \lln m^2}{8 \sqrt{\epsilon} m^3} + \frac{3 \lln m^2 - 6 \lln \epsilon \lln m^2 + 6 \lln^2 m^2-1}{ 12 m^4}+\mathcal{O}(\epsilon^{1/2}) \, ,
    \end{align}
    which yields an exact cancellation of the $1/\epsilon$ and $1/\sqrt{\epsilon}$ terms in the expression for $\frac{d}{dp^2}T_{11234}(p^2,\epsilon,\epsilon,m^2,m^2,m^2)$. The second term of \cref{eq:toy_subloop_A123_MSbar} is
    \begin{align}
        &B_0(m_2^2,m_1^2,m_3^2)\frac{\partial}{\partial m_2^2}B_0^\prime(m_3^2,m_1^2,m_2^2) = \nn\\
        &=\left(2-\lln m^2-\pi\sqrt{\frac{\epsilon}{m^2}}+\mathcal{O}(\epsilon)\right)\left(-\frac{\pi}{4 \sqrt{\epsilon} m^3} - \frac{\ln\frac{\epsilon}{m^2}}{2 m^4}+\mathcal{O}(\epsilon^{1/2})\right) = \nn\\
        &=-\frac{\pi (2 - \lln m^2)}{4\sqrt{\epsilon} m^3 } + \frac{\pi^2 - 4 \lln\epsilon + 4 \lln m^2 + 2 \lln\epsilon \lln m^2 - 2 \lln^2m^2}{4 m^4} \nn\\
        &\hphantom{=}+\mathcal{O}(\epsilon^{1/2})\,,
    \end{align}
    which yields a cancellation of the $1/\sqrt{\epsilon}$ term in $\frac{d}{dp^2}T_{11234}(p^2,m^2,m^2,\epsilon,m^2,\epsilon)$. Accordingly, we obtain
    \begin{align}
        & \frac{d}{dp^2}T_{11234}(p^2,\epsilon,\epsilon,m^2,m^2,m^2)\bigg|_{p^2=m^2}+B_0(m_1^2,m_2^2,m_3^2)\frac{\partial}{\partial m_1^2}B_0^\prime(m_3^2,m_1^2,m_2^2)\nn\\
         &= \frac{-50 + 6 \pi^2 + 3\lln \epsilon - 12 \lln m^2 + 18 \lln\epsilon \lln m^2 - 18 \lln^2 m^2}{36 m^4} \nn\\
         &\quad + \frac{3 \lln m^2 - 6 \lln \epsilon \lln m^2 + 6 \lln^2 m^2-1}{ 12 m^4}\nn\\
        &=\frac{1}{m^4}\bigg[\frac{1}{12}\ln\frac{\epsilon}{m^2}+\frac{\pi^2}{6}-\frac{53}{36}\bigg]\,,\nn\\
        & \frac{d}{dp^2}T_{11234}(p^2,m^2,m^2,\epsilon,m^2,\epsilon)+B_0(m_2^2,m_1^2,m_3^2)\frac{\partial}{\partial m_2^2}B_0^\prime(m_3^2,m_1^2,m_2^2)\nn\\
        &= \frac{-6 \lln\epsilon \lln m^2-3 \lln^2\epsilon+24 \lln\epsilon+9 \lln^2m^2-24 \lln m^2-\pi^2}{24 m^4}\nn\\
        &\quad+\frac{\pi^2 - 4 \lln\epsilon + 4 \lln m^2 + 2 \lln\epsilon \lln m^2 - 2 \lln^2m^2}{4 m^4}\nn\\
        &= \frac{-3 \ln^2\frac{\epsilon}{m^2}+5\pi^2}{24 m^4}\,.
    \end{align}
    We note that the renormalisation-scale dependence has dropped out in this expression. Thus, for the two-loop external-leg corrections at leading order in $A_{123}$, i.e.\ $\mathcal{O}((A_{123})^4)$, the incorporation of OS counterterms for the involved masses is already sufficient to obtain a result that is independent of the renormalisation scale. This fact can be explained by the finiteness of the $A_{123}$ counterterm at leading order in powers of $A_{123}$. Indeed at this order, vertex corrections --- in this toy model --- can only involve scalar triangle diagrams, resulting in the scalar integral $C_0$ (defined in \cref{EQ:defC0}) that is UV-finite, while field renormalisation contributions only involve derivatives of the $B_0$ function, which are also UV-finite (this is further illustrated in \cref{EQ:deltaA123_OS} below). In other words, even in the \msbar scheme, the contributions to the trilinear coupling $A_{123}$ are scale-independent at leading powers in $A_{123}$.

    The derivative of the total two-loop $\mathcal{O}((A_{123})^4)$ contribution to the $\phi_3$ self-energy hence reads
    \begin{align}
        &\hat\Sigma_{33}^{(2)\prime}(m^2)= \nn\\
        &=\ k^2(A_{123})^4\bigg[\frac{-53+6\pi^2+3\ln\frac{\epsilon}{m^2}}{36m^4}+\frac{-3 \ln^2\frac{\epsilon}{m^2}+5\pi^2}{24 m^4}\nn\\
        &\qquad\qquad\quad+\frac{\ln^2\frac{\epsilon}{m^2}- \ln \frac{\epsilon}{m^2}+2}{4 m^4}-\frac{\pi^2\ln2-3/2\zeta(3)}{m^4}\bigg]\nn\\
        &=\ \frac{k^2(A_{123})^4}{m^4}\bigg[\frac{1}{8}\ln^2\frac{\epsilon}{m^2}-\frac{1}{6}\ln\frac{\epsilon}{m^2}-\frac{35}{36}+\frac{3\pi^2}{8}-\pi^2\ln2+\frac{3}{2}\zeta(3)\bigg]\,.
    \end{align}
    Taking all pieces together, we can write the dominant contributions in powers of $A_{123}$ to the $\phi_3\to\chi\bar\chi$ decay width including external leg corrections up to two loops as
    \begin{align}
        \hat\Gamma(\phi_3\to\chi\bar\chi)=\Gamma^{(0)}(\phi_3\to\chi\bar\chi)\bigg\{&1-\frac{k(A_{123})^2}{m^2}\bigg[\frac12\ln\frac{m^2}{\epsilon}-1\bigg]\nn\\
        &+\frac{k^2(A_{123})^4}{m^4}\bigg[\frac18\ln^2\frac{m^2}{\epsilon}-\frac76\ln\frac{m^2}{\epsilon}+\frac{71}{36}-\frac{3\pi^2}{8}\nn\\
        &\hspace{4cm}+\pi^2\ln2-\frac32\zeta(3)\bigg]\bigg\}\,.
    \end{align}

    \item A second possible choice is to renormalise $A_{123}$ on-shell: for this we require that the on-shell-renormalised loop-corrected $\phi_2\to\phi_1\phi_3$ amplitude with on-shell momenta shall remain equal to the tree-level result. With this condition, we obtain
    \begin{align}
    \label{EQ:deltaA123_OS}
        &\delta^{(1)}_\text{OS} A_{123}=\nn\\
        &= k(A_{123})^3\bigg[C_0(p_1^2=m_2^2,p_2^2=m_3^2,(p_1+p_2)^2=m_1^2,m_3^2,m_1^2,m_2^2)\nn\\
        &\hspace{2.5cm}+\frac12\mathrm{Re}\big(B_0^\prime(m_1^2,m_2^2,m_3^2)+B_0^\prime(m_2^2,m_1^2,m_3^2)+B_0^\prime(m_3^2,m_1^2,m_2^2)\big)\bigg]\nn\\
        &=k(A_{123})^3\bigg[\frac{\partial}{\partial y}B_0(m^2,\epsilon,y)\bigg|_{y=m^2}+\frac12\mathrm{Re}\big(B_0^\prime(\epsilon,m^2,m^2)+2B_0^\prime(m^2,\epsilon,m^2)\big)\bigg]\,,
    \end{align}
    and the one-loop scalar integral $C_0$ is defined in \cref{EQ:defC0} below.
    Since this result contains only derivatives of the $B_0$ integral, it is manifestly UV-finite, in accordance with the above discussion concerning the scale-independence of $A_{123}$ (to leading powers in $A_{123}$). This result for $\delta^{(1)}_\text{OS}A_{123}$ yields additional terms in $\hat\Sigma_{33}^{(2)\text{, subloop}}$. Accordingly, one finds for the derivative of the total two-loop $\phi_3$ self-energy
    \begin{align}
        \hat\Sigma_{33}^{(2)\prime}(m^2)=&\ \frac{k^2(A_{123})^4}{m^4}\bigg[\frac{1}{8}\ln^2\frac{\epsilon}{m^2}+\frac{7}{24}\ln\frac{\epsilon}{m^2}-\frac{1}{18} +\frac{3\pi^2}{8}\nn\\
        &\hspace{2.1cm}-\pi^2\ln2+\frac{3}{2}\zeta(3)\bigg]\,.
    \end{align}
    This yields for the decay width
    \begin{align}
        \hat\Gamma(\phi_3\to\chi\bar\chi)=\Gamma^{(0)}(\phi_3\to\chi\bar\chi)\bigg\{&1-\frac{k(A_{123})^2}{m^2}\bigg[\frac12\ln\frac{m^2}{\epsilon}-1\bigg]\nn\\
        &+\frac{k^2(A_{123})^4}{m^4}\bigg[\frac18\ln^2\frac{m^2}{\epsilon}-\frac{31}{24}\ln\frac{m^2}{\epsilon}+\frac{19}{18}-\frac{3\pi^2}{8}\nn\\
        &\hspace{4cm}+\pi^2\ln2-\frac32\zeta(3)\bigg]\bigg\}\,,
    \end{align}
    which as discussed above is independent of the choice of the renormalisation scale.

    \item As a third possible choice, we explore whether it is possible to choose the finite part of the $A_{123}$ counterterm in such a way as to cancel the $\ln^2\epsilon$ term in $\hat\Gamma(\phi_3\to\chi\bar\chi)$ --- we will denote this version of the counterterm $\delta^{(1)}_\text{no-log-sq}A_{123}$. We should emphasise that this only shuffles the
    $\ln^2\epsilon$ term
    away from the calculation
    of the $\phi_3\to\chi\bar{\chi}$ cross-section and into the extraction of $A_{123}$ from a physical observable (e.g.\ related to the $\phi_3\to\phi_1\phi_2$ process) and thus it does not remove the
    potentially large logarithmic term entirely. However, if $A_{123}$ is only considered as a user-given input, this choice may
    be convenient. The piece of the subloop renormalisation related to $\delta^{(1)} A_{123}$ is given by
    \begin{align}
        \Delta \hat\Sigma_{33}^{(2\text{, subloop}) \prime}(m_3^2)=&\ 2kA_{123}\delta^{(1)}A_{123}B_0^\prime(m_3^2,m_1^2,m_2^2)\nn\\
        =&\ \frac{k A_{123}\delta^{(1)}A_{123}}{m^2}\bigg(2+\ln\frac{\epsilon}{m^2}\bigg)\,.
    \end{align}
    In this prescription we require that the term
    \begin{align}
        -\frac{k^2(A_{123})^4}{m^4}\bigg[\frac{1}{8}\ln^2\frac{\epsilon}{m^2}\bigg]
    \end{align}
    is cancelled. This yields
    \begin{align}
        \delta^{(1)}_\text{no-log-sq}A_{123}=\frac{k(A_{123})^{3}}{8m^2}\ln\frac{\epsilon}{m^2}\,.
    \end{align}
    Inserting this back into $\hat\Sigma_{33}^{(2)\prime}(m^2)$, we obtain
    \begin{align}
         \hat\Sigma_{33}^{(2)\prime}(m^2)
         =&\ \frac{k^2(A_{123})^4}{m^4}\bigg[\frac{1}{8}\ln^2\frac{\epsilon}{m^2}-\frac{1}{6}\ln\frac{\epsilon}{m^2}-\frac{35}{36}+\frac{3\pi^2}{8}-\pi^2\ln2+\frac{3}{2}\zeta(3)\bigg]\nn\\
         &+\frac{k^2(A_{123})^4}{m^4}\bigg[\frac{1}{8}\ln^2\frac{\epsilon}{m^2}+\frac{1}{4}\ln\frac{\epsilon}{m^2}\bigg]\nn\\
         =&\ \frac{k^2(A_{123})^4}{m^4}\bigg[\frac14\ln^2\frac{\epsilon}{m^2}+\frac{1}{12}\ln\frac{\epsilon}{m^2}-\frac{35}{36}+\frac{3\pi^2}{8}-\pi^2\ln2+\frac{3}{2}\zeta(3)\bigg]\,,
    \end{align}
    and in turn for the decay width,
    \begin{align}
        \hat\Gamma(\phi_3\to\chi\bar\chi)=\Gamma^{(0)}(\phi_3\to\chi\bar\chi)\bigg\{&1-\frac{k(A_{123})^2}{m^2}\bigg[\frac12\ln\frac{m^2}{\epsilon}-1\bigg]\nn\\
        +\frac{k^2(A_{123})^4}{m^4}&\bigg[-\frac{11}{12}\ln\frac{m^2}{\epsilon}+\frac{71}{36}-\frac{3\pi^2}{8}+\pi^2\ln2-\frac32\zeta(3)\bigg]\bigg\}\,.
    \end{align}
    It should be noted in this context that in general it is not possible to choose the counterterm $\delta^{(1)}A_{123}$ such that both the $\ln^2\epsilon$ and $\ln\epsilon$ terms are cancelled.

\end{enumerate}

\subsection{Numerical analysis}

%%%%%%%%%%% figure %%%%%%%%%%%
\begin{figure}
    \centering
    \includegraphics[width=0.75\textwidth]{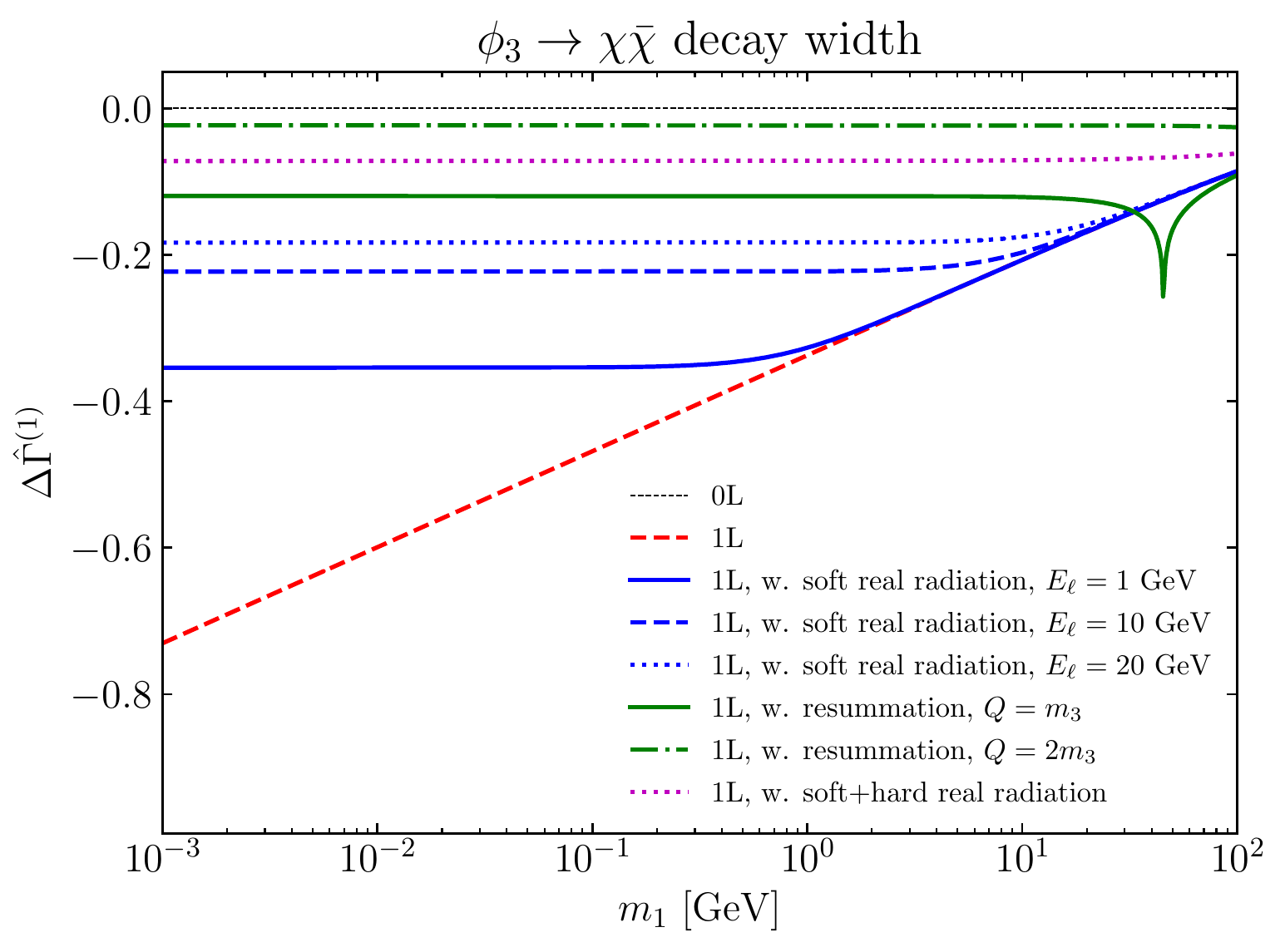}
    \caption{Size of the one-loop external $\phi_3$ leg correction to the $\phi_3\to\chi\bar\chi$ decay width in the toy model of \cref{sec:toy_model}, relative to the tree-level result, as a function of $m_1$. Black: tree-level result only; red: one-loop result; blue: one-loop result, including soft real $\phi_1$ radiation (with different choices for the detector energy resolution $E_\ell$); purple: one-loop result, including both soft and hard real $\phi_1$ radiation; green: one-loop result, including a resummation of $\phi_1$ contributions (with different choices of the renormalisation scale $Q$). For simplicity, we set all trilinear couplings to zero except $A_{123}$, which we take to be $A_{123}=3$ TeV. The other parameters of the model are taken to be $y_3=1$, $m_2=m_3=1$ TeV, $m_\chi=200$ GeV, $\lambda_{1122}=0.25$, $\lambda_{1133}=0.4$.}
    \label{fig:toymodel_vs_m1}
\end{figure}
%%%%%%%%%%% figure %%%%%%%%%%%

In this Section we present numerical investigations of the external leg corrections at one and two loops, in both mass configurations specified in \cref{EQ:B0p_m1zero,EQ:B0p_smallm1}. We start by showing in \cref{fig:toymodel_vs_m1} the relative modification of the $\phi_3\to\chi\bar\chi$ decay width due to the inclusion of one-loop external $\phi_3$ leg corrections as a function of the mass of the light scalar $\phi_1$. First, the red curve shows the one-loop result in the absence of any treatment of the infrared divergence, and therefore it diverges logarithmically if $m_1$ approaches zero, as expected from \cref{EQ:B0p_smallm1}. Next, we compare the impact of the different treatments of the IR problem discussed above. The blue curves are obtained by including soft real radiation of a $\phi_1$ scalar, displaying several values of the detector energy resolution $E_\ell$. While the IR divergence is cured independently of the numerical value of $E_\ell$, the one-loop contribution to $\hat\Gamma(\phi_3\to\chi\bar\chi)$ is sensitive to it: for $m_1=1$ MeV, one finds a one-loop contribution of approximately $-35$\% for $E_\ell=1$ GeV, compared to $-18$\% for $E_\ell=20$ GeV. If the hard part of the real radiation is also included (performing the phase-space integration numerically), one finds the magenta dotted curve, which is independent of the value of $E_\ell$. In this case the one-loop contribution amounts to about $-7$\% of the tree-level result. Finally, the two green curves indicate the result obtained after resummation of $\phi_1$ contributions, for two possible choices of the renormalisation scale, $Q=m_3$ and $Q=2m_3$, entering the loop functions in $\Delta m_1^2$, see \cref{EQ:delta_m1}. One can see that for small values of $m_1$ the resummation procedure also cures the IR divergence in the external leg corrections. The difference between the curves --- for $m_1=1$ MeV, we find a one-loop effect of about $-12$\% for the curve with $Q=m_3$, while it is only $-2.3$\% for $Q=2m_3$ --- can be interpreted as an indication of the theoretical uncertainty of this prescription. It can be associated in particular with diagrams containing higher-order subloop insertions of $\phi_1$ that are not included in the resummation performed here. This large theoretical uncertainty of the resummation procedure indicates a limitation of applying this approach for the purpose of obtaining reliable higher-order predictions for the $\phi_3\to\chi\bar\chi$ decay width. While the different predictions for the results including real radiation can be understood as corresponding to different experimental situations, i.e.\ they refer to different physical observables, there is no clear interpretation of which observable the resummed prediction for the decay width should be compared to --- especially as varying the renormalisation scale $Q$ leads to significantly modified results. It is furthermore apparent that the curve for $Q=m_3$ exhibits a spike around $m_1\simeq 42$ GeV arising from the fact that the sum $m_1^2+\Delta m_1^2$, appearing in logarithms of the loop functions, vanishes for this particular value of $m_1$. Since the resummation procedure is meant to be applied to the region where $m_1$ approaches zero, this numerical artifact at a relatively large non-zero value of $m_1$ has no particular physical significance.

%%%%%%%%%%% figure %%%%%%%%%%%
\begin{figure}
    \centering
    \includegraphics[width=\textwidth]{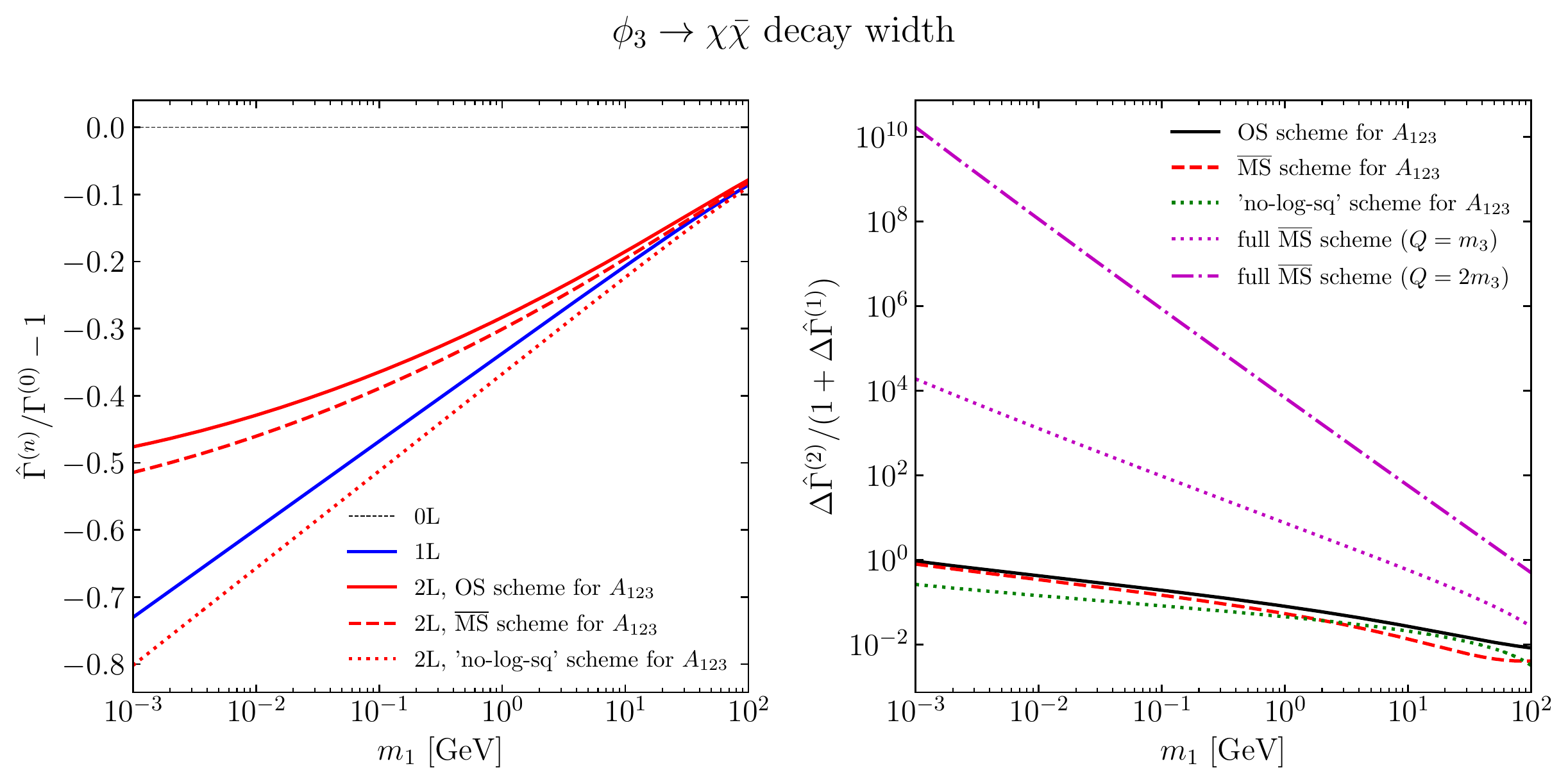}
    \caption{ The one- and two-loop external $\phi_3$ leg corrections to the $\phi_3\to\chi\bar\chi$ decay width as a function of $m_1$. In the left plot the one- and two-loop results for the decay width are shown relative to the tree-level result, while in the right plot the two-loop corrections are shown relative to the one-loop result. We have chosen $m_2=m_3=1$ TeV, $y_3=1$, $A_{123}=3$~TeV, while the other trilinear couplings are set to zero.}
    \label{fig:toymodel_vs_m12L}
\end{figure}
%%%%%%%%%%% figure %%%%%%%%%%%

Next, in \cref{fig:toymodel_vs_m12L} we investigate the impact of the two-loop external leg corrections for the different possible choices of renormalisation of the trilinear coupling $A_{123}$ and of the scalar masses that were discussed in the previous section. We set the mass of the heavy scalars to be $m_2=m_3=1\tev$ and take $A_{123}=3\text{ TeV}$. In the left plot, the one-loop (blue curve) and two-loop (red curves) results for the $\phi_3\to\chi\bar\chi$ decay width are shown relative to the tree-level result. For the two-loop results the OS scheme has been adopted for the scalar masses, and $A_{123}$ is renormalised in either the \msbar (dashed), the OS (solid), or the ``no-log-sq'' (dotted) scheme. The two-loop corrections employing either an OS or \msbar renormalisation for the trilinear coupling $A_{123}$ turn out to be numerically significant. For $m_1=10^{-3}\text{ GeV}$ they amount to $+25.4\%$ and $+21.6\%$ in the OS and \msbar schemes, respectively, as compared to $-73.0\%$ at one-loop order, entering $\hat\Gamma(\phi_3\to\chi\bar\chi)$ with the opposite sign than the one-loop corrections. On the other hand, in the ``no-log-sq'' scheme --- devised to eliminate the $\ln^2$ term in $\hat\Gamma(\phi_3\to\chi\bar\chi)$ --- the two-loop corrections are noticeably smaller (only $-7\%$ for $m_1=10^{-3}\text{ GeV}$) and have the same sign as their one-loop counterparts. It should be noted, however, that the red curves cannot be compared directly to each other, since they correspond to different physical situations as a consequence of the different renormalisation schemes used (we have not incorporated the shifts in the numerical values of $A_{123}$ that are induced by the different renormalisation schemes).

In the right plot of \cref{fig:toymodel_vs_m12L} the two-loop corrections are shown relative to the one-loop result for the considered three different choices of the renormalisation of $A_{123}$ as well as for a complete \msbar renormalisation (of the masses and of $A_{123}$), using two different values of the renormalisation scale $Q$. While the three choices for the renormalisation of $A_{123}$ together with an on-shell renormalisation of the masses give rise to the moderate two-loop effects that are shown in more detail in the left plot, one can see that the two-loop corrections in the full \msbar scheme yield unphysically large effects. Those corrections are many orders of magnitude larger than at one loop and furthermore exhibit a huge scale dependence. These unphysically large effects arise as a consequence of the severe nature of the terms enhanced by $1/m_1$ and $1/m_1^2$ in the derivatives of the $T_{11234}$ integrals, see the expressions in \cref{app:deriv_loopfn}. Thus, our analytical and numerical results highlight the importance of adopting an on-shell renormalisation for the masses entering the computation at the one-loop order, in order to avoid unphysical power-enhanced terms at the two-loop level. This situation is similar e.g.\ to the known occurrence of non-decoupling effects in the context of Higgs-mass calculations in supersymmetric models that is encountered for scenarios with heavy gluinos if the parameters in the scalar top sector are renormalised in the $\overline{\text{DR}}$ scheme, see for instance \ccite{Heinemeyer:1998np, Degrassi:2001yf, Braathen:2016mmb, Bahl:2019hmm,Bahl:2019wzx}.

%%%%%%%%%%% figure %%%%%%%%%%%
\begin{figure}
    \centering
    \includegraphics[width=\textwidth]{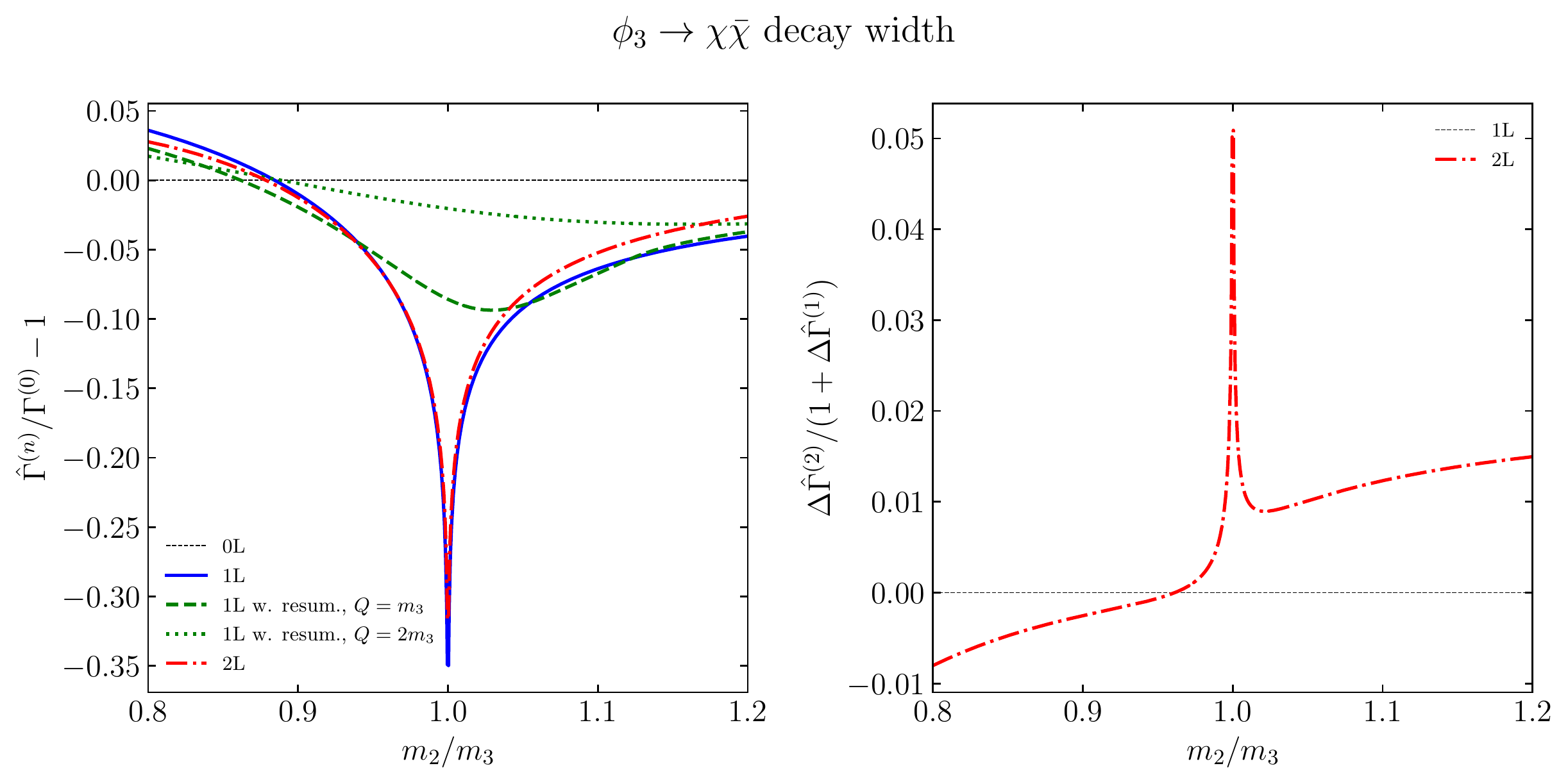}
    \caption{One- and two-loop external $\phi_3$ leg corrections to the $\phi_3\to\chi\bar\chi$ decay width in the toy model of \cref{sec:toy_model} as a function of the ratio $m_2/m_3$. In the left plot the one-loop result with and without resummation of the $\phi_1$ contributions (with different choices of the renormalisation scale $Q$) and the two-loop result for the decay width are shown relative to the tree-level result, while in the right plot the two-loop corrections are shown relative to the one-loop result. The parameters are $m_3=500\text{ GeV}$, $m_\chi=200\text{ GeV}$, $\lambda_{1122}=1$, $\lambda_{1133}=1.2$, and $A_{123}=1500$~GeV. An \msbar renormalisation scheme is employed for $A_{123}$ in the two-loop decay width results.}
    \label{fig:toymodel_vs_m2/m3}
\end{figure}
%%%%%%%%%%% figure %%%%%%%%%%%

As a final step of our numerical analysis of the toy model we show in \cref{fig:toymodel_vs_m2/m3} our results for the other mass configuration where $m_1=0$ and there is a mass splitting $\epsilon=m_3^2-m_2^2$. In our one-loop analysis we had seen that the mass splitting $\epsilon=m_3^2-m_2^2$ acts as an IR regulator for this mass configuration. The results presented in \cref{fig:toymodel_vs_m2/m3} are obtained using the analytical expressions for the derivatives of two-loop self-energies given in \cref{app:deriv_case2}, and adopting an OS scheme for all scalar masses. Because both the two-loop integral $T_{11234}(p^2,m_1^2,m_1^2,m^2+\epsilon,m^2+\epsilon,m^2)$ and its derivative with respect to $p^2$ are always IR divergent for $m_1=0$ (see \cref{eq:loopfn_exp_caseII2} below), $\epsilon$ does not suffice to regulate all IR divergences at two loops for this mass configuration. We therefore introduce an IR regulator mass squared of $m_\text{IR}^2=10 \text{ GeV}^2$ for $\phi_1$ in order to treat the otherwise divergent two-loop integral. We emphasise that this IR divergence that is caused by the occurrence of a squared massless propagator in $T_{11234}$ is different from the IR divergences that we had encountered in our one-loop analysis and also in the two-loop analysis for the other mass configuration. This can been seen from the fact that the divergence appears already in the self-energy and not only in its derivative. In fact, this is precisely the situation encountered in the so-called ``Goldstone boson catastrophe'' discussed above. In the prediction for a physical observable these IR divergences are expected to cancel with the corresponding IR divergences in the real radiation contributions at the two-loop level in an analogous way to the one-loop case that we had discussed above. While the sum of those contributions should be free of logarithms involving $m_\text{IR}$, it still contains logarithms involving the small but non-zero quantity $\epsilon$ that can be numerically large. On the left-hand side of \cref{fig:toymodel_vs_m2/m3}, we plot, as a function of the ratio $m_2/m_3$, the relative deviation of the $\phi_3\to\chi\bar\chi$ decay width from its tree-level prediction at the one-loop (blue and green curves) as well as at the two-loop level (red curve). We choose here a parameter point for which $m_3=500$~GeV, $m_\chi=200$ GeV, $A_{123}=3 m_3$, $\lambda_{1122}=1$, and $\lambda_{1133}=1.2$, while the other trilinear couplings are set to zero for simplicity. As expected, we find the external leg corrections to be divergent when $m_2/m_3\to 1$ at one- and two-loop order. The two green curves show the one-loop result incorporating a resummation of $\phi_1$ contributions for $Q=m_3$ (dashed green curve) and $Q=2m_3$ (dotted green curve). As for the other mass configuration the resummation regulates the IR divergence, but again we find that the resummation procedure exhibits a large dependence on the renormalisation scale $Q$. This difference originates from the quite significant difference in $\Delta m_1^2$ between the two cases. It stems from the dependence on $Q$ in $\Delta m_1^2$, or in other words from higher-order contributions from $\phi_1$ self-energy insertions. On the right-hand side of \cref{fig:toymodel_vs_m2/m3}, we illustrate the size of the two-loop correction relative to the one-loop result. The more rapid divergence of the two-loop corrections is a consequence of the $\ln^2\epsilon$ term.

To summarize our study of the toy model, we have identified --- at the one- and two-loop level --- the potentially dangerous logarithmic and squared-logarithmic contributions arising from external-leg contributions to a process involving a heavy scalar as an external state. We have shown that in the IR limit, in which these terms become divergent, the inclusion of real radiation cures the IR divergences and yields a finite and well-defined result. However, should the experimental resolution allow the separation of the real-radiation contributions (this could happen if either of the mass parameters that we defined as $\epsilon$ has a non-zero, but relatively small, value), large logarithmic and squared-logarithmic terms remain in the computed decay width. We have also illustrated how a resummation of higher-order contributions involving the light scalar can in principle alleviate the IR problem, although this method is in the present calculation plagued by large theoretical uncertainties arising from the problem of relating the prediction to a well-defined physical observable. Furthermore, we have discussed several possible choices of renormalisation schemes for the parameters entering this calculation, and their impact on the size of the remaining large logarithms. In this context, we pointed out the importance of renormalising the involved masses on-shell in order to avoid unphysically large power-enhanced corrections. At the same time, we found that, within the class of corrections that we considered, the renormalisation scale dependence of the contributions to $A_{123}$ vanishes, and the differences between the discussed options for renormalising this coupling are numerically moderate. While the two-loop effects remain smaller than those at one-loop order, their impact can become significant for large mass hierarchies, i.e.\ close to the IR limit. This is in particular the case if the \msbar or OS schemes are adopted for the renormalisation of $A_{123}$, while as expected a scheme that was specifically devised to absorb squared-logarithmic terms yields the smallest effects at two-loop order.

%% file: sec_applications.tex
In this Section, we discuss several concrete BSM models where large logarithms appear in the external leg corrections in an analogous way to the toy model as described in \cref{sec:wfr_ll}.

%%%%%%%%%%%

\subsection{MSSM}

%%%%%%%%%%% figure %%%%%%%%%%%
\begin{figure}
\centering
\includegraphics[scale=1]{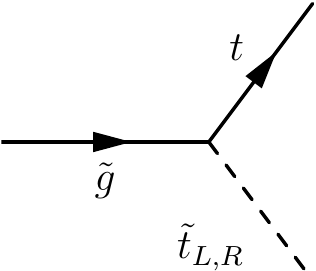}
\caption{Decay of a gluino into a stop and a top quark}
\label{fig:gluino_stop_top_decay}
\end{figure}
%%%%%%%%%%% figure %%%%%%%%%%%

As a first example, we consider corrections to processes involving external scalar top quarks, i.e.\ the superpartners of the top quark, in the MSSM. One such process is the decay of a gluino into a stop and a top quark as shown \cref{fig:gluino_stop_top_decay}. The following discussion is straightforwardly transferable to other processes involving scalar top quarks on an external leg.\footnote{The same type of corrections appears in the widely used OS scheme of the trilinear stop coupling $A_t$~\cite{Degrassi:2001yf,Brignole:2001jy,Dedes:2003km,Heinemeyer:2004xw}. These corrections induce large logarithms entering the prediction for the mass of the SM-like Higgs boson. This type of logarithms cannot be resummed by integrating out the heavy scalar top quarks~\cite{Bahl:2016brp,Sobolev:2020cjh,Slavich:2020zjv}.}

In the limit of vanishing electroweak gauge couplings, the stop mass matrix is given by
\begin{align}
\textbf{M}_{\tilde t} =
\begin{pmatrix}
m_{\tilde t_L}^2 + m_t^2  & m_t X_t                 \\
                 m_t X_t & m_{\tilde t_R}^2 + m_t^2
\end{pmatrix},\label{eq:stop_mass_matrix}
\end{align}
where $m_{\tilde t_L}$ and $m_{\tilde t_R}$ are the stop soft SUSY-breaking mass parameters,
$m_t$ is the top-quark mass and $X_t \equiv A_t - \mu/\tan\beta$ ($A_t$ is the trilinear stop coupling, $\mu$ is the Higgsino mass parameter, and $\tan\beta \equiv v_2/v_1$ is the ratio of the vacuum expectation values of the
two Higgs doublets). Here (and also for the rest of this Section), we assume all parameters to be real for simplicity.

%%%%%%%%%%%

\subsubsection{\texorpdfstring{$Y_t$}{Yt} terms}

First, we aim at calculating all corrections to the external stop leg that are leading in powers of the trilinear coupling $Y_t \equiv A_t + \mu \tan\beta$, which couples the heavy Higgs bosons $H$, $A$, and $H^\pm$ to the stops. We assume the heavy Higgs bosons to be much heavier than the electroweak scale but still lighter than the stops.

In order to simplify our discussion, we work in the approximation of vanishing electroweak gauge couplings. We neglect all terms that are proportional to the electroweak scale by setting the vacuum expectation value $v^2 \equiv v_1^2 + v_2^2$ to zero. In this approximation the stops do not mix, since the off-diagonal entry of the stop mass matrix is zero (i.e., $m_t = 0$ in this limit). Consequently, the left- and right-handed stop chirality eigenstates ($\tilde t_L$ and $\tilde t_R$) are also mass eigenstates.

In this approximation, all relevant couplings between the heavy Higgs bosons and the stops are given by
\begin{subequations}
\begin{align}
    c(H\tilde t_L \tilde t_L) &= c(H\tilde t_R \tilde t_R) = c(A\tilde t_L \tilde t_L) = c(A\tilde t_R \tilde t_R) = 0,\\
    c(H\tilde t_L \tilde t_R) &= - \frac{1}{\sqrt{2}} h_t c_\beta Y_t,\\
    c(A\tilde t_L \tilde t_R) &= - c(A\tilde t_R \tilde t_L) = \frac{1}{\sqrt{2}} h_t c_\beta Y_t,\\
    c(H^+\tilde t_R \tilde b_R) &= c(H^+\tilde t_L \tilde b_L) = c(H^+\tilde t_L \tilde b_R) = 0,\\
    c(H^+\tilde t_R \tilde b_L) &= - h_t c_\beta Y_t,
\end{align}
\end{subequations}
where $h_t$ is the MSSM top-Yukawa coupling, and $c_\beta \equiv \cos\beta$. Note that the couplings of the charged Higgs bosons involve only $\tilde t_R$.

%%%%%%%%%%% figure %%%%%%%%%%%
\begin{figure}
\centering
\begin{subfigure}[b]{.4\textwidth}\centering
\includegraphics[scale=1]{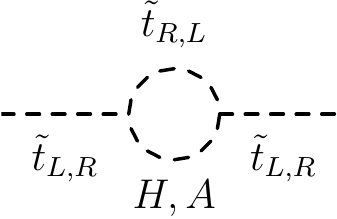}
\end{subfigure}
\begin{subfigure}[b]{.4\textwidth}\centering
\includegraphics[scale=1]{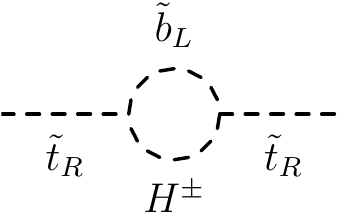}
\end{subfigure}
\caption{\textit{Left}: Neutral heavy Higgs contribution to the left- and right-handed stop self-energies. \textit{Right}: Charged Higgs contribution to the right-handed stop self-energies.}
\label{fig:B0_1L_MSSM}
\end{figure}
%%%%%%%%%%% figure %%%%%%%%%%%

As a consequence of the coupling structure, the off-diagonal stop self-energy is zero,
\begin{align}
\Sigma_{\tilde t_L\tilde t_R} = \Sigma_{\tilde t_R\tilde t_L} = 0.
\end{align}
The diagonal left- and right-handed stop self-energies receive contributions from the $H$ and $A$ bosons (see the left plot of \cref{fig:B0_1L_MSSM}). The diagonal right-handed stop self-energy receives in addition a contribution from the charged Higgs boson (see the right plot of \cref{fig:B0_1L_MSSM}).

The leading contribution in powers of $Y_t$ to the stop self-energies at the one-loop level is given by
\begin{subequations}
\begin{align}
\hat\Sigma_{\tilde t_L\tilde t_L}^{(1)}(p^2) &= k h_t^2 c_\beta^2 Y_t^2 B_0(p^2, m_A^2, \msusy^2),\\
%%%%%
\hat\Sigma_{\tilde t_R\tilde t_R}^{(1)}(p^2) &= 2 k h_t^2 c_\beta^2 Y_t^2 B_0(p^2, m_A^2, \msusy^2),
\end{align}
\end{subequations}
where we set $m_{\tilde t_L} = m_{\tilde t_R} = \msusy$, and $m_A$ is the mass of the heavy Higgs bosons. For the leading genuine two-loop corrections, we obtain
\begin{subequations}
\begin{align}
\hat\Sigma_{\tilde t_L\tilde t_L}^{(2,\text{genuine})}(p^2) &= k^2 h_t^4 c_\beta^4 Y_t^4 \Big(3 T_{11234}(p^2,m_A^2,m_A^2,\msusy^2,\msusy^2,\msusy^2) \nonumber\\
&\hspace{2.6cm} + 2 T_{11234}(p^2,\msusy^2,\msusy^2,m_A^2,\msusy^2,m_A^2)\Big),\\
%%%%%%%%%%%%
\hat\Sigma_{\tilde t_R\tilde t_R}^{(2,\text{genuine})}(p^2) &= 2 k^2 h_t^4 c_\beta^4 Y_t^4 \Big(3 T_{11234}(p^2,m_A^2,m_A^2,\msusy^2,\msusy^2,\msusy^2) \nonumber\\
&\hspace{2.7cm} + T_{11234}(p^2,\msusy^2,\msusy^2,m_A^2,\msusy^2,m_A^2)\Big).
\end{align}
\end{subequations}
The subloop renormalization is given by
\begin{subequations}
\begin{align}
\hat\Sigma_{\tilde t_L\tilde t_L}^{(2,\text{subloop})}(p^2) &= k h_t^2 c_\beta^2 Y_t^2 \bigg[C_0(0,p^2,p^2,m_A^2,m_A^2,\msusy^2) \delta^{(1)}m_A^2 \nonumber\\
&\hspace{2.4cm}+ C_0(0,p^2,p^2,m_A^2,\msusy^2,\msusy^2) \delta^{(1)}m_{\tilde t_R}^2   \nonumber\\
&\hspace{2.4cm} + B_0(p^2, m_A^2, \msusy^2) \Big(\frac{2\delta^{(1)}(h_t c_\beta Y_t)}{h_t c_\beta Y_t} + \delta^{(1)} Z_{\tilde t_L}\Big)\bigg],\\
%%%%%%%%%%%%
\hat\Sigma_{\tilde t_R\tilde t_R}^{(2,\text{subloop})}(p^2) &= k h_t^2 c_\beta^2 Y_t^2 \bigg[2 C_0(0,p^2,p^2,m_A^2,m_A^2,\msusy^2) \delta^{(1)}m_A^2 \nonumber\\
&\hspace{2.4cm}+ C_0(0,p^2,p^2,m_A^2,\msusy^2,\msusy^2) (\delta^{(1)}m_{\tilde t_L}^2 + \delta^{(1)}m_{\tilde b_L}^2)  \nonumber\\
&\hspace{2.4cm} + 2 B_0(p^2, m_A^2, \msusy^2) \Big(2\frac{\delta^{(1)}(h_t c_\beta Y_t)}{h_t c_\beta Y_t} + \delta^{(1)} Z_{\tilde t_R}\Big)\bigg],
\end{align}
\end{subequations}
where $\delta^{(1)}m_A^2$ is the one-loop mass counterterm of the heavy Higgses, $\delta^{(1)}m_{\tilde t_L,\tilde b_R,\tilde t_R}^2$ are the one-loop mass counterterms of the stops and sbottoms, $\delta^{(1)} Z_{\tilde t_{L,R}}$ are the one-loop field renormalisation constants of the stops, and $\delta^{(1)}(h_t c_\beta Y_t)$ is the one-loop counterterm of the Higgs--stop--stop interaction.

Renormalising all counterterms in the OS scheme (apart from the field renomalisation), we obtain
\begin{subequations}
\begin{align}
\delta^{(1)}m_A^2 &= 3 k h_t^2 c_\beta^2 Y_t^2 \Re B_0(m_A^2,\msusy^2,\msusy^2), \label{eq:MSSM_Yt2_ct1}\\
\delta^{(1)}m_{\tilde t_L}^2 &= \delta^{(1)}m_{\tilde b_L}^2 = k h_t^2 c_\beta^2 Y_t^2 \Re B_0(\msusy^2,m_A^2,\msusy^2), \\
\delta^{(1)}m_{\tilde t_R}^2 &= 2 k h_t^2 c_\beta^2 Y_t^2 \Re B_0(\msusy^2,m_A^2,\msusy^2), \\
\frac{\delta^{(1)}(h_t c_\beta Y_t)}{h_t c_\beta Y_t} &= \frac{1}{2}\big(\Re\Sigma_{HH}^{(1)\prime}(m_H^2) + \Re\Sigma_{\tilde t_L\tilde t_L}^{(1)\prime}(m_{\tilde t_L}^2) + \Re\Sigma_{\tilde t_R\tilde t_R}^{(1)\prime}(m_{\tilde t_R}^2)\big),\label{eq:MSSM_Yt2_ct2}
\end{align}
\end{subequations}
with
\begin{subequations}
\begin{align}
\Re\hat\Sigma_{HH}^{(1)\prime}(m_H^2) &= 3 k h_t^2 c_\beta^2 Y_t^2 \Re B_0^\prime(m_A^2,\msusy^2,\msusy^2), \\
\Re\hat\Sigma_{\tilde t_L\tilde t_L}^{(1)\prime}(m_{\tilde t_L}^2) &= k h_t^2 c_\beta^2 Y_t^2 \Re B_0^\prime(\msusy^2,m_A^2,\msusy^2), \\
\Re\hat\Sigma_{\tilde t_R\tilde t_R}^{(1)\prime}(m_{\tilde t_R}^2)&= 2 k h_t^2 c_\beta^2 Y_t^2 \Re B_0^\prime(\msusy^2,m_A^2,\msusy^2)\,.
\end{align}
\end{subequations}
The counterterm for the coupling $h_t c_\beta Y_t$ is fixed by demanding that the amplitude for the process $\tilde t_L \tilde t_R \to H$ up to the one-loop level should coincide with the tree-level result (taking into account contributions at leading order in $Y_t$). All genuine vertex corrections to this process vanish at this order, such that only the external leg corrections enter the definition of the counterterm. Using instead for the renormalisation of $Y_t$ e.g.\ the \DR scheme would not significantly change the numerical size of the two-loop corrections (see the discussion in \cref{sec:wfr_ll_2L}). We employ the \DR scheme for all field renormalisation constants. As already mentioned above, this choice is insignificant, as the \DR field renormalisation constants drop out in the sum of the vertex corrections and the LSZ factor.

Using these expressions, we obtain the following results for the two-loop stop self-energies ($\hat\Sigma^{(2)} = \hat\Sigma^{(2,\text{genuine})} + \hat\Sigma^{(2,\text{subloop})}$),
\begin{subequations}
\begin{align}
\hat\Sigma_{\tilde t_L\tilde t_L}^{(2)}(p^2) ={}& k^2 h_t^4 c_\beta^4 Y_t^4 \Big[3 B_0(p^2,m_A^2,\msusy^2) B_0^\prime(m_A^2,\msusy^2,\msusy^2) \nonumber\\
&\hspace{1.9cm} + 3 B_0(p^2,m_A^2,\msusy^2) B_0^\prime(\msusy^2,m_A^2,\msusy^2) \nonumber\\
&\hspace{1.9cm} + 3 B_0(m_A^2,\msusy^2,\msusy^2) C_0(0,p^2,p^2,m_A^2,m_A^2,\msusy^2) \nonumber\\
&\hspace{1.9cm} + 2 B_0(\msusy^2,m_A^2,\msusy^2) C_0(0,p^2,p^2,m_A^2,\msusy^2,\msusy^2) \nonumber\\
&\hspace{1.9cm} + 3 T_{11234}(p^2,m_A^2,m_A^2,\msusy^2,\msusy^2,\msusy^2) \nonumber\\
&\hspace{1.9cm} + 2 T_{11234}(p^2,\msusy^2,\msusy^2,m_A^2,\msusy^2,m_A^2)\Big], \\
%%%%%
\hat\Sigma_{\tilde t_R\tilde t_R}^{(2)}(p^2) ={}& 2 k^2 h_t^4 c_\beta^4 Y_t^4 \Big[3 B_0(p^2,m_A^2,\msusy^2) B_0^\prime(m_A^2,\msusy^2,\msusy^2) \nonumber\\
&\hspace{2.1cm} + 3 B_0(p^2,m_A^2,\msusy^2) B_0^\prime(\msusy^2,m_A^2,\msusy^2) \nonumber\\
&\hspace{2.1cm} + 3 B_0(m_A^2,\msusy^2,\msusy^2) C_0(0,p^2,p^2,m_A^2,m_A^2,\msusy^2) \nonumber\\
&\hspace{2.1cm} + B_0(\msusy^2,m_A^2,\msusy^2) C_0(0,p^2,p^2,m_A^2,\msusy^2,\msusy^2) \nonumber\\
&\hspace{2.1cm} + 3 T_{11234}(p^2,m_A^2,m_A^2,\msusy^2,\msusy^2,\msusy^2) \nonumber\\
&\hspace{2.1cm} + T_{11234}(p^2,\msusy^2,\msusy^2,m_A^2,\msusy^2,m_A^2)\Big],
\end{align}
\end{subequations}
where we dropped the ``Re'' in the counterterm expressions of \cref{eq:MSSM_Yt2_ct1,eq:MSSM_Yt2_ct2} since all involved loop functions are real for the considered mass hierarchy.

The leading corrections
in powers of $Y_t$ to the gluino decay width are then obtained by taking into account the external stop LSZ factor (see \cref{sec:wfr_ll}). For an external $\tilde t_{L}$ we obtain
\begin{align}
\hat\Gamma_{\tilde g \to t + \tilde t_{L}} &= \Gamma_{\tilde g \to t + \tilde t_{L}}^{(0)}\bigg\{ 1 - \Re\hat\Sigma_{\tilde t_{L}\tilde t_{L}}^{(1)\prime}(m^2_{\tilde t_{L}}) - \Re \hat\Sigma_{\tilde t_{L}\tilde t_{L}}^{(2)\prime}(m^2_{\tilde t_{L}}) \nn\\
&\hspace{2.55cm} + \left(\Re\hat\Sigma_{\tilde t_{L}\tilde t_{L}}^{(1)\prime}(m^2_{\tilde t_{L}})\right)^2 - \frac{1}{2} \left(\Im\hat\Sigma_{\tilde t_{L}\tilde t_{L}}^{(1)\prime}(m^2_{\tilde t_{L}})\right)^2 + \nn\\
&\hspace{2.55cm}+\Im\hat\Sigma_{\tilde t_{L}\tilde t_{L}}^{(1)}(m^2_{\tilde t_{L}})\Im\hat\Sigma_{\tilde t_{L}\tilde t_{L}}^{(1)\prime\prime}(m^2_{\tilde t_{L}})  + \mathcal{O}(k^3)\bigg\} \nn\\
&= \Gamma_{\tilde g \to t + \tilde t_{L}}^{(0)}\bigg\{1 - k h_t^2 c_\beta^2 \yt^2 \bigg[\frac{1}{2}\ln\frac{\msusy^2}{m_A^2}  - 1\bigg] \nn\\
&\hspace{2.55cm} - k^2 h_t^4 c_\beta^4 \yt^4 \bigg[\frac{1}{4}\ln^2\frac{\msusy^2}{m_A^2} - 2 \ln\frac{\msusy}{m_A^2}  + \frac{11}{12}\pi^2 - \frac{35}{12}\bigg] \nn\\
&\hspace{2.55cm} + \mathcal{O}\left(\frac{m_A}{\msusy}\right) + \mathcal{O}(k^3)\bigg\}\,, \label{eq:MSSM_Yt_GammaL}
\end{align}
where on the right-hand side of the second equality we performed an expansion in $m_A/\msusy$ and $\yt \equiv Y_t/\msusy$. For an external $\tilde t_{R}$
we obtain
\begin{align}
\hat\Gamma_{\tilde g \to t + \tilde t_{R}} &= \Gamma_{\tilde g \to t + \tilde t_{R}}^{(0)}\bigg\{ 1 - \Re\hat\Sigma_{\tilde t_{R}\tilde t_{R}}^{(1)\prime}(m^2_{\tilde t_{R}}) - \Re \hat\Sigma_{\tilde t_{R}\tilde t_{R}}^{(2)\prime}(m^2_{\tilde t_{R}}) \nn\\
&\hspace{2.55cm} + \left(\Re\hat\Sigma_{\tilde t_{R}\tilde t_{R}}^{(1)\prime}(m^2_{\tilde t_{R}})\right)^2 - \frac{1}{2}\left(\Im\hat\Sigma_{\tilde t_{R}\tilde t_{R}}^{(1)\prime}(m^2_{\tilde t_{R}})\right)^2  + \nn\\
&\hspace{2.55cm}+\Im\hat\Sigma_{\tilde t_{R}\tilde t_{R}}^{(1)}(m^2_{\tilde t_{R}})\Im\hat\Sigma_{\tilde t_{R}\tilde t_{R}}^{(1)\prime\prime}(m^2_{\tilde t_{R}}) + \mathcal{O}(k^3)\bigg\} \nn\\
&= \Gamma_{\tilde g \to t + \tilde t_{R}}^{(0)}\bigg\{1 - k h_t^2 c_\beta^2 \yt^2 \bigg[\ln\frac{\msusy^2}{m_A^2}  - 2\bigg] \nn\\
&\hspace{2.55cm} - k^2 h_t^4 c_\beta^4 \yt^4 \bigg[\frac{1}{4}\ln^2\frac{\msusy^2}{m_A^2} - 2 \ln\frac{\msusy^2}{m_A^2} + \frac{17}{12}\pi^2 - \frac{47}{6}\bigg]\bigg] \nn\\
&\hspace{2.55cm} + \mathcal{O}\left(\frac{m_A}{\msusy}\right) + \mathcal{O}(k^3)\bigg\}\,. \label{eq:MSSM_Yt_GammaR}
\end{align}
Note that even though the one-loop corrections to the decay widths with $\tilde t_{L}$ and $\tilde t_{R}$ differ from each other, the logarithmic terms in the two-loop corrections are identical.

%%%%%%%%%%% figure %%%%%%%%%%%
\begin{figure}
\centering
\includegraphics[width=0.7\textwidth]{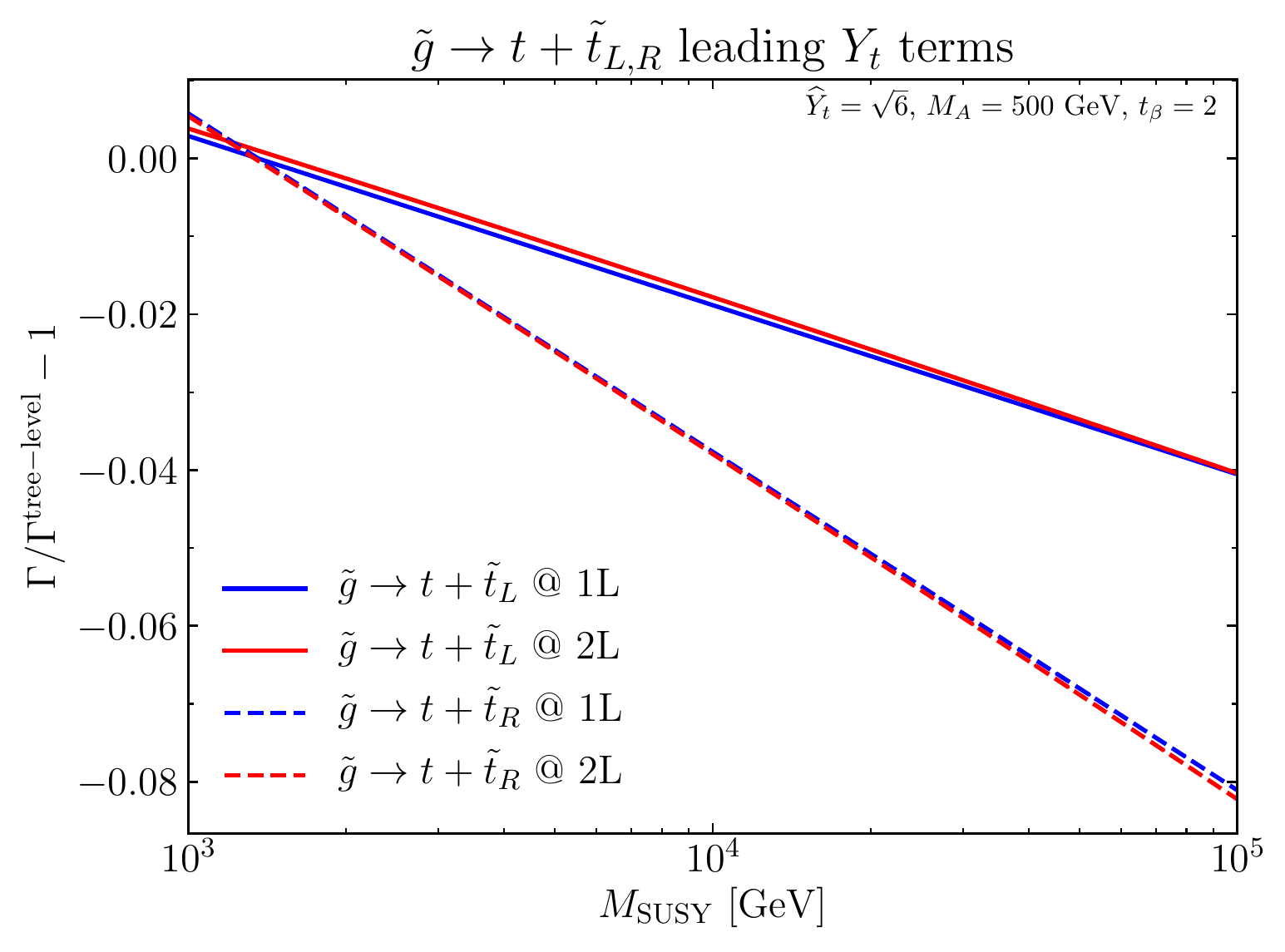}
\caption{The gluino decay widths $\tilde g \to t + \tilde t_{L,R}$ as a function of \msusy, where the final state $t + \tilde t_L$ is displayed by the solid curves and $t + \tilde t_R$ by the dashed curves. The one-loop results including all corrections in leading powers of $Y_t$ are shown relative to the tree-level decay width by the blue curves, while the corresponding two-loop results are displayed by the red curves.}
\label{fig:MSSM_Yt_terms}
\end{figure}
%%%%%%%%%%% figure %%%%%%%%%%%

We show an exemplary numerical evaluation for these corrections in \cref{fig:MSSM_Yt_terms}. For this example, we fix $\yt = \sqrt{6}$, $m_A = 500\gev$, $t_\beta \equiv \tan\beta = 2$, and $m_{\tilde g} = 1.2 \msusy$. We vary \msusy and show the relative deviation from the tree-level results for the gluino decay widths including the one- (blue curves) and two-loop (red curves) corrections in leading powers of $Y_t$. For the decay into $\tilde t_L$ and a top quark (solid curves), the one-loop corrections lower the tree-level decay width by up to $\sim 4\%$ in the considered parameter region. We find the two-loop corrections to have a significantly smaller effect. The inclusion of the one-loop corrections has a larger effect for the decay into $\tilde t_R$ and a top quark (dashed curves) reaching values of up to $\sim 8\%$. This is a consequence of the additional factor of two appearing in the one-loop corrections of \cref{eq:MSSM_Yt_GammaR} with respect to \cref{eq:MSSM_Yt_GammaL}. Also for the gluino decay into $\tilde t_R$ and a top quark, the two-loop corrections have much smaller effect of $\lesssim 0.5\%$ in this case.

The example of the corrections in the MSSM that are enhanced by powers of the trilinear coupling $Y_t$ is a realisation of the situation of the toy model in \cref{sec:wfr_ll_2L} for $m_2^2=m_3^2=m^2$ and $m_1^2=\epsilon$. In the MSSM example the mass $m_1$ of the light scalar in the toy model corresponds to the common mass scale of the MSSM Higgs bosons $H$, $A$, and $H^\pm$, which has been set to $\sim 500\gev$ in the present example. The heavy scale $m$ of the toy model corresponds to $m_{\tilde t_L} = m_{\tilde t_R} = \msusy$. If a gluino is detected in future searches in the decay modes $\tilde g \to t + \tilde t_{L,R}$, there should be significant experimental sensitivity for distinguishing the cases with and without additional $H$, $A$, or $H^\pm$ radiation for this process. Thus, the appropriate theoretical description would not be the IR limit discussed above but rather the case illustrated here where the prediction for a measurable physical observable contains potentially large logarithms containing the ratio of the widely separated scales $\msusy$ and $m_A$.

Regarding the encountered numerical effects, we find these potentially large logarithms to have a sizeable impact at the one-loop level. At the two-loop level, the size of the corrections is moderate for the considered example suggesting that a fixed-order treatment is sufficient and that no resummation of large logarithmic corrections is needed.

%%%%%%%%%%%

\subsubsection{\texorpdfstring{$X_t$}{Xt} terms --- case~1}
\label{sec:MSSM_Xt_case1}

Next, we look at the corrections leading in powers of $X_t$ with $X_t = A_t - \mu/t_\beta$. As for corrections leading in $Y_t$, we work in the unbroken phase of the theory (i.e., $v=0$). In this limit, $X_t$ only appears in the couplings of the light \cp-even Higgs boson $h$, the neutral Goldstone boson $G$, and the charged Goldstone bosons $G^\pm$ to stops and sbottoms,
\begin{subequations}
\begin{align}
    c(h\tilde t_L \tilde t_L) &= c(h\tilde t_R \tilde t_R) = c(G\tilde t_L \tilde t_L) = c(G\tilde t_R \tilde t_R) = 0,\\
    c(h\tilde t_L \tilde t_R) &= \frac{1}{\sqrt{2}} h_t s_\beta X_t,\\
    c(G\tilde t_L \tilde t_R) &= - c(G\tilde t_R \tilde t_L) = \frac{1}{\sqrt{2}} h_t s_\beta X_t,\\
    c(G^+\tilde t_R \tilde b_R) &= c(G^+\tilde t_L \tilde b_L) = c(G^+\tilde t_L \tilde b_R) = 0,\\
    c(G^+\tilde t_R \tilde b_L) &= - h_t s_\beta X_t.
\end{align}
\end{subequations}
As a direct consequence of this coupling structure, we obtain as for the case above
\begin{align}
\Sigma_{\tilde t_L\tilde t_R} = \Sigma_{\tilde t_R\tilde t_L} = 0.
\end{align}
The limit $v=0$ also implies that $h$, $G$, and $G^\pm$ are massless. This situation corresponds to the second mass configuration discussed in \cref{sec:wfr_ll}, where the two heavy particles $\tilde t_L$ and $\tilde t_R$ are separated by a small mass difference $\epsilon$, while the light particle is massless. Accodingly, we take $m_{\tilde t_L} = \msusy$, $m_{\tilde t_R}^2 = \msusy^2 + \epsilon$ and then consider the limit $\epsilon\to 0$.

The stop self-energies obtain corrections proportional to $X_t^2$ from diagrams analogous to the ones in \cref{fig:B0_1L_MSSM} with $H$, $A$, and $H^\pm$ replaced by $h$, $G$, and $G^\pm$, respectively. These corrections are given by
\begin{subequations}
\begin{align}
  \hat\Sigma_{\tilde t_L\tilde t_L}^{(1)}(p^2) &= k h_t^2 s_\beta^2 X_t^2 B_0(p^2,0,m_{\tilde t_R}^2), \\
  \hat\Sigma_{\tilde t_R\tilde t_R}^{(1)}(p^2) &= 2 k h_t^2 s_\beta^2 X_t^2 B_0(p^2,0,m_{\tilde t_L}^2).
\end{align}
\end{subequations}
The genuine two-loop corrections are given by
\begin{subequations}
\begin{align}
\hat\Sigma_{\tilde t_L\tilde t_L}^{(2,\text{genuine})}(p^2) &= k^2 h_t^4 s_\beta^4 X_t^4 \Big(3 T_{11234}(p^2,\mir^2,\mir^2,m_{\tilde t_R}^2,m_{\tilde t_L}^2,m_{\tilde t_R}^2) \nonumber\\
&\hspace{2.6cm} + 2 T_{11234}(p^2,m_{\tilde t_R}^2,m_{\tilde t_R}^2,0,m_{\tilde t_L}^2,0)\Big),\\
%%%%%%%%%%%%
\hat\Sigma_{\tilde t_R\tilde t_R}^{(2,\text{genuine})}(p^2) &= 2 k^2 h_t^4 s_\beta^4 X_t^4 \Big(3 T_{11234}(p^2,\mir^2,\mir^2,m_{\tilde t_L}^2,m_{\tilde t_R}^2,m_{\tilde t_L}^2) \nonumber\\
&\hspace{2.6cm} + T_{11234}(p^2,m_{\tilde t_L}^2,m_{\tilde t_L}^2,0,m_{\tilde t_R}^2,0)\Big).
\end{align}
\end{subequations}
As for the $Y_t$ terms, all $T_{12345}$ integrals cancel. Here, we have introduced $\mir$ as an infrared regulator mass for $h$, $G$, and $G^\pm$ to regulate the partially infrared-divergent two-loop corrections where necessary. All explicit expressions given below are expanded in
$\mir/\msusy$.

For the subloop renomalisation, we obtain
\begin{subequations}
\begin{align}
\hat\Sigma_{\tilde t_L\tilde t_L}^{(2,\text{subloop})}(p^2) &= k h_t^2 s_\beta^2 X_t^2 \bigg[C_0(p^2,0,p^2,0,m_{\tilde t_R}^2 ,m_{\tilde t_R}^2 ) \delta^{(1)}m_{\tilde t_R}^2  \nonumber\\
&\hspace{2.4cm} + C_0(0,p^2,p^2,\mir^2,\mir^2 ,m_{\tilde t_R}^2 ) \delta^{(1)}m_{h,G,G^\pm}^2  \nonumber\\
&\hspace{2.4cm} + B_0(p^2, 0, m_{\tilde t_R}^2) \Big(\frac{2\delta^{(1)}(h_t s_\beta X_t)}{h_t s_\beta X_t} + \delta^{(1)} Z_{\tilde t_L}\Big)\bigg],\\
%%%%%%%%%%%%
\hat\Sigma_{\tilde t_R\tilde t_R}^{(2,\text{subloop})}(p^2) &= 2 k h_t^2 s_\beta^2 X_t^2 \bigg[C_0(p^2,0,p^2,0,m_{\tilde t_L}^2 ,m_{\tilde t_L}^2 ) \delta^{(1)}m_{\tilde t_L}^2  \nonumber\\
&\hspace{2.4cm} + C_0(0,p^2,p^2,\mir^2,\mir^2 ,m_{\tilde t_R}^2 ) \delta^{(1)}m_h^2  \nonumber\\
&\hspace{2.4cm} + B_0(p^2, 0, m_{\tilde t_L}^2) \Big(\frac{2\delta^{(1)}(h_t s_\beta X_t)}{h_t s_\beta X_t} + \delta^{(1)} Z_{\tilde t_R}\Big)\bigg].
\end{align}
\end{subequations}
Here, we also introduce mass counterterms for the light scalars. In contrast to the broken phase of the MSSM, in which the mass counterterms of the light Higgs boson $h$ (and also the Goldstone bosons) are dependent quantities that can be expressed in terms of the counterterms of the independent parameters $m_A$, $\tan\beta$ and $m_Z$, in the unbroken phase, independent mass counterterms can be introduced (by introducing counterterms for the bi-linear potential parameters). It should also be noted that in the unbroken phase the Goldstone bosons are physical particles with a finite mass.

We fix all counterterms in the OS scheme. The counterterm for the coupling combination $h_t s_\beta X_t$ is fixed by demanding that the one-loop matrix element for the $h \to \tilde t_L\tilde t_R$ process is equal to its tree-level value. This yields
\begin{subequations}
\begin{align}
\delta^{(1)}m_{\tilde t_L}^2 &= \delta^{(1)}m_{\tilde b_L}^2 = k h_t^2 s_\beta^2 X_t^2 \Re B_0(m_{\tilde t_L}^2, 0, m_{\tilde t_R}^2), \\
\delta^{(1)}m_{\tilde t_R}^2 &= 2 k h_t^2 s_\beta^2 X_t^2 \Re B_0(m_{\tilde t_R}^2, 0, m_{\tilde t_L}^2), \\
\delta^{(1)}m_{h,G,G^\pm}^2  &= 3 k h_t^2 s_\beta^2 X_t^2 \Re B_0(\mir^2, m_{\tilde t_L}^2, m_{\tilde t_R}^2), \\
\frac{\delta^{(1)}(h_t s_\beta X_t)}{h_t s_\beta X_t} &= \frac{1}{2}\big(\Sigma_{hh}^{(1)\prime}(m_h^2) + \Sigma_{\tilde t_L\tilde t_L}^{(1)\prime}(m_{\tilde t_L}^2) + \Sigma_{\tilde t_R\tilde t_R}^{(1)\prime}(m_{\tilde t_R}^2)\big),\label{eq:MSSM_Xt_CTs}
\end{align}
\end{subequations}
with
\begin{subequations}
\begin{align}
\hat\Sigma_{hh}^{(1)\prime}(m_h^2) &= 3 k h_t^2 s_\beta^2 X_t^2 \Re B_0^\prime(\mir^2, m_{\tilde t_L}^2, m_{\tilde t_R}^2), \\
\hat\Sigma_{\tilde t_L\tilde t_L}^{(1)\prime}(m_{\tilde t_L}^2) &= k h_t^2 s_\beta^2 X_t^2 \Re B_0^\prime(m_{\tilde t_L}^2, 0, m_{\tilde t_R}), \\
\hat\Sigma_{\tilde t_R\tilde t_R}^{(1)\prime}(m_{\tilde t_R}^2)&= 2 k h_t^2 s_\beta^2 X_t^2 \Re B_0^\prime(m_{\tilde t_R}^2, 0, m_{\tilde t_L}^2)\,.
\end{align}
\end{subequations}
It is important to remark that the counterterm $\delta^{(1)}m_{h,G,G^\pm}^2$ is of $\mathcal{O}(\msusy^2)$. This is in contrast to the broken phase of the MSSM in which the loop corrections to the $h$ tree-level mass are of $\mathcal{O}(v^2)$.\footnote{In the broken phase, the $\mathcal{O}(\msusy^2)$ contributions of the $hh$ self energy are cancelled as a consequence of the renormalisation procedure.} Since we assume that the physical masses of $h$, $G$, and $G^\pm$ are zero, we need to cancel the $\mathcal{O}(\msusy^2)$ one-loop corrections by choosing the mass counterterm appropriately.

Summing the genuine and the subloop two-loop contributions, we get
\begin{subequations}
\begin{align}
\hat\Sigma_{\tilde t_L\tilde t_L}^{(2)}(p^2) ={}& k^2 h_t^4 s_\beta^4 X_t^4 \Big[3 B_0(p^2, 0, m_{\tilde t_R}^2) B_0^\prime(0, m_{\tilde t_L}^2, m_{\tilde t_R}^2) \nonumber\\
&\hspace{1.9cm} + 3 B_0(p^2, 0, m_{\tilde t_R}^2) B_0^\prime(m_{\tilde t_R}^2, 0, m_{\tilde t_L}^2) \nonumber\\
&\hspace{1.9cm} + 2 B_0(m_{\tilde t_R}^2, 0, m_{\tilde t_L}^2) C_0(p^2, 0, p^2, 0, m_{\tilde t_R}^2, m_{\tilde t_R}^2) \nonumber\\
&\hspace{1.9cm} + 3 B_0(\mir^2, m_{\tilde t_R}^2, m_{\tilde t_L}^2) C_0(0, p^2, p^2, \mir^2, \mir^2, m_{\tilde t_R}^2) \nonumber\\
&\hspace{1.9cm} + 3 T_{11234}(p^2, \mir^2, \mir^2, m_{\tilde t_R}^2, m_{\tilde t_L}, m_{\tilde t_R}^2) \nonumber\\
&\hspace{1.9cm} + 2 T_{11234}(p^2, m_{\tilde t_R}^2, m_{\tilde t_R}^2, 0, m_{\tilde t_L}^2, 0)\Big], \\
%%%%%
\hat\Sigma_{\tilde t_R\tilde t_R}^{(2)}(p^2) ={}& 2 k^2 h_t^4 s_\beta^4 X_t^4 \Big[3 B_0(p^2, 0, m_{\tilde t_L}^2) B_0^\prime(0, m_{\tilde t_R}^2, m_{\tilde t_L}^2) \nonumber\\
&\hspace{2.2cm} + 3 B_0(p^2, 0, m_{\tilde t_L}^2) B_0^\prime(m_{\tilde t_L}^2, 0, m_{\tilde t_R}^2) \nonumber\\
&\hspace{2.2cm} + B_0(m_{\tilde t_L}^2, 0, m_{\tilde t_R}^2) C_0(p^2, 0, p^2, 0, m_{\tilde t_L}^2, m_{\tilde t_L}^2) \nonumber\\
&\hspace{2.2cm} + 3 B_0(\mir^2, m_{\tilde t_R}^2, m_{\tilde t_L}^2) C_0(0, p^2, p^2, \mir^2, \mir^2, m_{\tilde t_R}^2) \nonumber\\
&\hspace{2.2cm} + 3 T_{11234}(p^2, \mir^2, \mir^2, m_{\tilde t_L}^2, m_{\tilde t_R}, m_{\tilde t_L}^2) \nonumber\\
&\hspace{2.2cm} + T_{11234}(p^2, m_{\tilde t_L}^2, m_{\tilde t_L}^2, 0, m_{\tilde t_R}^2, \mir^2)\Big].
\end{align}
\end{subequations}
The leading $X_t$ corrections to the gluino decay widths up to the two-loop level are then given by (in the limit $\epsilon\to 0$ and $\mir/\msusy \to 0$),
\begin{align}
\hat\Gamma_{\tilde g \to t + \tilde t_{L}} &= \Gamma_{\tilde g \to t + \tilde t_{L}}^{(0)}\bigg\{ 1 - \Re\hat\Sigma_{\tilde t_{L}\tilde t_{L}}^{(1)\prime}(m^2_{\tilde t_{L}}) - \Re \hat\Sigma_{\tilde t_{L}\tilde t_{L}}^{(2)\prime}(m^2_{\tilde t_{L}}) \nn\\
&\hspace{2.55cm} + \left(\Re\hat\Sigma_{\tilde t_{L}\tilde t_{L}}^{(1)\prime}(m^2_{\tilde t_{L}})\right)^2 - \frac{1}{2}\left(\Im\hat\Sigma_{\tilde t_{L}\tilde t_{L}}^{(1)\prime}(m^2_{\tilde t_{L}})\right)^2   + \nn\\
&\hspace{2.55cm}+\Im\hat\Sigma_{\tilde t_{L}\tilde t_{L}}^{(1)}(m^2_{\tilde t_{L}})\Im\hat\Sigma_{\tilde t_{L}\tilde t_{L}}^{(1)\prime\prime}(m^2_{\tilde t_{L}}) + \mathcal{O}(k^3)\bigg\} = \nn\\
&= \Gamma_{\tilde g \to t + \tilde t_{L}}^{(0)}\bigg\{1 - k h_t^2 s_\beta^2 \xt^2 \left[\ln\frac{\msusy^2}{\epsilon} - 1\right] \nn\\
&\hspace{2.55cm} - k^2 h_t^4 s_\beta^4 \xt^4 \bigg[\ln^2\frac{\msusy^2}{\epsilon} - \frac{15}{4} \ln\frac{\msusy^2}{\epsilon} + \frac{1}{2}\ln\frac{\mir^2}{\epsilon} + \frac{1}{6}\pi^2 - \frac{35}{12}\bigg] \nn\\
&\hspace{2.55cm} + \mathcal{O}\left(\frac{\epsilon}{\msusy}\right) + \mathcal{O}(k^3)\bigg\}\,, \label{eq:MSSM_Xt_GammaL}
\end{align}
and
\begin{align}
\hat\Gamma_{\tilde g \to t + \tilde t_{R}} &= \Gamma_{\tilde g \to t + \tilde t_{R}}^{(0)}\bigg\{ 1 - \Re\hat\Sigma_{\tilde t_{R}\tilde t_{R}}^{(1)\prime}(m^2_{\tilde t_{R}}) - \Re \hat\Sigma_{\tilde t_{R}\tilde t_{R}}^{(2)\prime}(m^2_{\tilde t_{R}}) \nn\\
&\hspace{2.55cm} + \left(\Re\hat\Sigma_{\tilde t_{R}\tilde t_{R}}^{(1)\prime}(m^2_{\tilde t_{R}})\right)^2 - \frac{1}{2} \left(\Im\hat\Sigma_{\tilde t_{R}\tilde t_{R}}^{(1)\prime}(m^2_{\tilde t_{R}})\right)^2 + \nn\\
&\hspace{2.55cm}+\Im\hat\Sigma_{\tilde t_{R}\tilde t_{R}}^{(1)}(m^2_{\tilde t_{R}})\Im\hat\Sigma_{\tilde t_{R}\tilde t_{R}}^{(1)\prime\prime}(m^2_{\tilde t_{R}}) + \mathcal{O}(k^3)\bigg\} = \nn\\
&= \Gamma_{\tilde g \to t + \tilde t_{R}}^{(0)}\bigg\{1 - 2 k h_t^2 s_\beta^2 \xt^2 \left[\ln\frac{\msusy^2}{\epsilon} - 1\right] \nn\\
&\hspace{2.55cm} - k^2 h_t^4 s_\beta^4 \xt^4 \bigg[\ln^2\frac{\msusy^2}{\epsilon} - \frac{7}{2} \ln\frac{\msusy^2}{\epsilon} + \frac{1}{4} \ln\frac{\mir^2}{\epsilon} + \frac{5}{3}\pi^2 - \frac{47}{6}\bigg] \nn\\
&\hspace{2.55cm} + \mathcal{O}\left(\frac{\epsilon}{\msusy}\right) + \mathcal{O}(k^3)\bigg\}\,, \label{eq:MSSM_Xt_GammaR}
\end{align}
where $\xt \equiv X_t/\msusy$. Taking into account all two-loop corrections in the considered approximation, all infrared divergences $\propto 1/\mir$ or $\propto 1/\mir^2$ cancel.
In order to cancel the remaining $\ln \mir^2$ terms, additional real light scalar radiation contributions would have to be taken into account, as explained in \cref{sec:wfr_ll_2L}. This real radiation contribution would not introduce any additional large logarithms. Therefore, we set $\mir^2 = \epsilon$ for our numerical results presented below.

%%%%%%%%%%% figure %%%%%%%%%%%
\begin{figure}
\centering
\includegraphics[width=0.7\textwidth]{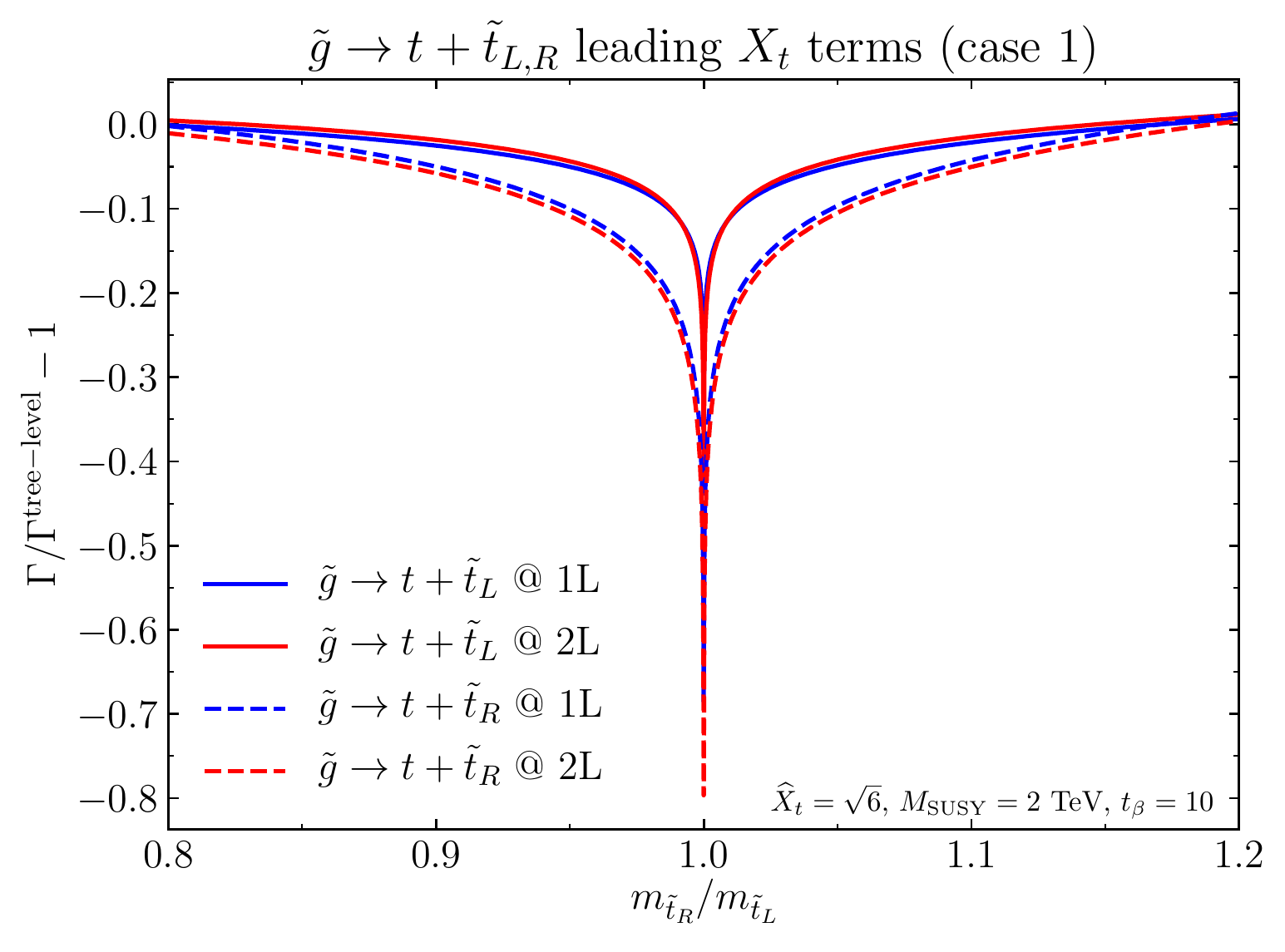}
\caption{ The gluino decay widths $\tilde g \to t + \tilde t_{L,R}$ as a function of $m_{\tilde t_R}/m_{\tilde t_L}$, where the final state $t + \tilde t_L$ is displayed by the solid curves and $t + \tilde t_R$ by the dashed curves. The one-loop results including all corrections in leading powers of $X_t$ are shown relative to the tree-level decay width by the blue curves, while the corresponding two-loop results are displayed by the red curves.}
\label{fig:MSSM_Xt_terms_case1}
\end{figure}
%%%%%%%%%%% figure %%%%%%%%%%%

We assess the numerical size of these corrections in \cref{fig:MSSM_Xt_terms_case1} showing the relative deviation from the tree-level results for the gluino decay widths including the one- (blue curves) and two-loop (red curves) corrections in leading powers of $X_t$ as a function of $m_{\tilde t_R}/m_{\tilde t_L}$. For the gluino decay into $\tilde t_L$ and a top quark (solid curves) the effect of the loop corrections is relatively modest ($\sim 5\%$) for $|1 - m_{\tilde t_R}/m_{\tilde t_L}| \gtrsim 0.05$. If the two stops are nearly mass degenerate --- corresponding to $\epsilon \sim 0$ --- the external leg corrections are numerically large. As discussed in \cref{sec:infrared_limit}, the real radiation contributions should be included if the additional final state particles cannot be resolved experimentally. Close to $m_{\tilde t_L}\sim m_{\tilde t_R}$ also the two-loop corrections have a sizeable impact, reducing the overall size of the loop corrections by up to $\sim 10\%$. For the gluino decay into $\tilde t_R$ and a top quark (dashed curves) the overall size of the one-loop corrections is enhanced because of the additional factor of two in the one-loop part of \cref{eq:MSSM_Xt_GammaL} with respect to \cref{eq:MSSM_Xt_GammaR}. As for the previous MSSM example we find that large logarithms occur in the prediction for the decay width of the gluino, which apart from the parameter region where real radiation contributions should be included are theoretically well under control in the two-loop fixed-order result.

%%%%%%%%%%%

\subsubsection{\texorpdfstring{$X_t$}{Xt} terms --- case~2}
\label{sec:MSSM_Xt_case2}

As a final MSSM example, we show that large logarithms also appear in the broken phase of the theory (i.e., if $v \neq 0$). Here, we assume that $m_{\tilde t_L} = m_{\tilde t_R}$.
In the broken phase, the off-diagonal terms of the stop mass matrix (see \cref{eq:stop_mass_matrix}) are non-zero. After diagonalisation, we denote the mass eigenstates by $\tilde t_1$ and $\tilde t_2$ with $m_{\tilde t_2} \ge m_{\tilde t_1}$. Their couplings involving $X_t$ are given by\footnote{These expressions are valid in the limit $m_A\to\infty$ implying that the mixing angle $\alpha$ of the \cp-even Higgs bosons approaches $\alpha \to \beta - \pi/2$, while in the case of the unbroken phase discussed above we had $\alpha \to 0$.}
\begin{subequations}
\begin{align}
    c(h\tilde t_1 \tilde t_1) &= - c(h\tilde t_2 \tilde t_2) = \frac{1}{\sqrt{2}} h_t s_\beta X_t,\label{eq:MSSM_Xt_case2_couplings1}\\
    c(h\tilde t_1 \tilde t_2) &= c(h\tilde t_2 \tilde t_1) = 0,\\
    c(G\tilde t_1 \tilde t_1) &= c(G\tilde t_2 \tilde t_2) = 0,\\
    c(G\tilde t_1 \tilde t_2) &= - c(G\tilde t_2 \tilde t_1) = \frac{1}{\sqrt{2}} h_t s_\beta X_t,\\
    c(G^+\tilde t_1 \tilde b_1) &= c(G^+\tilde t_2 \tilde b_1) = -\frac{1}{\sqrt{2}} h_t s_\beta X_t,\\
    c(G^+\tilde t_1 \tilde b_2) &= c(G^+\tilde t_2 \tilde b_2) = 0 \label{eq:MSSM_Xt_case2_couplings2}.
\end{align}
\end{subequations}
In contrast to the discussion in \cref{sec:MSSM_Xt_case1}, for the case considered here the couplings are not the only source of the $X_t$ dependence. This is caused by the fact that also the stop masses depend on $X_t$. The difference between the two stop masses is given by $m_{\tilde t_2}^2 - m_{\tilde t_1}^2 = 2 m_t X_t$. Consequently, the stops become mass-degenerate in the limit $v/\msusy\to 0$, and large logarithms appear in analogy to \cref{sec:MSSM_Xt_case1}.

We restrict ourselves here to a discussion at the one-loop level. The stop self-energy contributions containing $h$, $G$ and $G^\pm$ yield at leading order in $X_t$
\begin{subequations}
\begin{align}
\hat\Sigma_{\tilde t_1\tilde t_1}^{(1)}(p^2) = \hat\Sigma_{\tilde t_2\tilde t_2}^{(1)}(p^2) &= \frac{1}{2} k h_t^2 s_\beta^2 X_t^2 \bigg[B_0(p^2,\mir^2,\msusy^2) \nn\\
&\hspace{2.6cm} + B_0(p^2,\mir^2,\msusy^2 - m_t X_t + m_t^2)  \nn\\
&\hspace{2.6cm} + B_0(p^2,\mir^2,\msusy^2 + m_t X_t + m_t^2)\bigg],  \label{eq:MSSM_Xt_case2_diag_ses}\\
%%%%%
\hat\Sigma_{\tilde t_1\tilde t_2}^{(1)}(p^2) = \hat\Sigma_{\tilde t_2\tilde t_1}^{(1)}(p^2) &= \frac{1}{2} k h_t^2 s_\beta^2 X_t^2 B_0(p^2,\mir^2,\msusy^2)\;,
\end{align}
\end{subequations}
where we introduced an infrared regulator mass as in \cref{sec:MSSM_Xt_case1}. In contrast to \cref{sec:MSSM_Xt_case1}, the off-diagonal self energies are non-zero.

We then obtain for the derivative of the diagonal self energies with respect to the external momentum expanded in the limit $v/\msusy\to 0$ (and in the limit $\mir/v \to 0$)
\begin{align}
\Re&\frac{\partial}{\partial p^2}\hat\Sigma_{\tilde t_1\tilde t_1}^{(1)}(p^2)\bigg|_{p^2=m_{\tilde t_1}^2} = \Re\frac{\partial}{\partial p^2}\hat\Sigma_{\tilde t_2\tilde t_2}^{(1)}(p^2)\bigg|_{p^2=m_{\tilde t_2}^2} = \nn\\
=& \frac{1}{2} k h_t^2 s_\beta^2 \xt^2 \bigg[\ln\frac{\msusy^2}{m_t^2} + \frac{1}{2}\ln\frac{\msusy^2}{\mir^2}  - 3 - \ln 2 - 2 \ln|\xt|\bigg] \nonumber\\
& + \mathcal{O}\left(m_t/\msusy\right)\, \label{eq:MSSM_Xt_case2_diag_dses}.
\end{align}
The off-diagonal self energies appear on the external leg without a derivative with respect to the external momentum but divided by the stop mass difference. For them, we obtain
\begin{align}
&\frac{\Re \hat\Sigma_{\tilde t_1\tilde t_2}^{(1)}(m_{\tilde t_1}^2)}{m_{\tilde t_1}^2 - m_{\tilde t_2}^2} = \frac{\Re \hat\Sigma_{\tilde t_2\tilde t_1}^{(1)}(m_{\tilde t_2}^2)}{m_{\tilde t_2}^2 - m_{\tilde t_1}^2} = \nonumber\\
&= \frac{1}{8} k h_t^2 s_\beta^2 \xt^2\left(\ln\frac{\msusy^2}{m_t^2} - 2 \ln|\xt|\right) + \mathcal{O}\left(m_t/\msusy\right)\,.
\end{align}
Also the vertex corrections can in principle generate $X_t^2$ terms even though only one Higgs--stop--stop coupling can appear (see e.g.~\ccite{Bahl:2020jaq}). We checked, however, that this is not the case for the one-loop corrections to the gluino decay into one of the stops and the top quark.

The gluino decay widths taking into account the one-loop corrections in leading powers of $X_t$ are then given by
\begin{align}
\hat\Gamma_{\tilde g \to t + \tilde t_1} ={}& \Gamma_{\tilde g \to t + \tilde t_1}^{(0)}\left[ 1 - \Re \frac{\partial}{\partial p^2}\hat\Sigma_{\tilde t_1\tilde t_1}^{(1)}(p^2)\big|_{p^2=m_{\tilde t_1^2}}\right] - 2 \frac{\Re \hat\Sigma_{\tilde t_1\tilde t_2}^{(1)}(m_{\tilde t_1}^2)}{m_{\tilde t_1}^2 - m_{\tilde t_2}^2}\cdot \Gamma_{\tilde g \to t + \tilde t_2}^{(0)} = \nonumber\\
={}& \Gamma_{\tilde g \to t + \tilde t_1}^{(0)} \left[1 - \frac{1}{2} k h_t^2 s_\beta^2 \xt^2\left(\ln\frac{\msusy^2}{m_t^2} + \frac{1}{2}\ln\frac{\msusy^2}{\mir^2}  - 3 - \ln 2 - 2 \ln|\xt|\right) \right]\nonumber\\
& - \frac{1}{4} k h_t^2 s_\beta^2 \xt^2\left(\ln\frac{\msusy^2}{m_t^2} - 2 \ln|\xt|\right)\cdot \Gamma_{\tilde g \to t + \tilde t_2}^{(0)}\,, \\
\hat\Gamma_{\tilde g \to t + \tilde t_2} ={}& \Gamma_{\tilde g \to t + \tilde t_2}^{(0)}\left[ 1 - \Re \frac{\partial}{\partial p^2}\hat\Sigma_{\tilde t_2\tilde t_2}^{(1)}(p^2)\big|_{p^2=m_{\tilde t_2^2}}\right] - 2 \frac{\Re \hat\Sigma_{\tilde t_2\tilde t_1}^{(1)}(m_{\tilde t_2}^2)}{m_{\tilde t_2}^2 - m_{\tilde t_1}^2}\cdot \Gamma_{\tilde g \to t + \tilde t_1}^{(0)} = \nonumber\\
={}& \Gamma_{\tilde g \to t + \tilde t_2}^{(0)} \left[1 - \frac{1}{2} k h_t^2 s_\beta^2 \xt^2\left(\ln\frac{\msusy^2}{m_t^2} + \frac{1}{2}\ln\frac{\msusy^2}{\mir^2}  - 3 - \ln 2 - 2 \ln|\xt|\right) \right]\nonumber\\
& - \frac{1}{4} k h_t^2 s_\beta^2 \xt^2\left(\ln\frac{\msusy^2}{m_t^2} - 2 \ln|\xt|\right)\cdot \Gamma_{\tilde g \to t + \tilde t_1}^{(0)}\,.
\end{align}
It is important to remark that here the infrared divergences related to the occurrence of the massless Higgs boson $h$ do not cancel (see below regarding the Goldstone contributions). This issue can be addressed by one of the strategies discussed in \cref{sec:infrared_limit}: either external $h$ radiation could be included summing over different final states, or the higher-order mass corrections for the Higgs boson $h$ could be included in analogy to the Goldstone boson resummation.

We focus here on the inclusion of real radiation. Having a close look at the coupling structure in \cref{eq:MSSM_Xt_case2_couplings1,eq:MSSM_Xt_case2_couplings2} and at the diagonal stop self energies in \cref{eq:MSSM_Xt_case2_diag_ses}, it is easy to identify the diagrams with an internal $h$ boson and the same internal stop as the external stops as origin of the infrared divergence. Following the discussion in \cref{sec:infrared_limit}, we obtain for the gluino decay into a top quark,
a stop, and the Higgs boson $h$ in lowest order
\begin{align}
\Gamma^{(0)}_{\tilde g \to t + \tilde t_{1,2} + h} = \Gamma^{(0)}_{\tilde g \to t + \tilde t_{1,2}} \cdot \frac{1}{2}k h_t s_\beta \xt^2\left[\frac{1}{2}\ln\frac{E_\ell^2}{\mir^2} - 1 + \ln 2\right] \, .
\end{align}
The infrared-divergent term exactly cancels with the corresponding term in \cref{eq:MSSM_Xt_case2_diag_dses} rendering the combination $\hat\Gamma_{\tilde g\to t + \tilde t_{1,2}} + \Gamma^{(0)}_{\tilde g\to t + \tilde t_{1,2} + h}$ free of infrared divergences.

It should be noted that the inclusion of real radiation of the neutral and charged Goldstone bosons would not yield additional infrared divergences. Accordingly, the virtual contributions of the neutral and charged Goldstone bosons also do not give rise to large logarithms (see \cref{eq:toy_model_real_rad_m1zero} with $\epsilon = 2 m_t X_t$ and $m_3 = \msusy$).

%%%%%%%%%%% figure %%%%%%%%%%%
\begin{figure}
\centering
\includegraphics[width=0.7\textwidth]{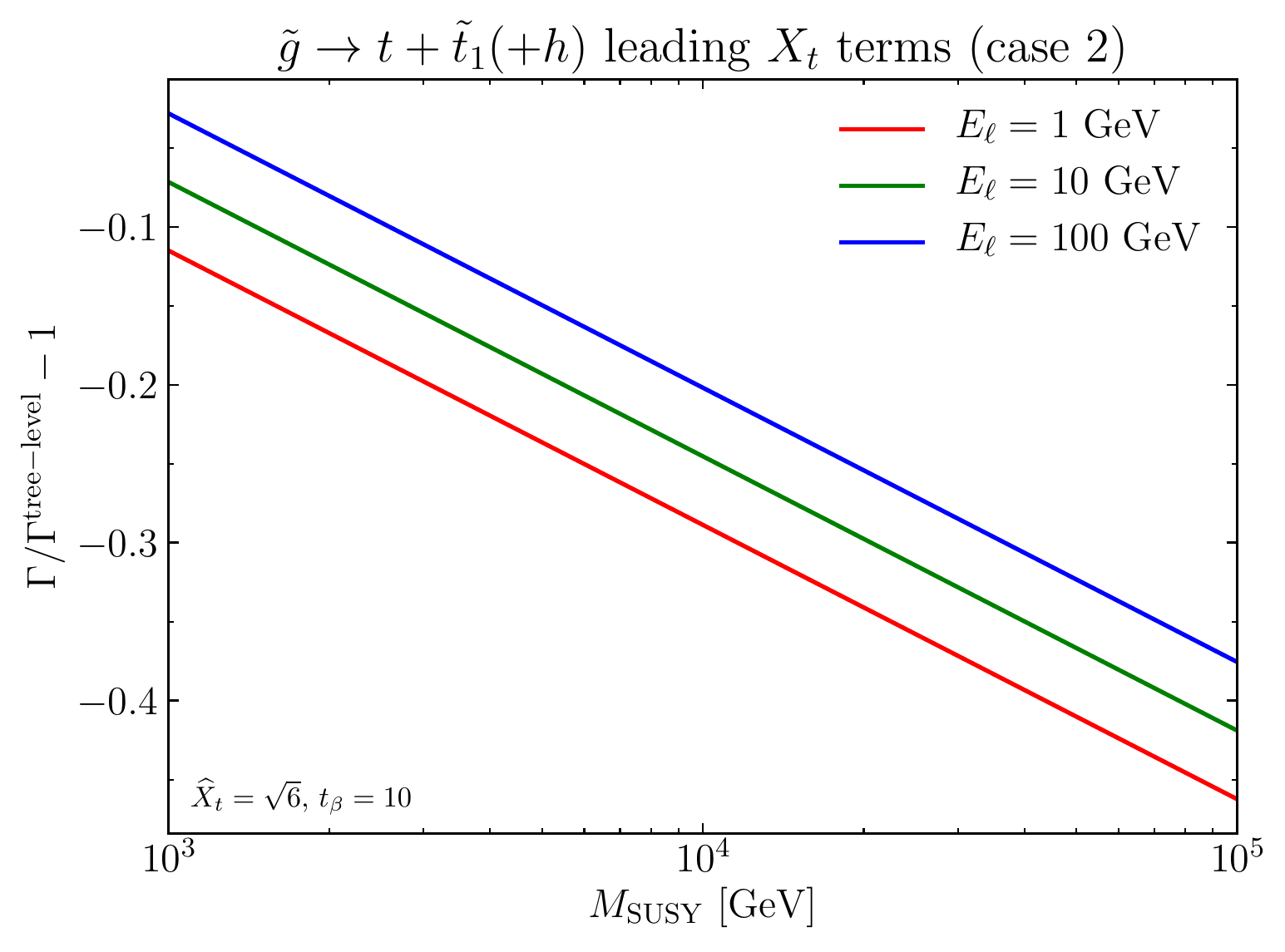}
\caption{ The gluino decay width $\tilde g \to t + \tilde t_1 (+h)$ at one-loop order including the corrections in leading powers of $X_t$ is shown relative to the tree-level decay width as a function of $\msusy$. The result is shown for different detector resolutions: $E_\ell = 1\gev$ (red), $E_\ell = 10\gev$ (green), and $E_\ell = 100\gev$ (blue).}
\label{fig:MSSM_Xt_terms_case2}
\end{figure}
%%%%%%%%%%% figure %%%%%%%%%%%

Taking into account $h$ radiation, we show the numerical result for the decay width of the gluino into to the lighter stop and a top quark in \cref{fig:MSSM_Xt_terms_case2}. We choose $\xt = \sqrt{6}$ and $t_\beta = 10$. Three different values are chosen for the detector resolution: $E_\ell = 1\gev$ (red), $E_\ell = 10\gev$ (green), and $E_\ell = 100\gev$ (blue). The
relative difference between the one-loop corrected and the tree-level decay width increases logarithmically for increasing \msusy, reaching values of up to $\sim -40\%$. Varying the detector resolution by two orders of magnitude corresponds to a variation of the decay width of $\sim 10\%$.

As an alternative to including real radiation diagrams, one could consider to take into account terms proportional to the electroweak gauge couplings. This would render the Higgs and Goldstone boson masses non-zero. While in this case no infrared divergences would appear, large logarithms would still occur in the limit $v/\msusy \to 0$, which corresponds to taking simultaneously the limits $m_1 \to 0$ and $m_2\to m_3$ in \cref{EQ:B0p_m1zero}.

Similarly to the examples discussed above, we expect the two-loop corrections to be much smaller than the one-loop corrections evaluated here and thus a fixed-order treatment should be sufficient for obtaining a percent-level precision.

%%%%%%%%%%%%%%%%%%%%%%%%%%%%%%%%%%%%%%%%%%%%%%%%%%%%%%%

\subsection{N2HDM}
\label{sec:N2HDM}

We now turn to a second example of a frequently discussed BSM model in which potentially large logarithms arising from wave-function normalisation contributions may appear, namely the N2HDM, i.e.\ a real-singlet extension of the Two-Higgs-Doublet Model. We consider once again corrections to processes involving a heavy scalar on an external leg. Specifically, we discuss the decay process $h_3\to\tau^+\tau^-$, where $h_3$ is the heaviest of the three CP-even mass eigenstates (which we will assume to be mostly doublet-like).

%%%%%%%%%%%

\subsubsection{The model}

We start with a \cp-invariant tree-level potential written in terms of two $SU(2)_L$ doublets $\Phi_1,\ \Phi_2$ of hypercharge $1/2$ and a singlet $\Phi_S$ as
\begin{align}
    V^{(0)}=&\ m_{11}^2|\Phi_1|^2+ m_{22}^2|\Phi_2|^2-m_{12}^2\big(\Phi_1^\dagger\Phi_2+\text{h.c.}\big)\nn\\
    &+\frac12\lambda_1|\Phi_1|^4+\frac12\lambda_2|\Phi_2|^4+\lambda_3|\Phi_1|^2|\Phi_2|^2+\lambda_4|\Phi_2^\dagger\Phi_1|^2+\frac12\lambda_5\big((\Phi_1^\dagger\Phi_2)^2+\text{h.c.}\big)\nn\\
    &+\frac12m_S^2\Phi_S^2+\frac16a_S\Phi_S^3+\frac1{24}\lambda_S|\Phi_S|^4+\frac12a_{1S}|\Phi_1|^2\Phi_S+\frac12a_{2S}|\Phi_2|^2\Phi_S\nn\\
    &+\frac16\lambda_{1S}|\Phi_1|^2\Phi_S^2+\frac12\lambda_{2S}|\Phi_2|^2\Phi_S^2\,.
\end{align}
For this model, we derived a \texttt{FeynArts} model file using \texttt{SARAH}~\cite{Staub:2009bi,Staub:2010jh,Staub:2010jh,Staub:2013tta}.

In the following we will use the shorthand notations
\begin{align}
\label{EQ:n2hdm_coupsdef}
 X_a&\equiv \frac{a_{1S}-a_{2S}}{4}\, ,\nn\\
 Y_a&\equiv \frac{a_{1S}s_\beta^2+a_{2S}c_\beta^2}{4}\, ,\nn\\
 Z_a&\equiv \frac{a_S}{4}-Y_a
\end{align}
for combinations of the trilinear couplings that appear in the calculated self-energies, where $\tan\beta$ denotes the ratio of the vacuum expectation values of the two Higgs doublets.

%%%%%%%%%%%

\subsubsection{\texorpdfstring{$X_a^4$}{Xa\^{}4} external leg corrections to the \texorpdfstring{$h_3\to\tau^+\tau^-$}{h3 -> tau tau} process}

Restricting ourselves to contributions involving only powers of $X_a$ --- defined in terms of Lagrangian trilinear couplings in \cref{EQ:n2hdm_coupsdef} ---
we find for the contributions to the $h_3$ self-energy at one-loop order
\begin{align}
 \hat\Sigma^{(1)}_{h_3h_3}(p^2)=&\ k X_a^2 c_{\alpha_3}^2s_{2\beta}^2 \big[B_0(p^2, m_A^2, m_G^2) + 2 B_0(p^2, m_{H^\pm}^2, m_{G^\pm}^2) \nn\\
 &\hspace{3cm}+ 4 s_{\alpha_3}^2B_0(p^2, m_{h_3}^2, m_{h_1}^2)\big] \,,
\end{align}
where $\alpha_3$ is the third \cp-even mixing angle (see e.g.\ Eq. (11) in Ref.\cite{Engeln:2018mbg} for its definition). At two-loop order we find
\begin{align}
 \hat\Sigma^{(2)\text{, genuine}}_{h_3h_3}(p^2) =&\ k^2 X_a^4 c_{\alpha_3}^4s_{2\beta}^4 \big\{T_{12345}(p^2, m_A^2, m_G^2, m_{h_3}^2, m_G^2, m_A^2) \nn\\
  &\hspace{2.5cm}+ 2T_{12345}(p^2, m_{H^\pm}^2, m_{G^\pm}^2, m_{h_3}^2, m_{G^\pm}^2, m_{H^\pm}^2)\nn\\
  &\hspace{2.5cm}+ 16 s_{\alpha_3}^4 T_{12345}(p^2, m_{h_3}^2, m_{h_1}^2, m_{h_3}^2, m_{h_1}^2, m_{h_3}^2) \nn\\
  &\hspace{2.5cm}+T_{11234}(p^2, m_A^2, m_A^2, m_G^2, m_G^2, m_{h_3}^2) \nn\\
  &\hspace{2.5cm}+
 2 T_{11234}(p^2, m_{H^\pm}^2, m_{H^\pm}^2, m_{G^\pm}^2, m_{G^\pm}^2, m_{h_3}^2)\nn\\
 &\hspace{2.5cm} + 4 s_{\alpha_3}^2 \big[T_{11234}(p^2, m_{h_3}^2, m_{h_3}^2, m_{h_1}^2, m_A^2, m_G^2) \nn\\
 &\hspace{3.25cm}+ 2T_{11234}(p^2, m_{h_3}^2, m_{h_3}^2, m_{h_1}^2, m_{H^\pm}^2, m_{G^\pm}^2)\big]\nn\\
 &\hspace{2.5cm} + 16 s_{\alpha_3}^4 T_{11234}(p^2, m_{h_3}^2, m_{h_3}^2, m_{h_1}^2, m_{h_3}^2, m_{h_1}^2)\nn\\
 &\hspace{2.5cm}+T_{11234}(p^2, m_G^2, m_G^2, m_A^2, m_A^2, m_{h_3}^2)\nn\\
  &\hspace{2.5cm}+  2 T_{11234}(p^2, m_{G^\pm}^2, m_{G^\pm}^2, m_{H^\pm}^2, m_{H^\pm}^2, m_{h_3}^2) \nn\\
 &\hspace{2.5cm}+8 s_{\alpha_3}^4 T_{11234}(p^2, m_{h_1}^2, m_{h_1}^2, m_{h_3}^2, m_{h_3}^2, m_{h_3}^2)\big\} \,,
\end{align}
and
\begin{align}
 \hat\Sigma^{(2)\text{, subloop}}_{h_3h_3}(p^2)=\ kX_a^2c_{\alpha_3}^2s_{2\beta}^2 \bigg\{&C_0(0,p^2,p^2,m_A^2,m_A^2, m_G^2)\delta^{(1)}m_A^2\nn\\
 &+C_0(0,p^2,p^2,m_A^2,m_G^2, m_G^2)\delta^{(1)}m_G^2\nn\\
 &+2C_0(0,p^2,p^2,m_{H^\pm}^2,m_{H^\pm}^2, m_{G^\pm}^2)\delta^{(1)}m_{H^\pm}^2\nn\\
 &+2C_0(0,p^2,p^2,m_{H^\pm}^2,m_{G^\pm}^2, m_{G^\pm}^2)\delta^{(1)}m_{G^\pm}^2\nn\\
 &+4s_{\alpha_3}^2\big[C_0(0,p^2,p^2,m_{h_3}^2,m_{h_3}^2, m_{h_1}^2)\delta^{(1)}m_{h_3}^2\nn\\
 &\quad\quad+C_0(0,p^2,p^2,m_{h_3}^2,m_{h_1}^2, m_{h_1}^2)\delta^{(1)}m_{h_1}^2\big]\nn\\
 &+\big[B_0(p^2, m_G^2, m_A^2) + 2 B_0(p^2, m_{G^\pm}^2, m_{H^\pm}^2) \nn\\
 &\hspace{3cm}+ 4 s_{\alpha_3}^2B_0(p^2, m_{h_1}^2, m_{h_3}^2)\big]\nn\\
 &\quad\times\bigg(2\frac{\delta^{(1)}(X_ac_{\alpha_3}s_{2\beta})}{X_ac_{\alpha_3}s_{2\beta}}+\delta^{(1)}Z_{h_3}\bigg)\bigg\}\,.
\end{align}
We adopt an on-shell renormalisation scheme for the masses and the trilinear coupling, together with an \msbar renormalisation of the $h_3$ field
(which drops out in the sum of the vertex corrections and the LSZ factor).
We then obtain
\begin{subequations}
\begin{align}
 \delta^{(1)}m_{h_1}^2&=2 k X_a^2c_{\alpha_3}^2s_{2 \beta}^2s_{\alpha_3}^2\mathrm{Re} B_0(m_{h_1}^2, m_{h_3}^2, m_{h_3}^2)  \,,\\
 \delta^{(1)}m_G^2&=k X_a^2c_{\alpha_3}^2s_{2\beta}^2 \mathrm{Re}B_0(m_G^2, m_A^2,m_{h_3}^2)\,,\\
 \delta^{(1)}m_{G^\pm}^2&=k X_a^2c_{\alpha_3}^2s_{2 \beta}^2 \mathrm{Re}B_0(m_{G^\pm}^2, m_{h_3}^2, m_{H^\pm}^2)\,,\\
 \delta^{(1)}m_{h_3}^2&=k X_a^2 c_{\alpha_3}^2s_{2\beta}^2 \mathrm{Re}\big[ 2 B_0(m_{h_3}^2, m_{G^\pm}^2, m_{H^\pm}^2) + B_0(m_{h_3}^2, m_G^2, m_A^2)\nn\\
 &\hspace{3cm} + 4 s_{\alpha_3}^2B_0(m_{h_3}^2, m_{h_1}^2, m_{h_3}^2)\big]\,,\\
 \delta^{(1)}m_A^2&=k X_a^2c_{\alpha_3}^2s_{2\beta}^2  \mathrm{Re}B_0(m_A^2, m_G^2, m_{h_3}^2) \,,\\
 \delta^{(1)}m_{H^\pm}^2&=k X_a^2c_{\alpha_3}^2s_{2 \beta}^2 \mathrm{Re}B_0(m_{H^\pm}^2, m_{h_3}^2, m_{G^\pm}^2)\,,\\
 \delta^{(1)}Z_{h_1}&=0\,,\\
 \delta^{(1)}Z_{h_3}&=0\,,\\
 \frac{\delta^{(1)}(X_ac_{\alpha_3}s_{2\beta})}{X_ac_{\alpha_3}s_{2\beta}}&=kX_a^2c_{\alpha_3}^2s_{2\beta}^2s_{\alpha_3}^2C_0(k_1^2=m_{h_3}^2,k_2^2=m_{h_1}^2,(k_1+k_2)^2=m_{h_3}^2,m_{h_3}^2,m_{h_3}^2,m_{h_1}^2)\nn\\
 &\hspace{.5cm}+k X_a^2 c_{\alpha_3}^2s_{2\beta}^2 \mathrm{Re}\Big[ 2 B_0^\prime(m_{h_3}^2, m_{G^\pm}^2, m_{H^\pm}^2) + B_0^\prime(m_{h_3}^2, m_G^2, m_A^2)\\
 &\hspace{3cm} + s_{\alpha_3}^2\big(4B_0^\prime(m_{h_3}^2, m_{h_1}^2, m_{h_3}^2)+B_0^\prime(m_{h_1}^2, m_{h_3}^2, m_{h_3}^2)\big)\Big]\nn\,,
\end{align}
\end{subequations}
where we have only included terms that give rise to contributions of order $X_a^4c_{\alpha_3}^4s_{2\beta}^4$ in $\hat\Sigma^{(2)\text{, subloop}}_{h_3h_3}$ and that involve heavy states (we drop terms involving only light states or involving $c_{2\beta}^2$).

In order to investigate the potentially large logarithms arising from the external leg correction in the prediction for the $h_3\to\tau^+\tau^-$ decay width we consider a mass hierarchy where the first two CP-even mass eigenstates as well as the Goldstone bosons are light,\footnote{Note that mass differences between $h_1$, $h_2$, $G$, $G^\pm$ that are much smaller than $\epsilon$ do not impact the results in the following.} with $m_{h_1}^2\sim m_{h_2}^2\sim m_G^2\sim m_{G^\pm}^2\sim \epsilon$, while the remaining scalars are heavy, $m_{h_3}=m_A=m_{H^\pm}=m$. At one-loop order we obtain
\begin{align}
    \hat\Sigma^{(1)\prime }_{h_3h_3}(m^2)=\frac{kX_a^2c_{\alpha_3}^2s_{2\beta}^2}{m^2}(3+4s_{\alpha_3}^2)\bigg[\frac12\ln\frac{m^2}{\epsilon}-1\bigg]\,.
\end{align}
At the two-loop level, for reasons of clarity we study the terms with different powers of $s_{\alpha_3}$ separately. Starting with terms of order $s_{\alpha_3}^4$, we have
\begin{align}
 \hat\Sigma^{(2)\prime}_{h_3h_3}(m^2)\bigg|^{\ \mathcal{O}(s_{\alpha_3}^4)}=
 &\nn\\
 = k^2 X_a^4 c_{\alpha_3}^4s_{2\beta}^4s_{\alpha_3}^4\bigg\{&16\frac{\partial}{\partial p^2}T_{12345}(p^2,m^2,\epsilon,m^2,\epsilon,m^2)\nn\\
 &+16\frac{\partial}{\partial p^2}T_{11234}(p^2,m^2,m^2,\epsilon,m^2,\epsilon)\nn\\
 &+8\frac{\partial}{\partial p^2}T_{11234}(p^2,\epsilon,\epsilon,m^2,m^2,m^2)\nn\\
 &+16\frac{\partial}{\partial p^2}C_0(0,p^2,p^2,m^2,m^2, \epsilon)B_0(m^2,\epsilon,m^2)\nn\\
 &+8\frac{\partial}{\partial p^2}C_0(0,p^2,p^2,m^2,\epsilon, \epsilon)B_0(\epsilon,m^2,m^2)\nn\\
 &+8B_0^\prime(p^2,m^2,\epsilon)\times\bigg[C_0(m^2,\epsilon,m^2,m^2,m^2,\epsilon)+4B_0^\prime(m^2,\epsilon,m^2)\nn\\
 &\hspace{5cm}+B_0^\prime(\epsilon,m^2,m^2)\bigg]\bigg\}\bigg|_{p^2=m^2}\nn\\
 =\frac{2k^2 X_a^4 c_{\alpha_3}^4s_{2\beta}^4s_{\alpha_3}^4}{m^4}\bigg[&\frac{121}{9}+4\sqrt{3}\pi + \frac{7\pi^2}{3} - 8 \pi^2 \ln2 + \frac{2}{3}\left(21+\sqrt{3}\pi\right) \ln\frac{\epsilon}{m^2}\nn\\
 &\hspace{5cm}+ 5 \ln^2\frac{\epsilon}{m^2} + 12 \zeta(3)\bigg]\,.
\end{align}
At order $s_{\alpha_3}^2$, we find
\begin{align}
 \hat\Sigma^{(2)\prime}_{h_3h_3}(m^2)\bigg|^{\ \mathcal{O}(s_{\alpha_3}^2)}=
 &\nn\\
 = k^2 X_a^4 c_{\alpha_3}^4s_{2\beta}^4s_{\alpha_3}^2\bigg\{&12\frac{\partial}{\partial p^2}T_{11234}(p^2,m^2,m^2,\epsilon,m^2,\epsilon)\nn\\
 &+12\frac{\partial}{\partial p^2}C_0(0,p^2,p^2,m^2,m^2, \epsilon)B_0(m^2,\epsilon,m^2)\nn\\
 &+6B_0^\prime(p^2,m^2,\epsilon)\times\bigg[C_0(m^2,\epsilon,m^2,m^2,m^2,\epsilon)+8B_0^\prime(m^2,\epsilon,m^2)\nn\\
 &\hspace{5cm}+B_0^\prime(\epsilon,m^2,m^2)\bigg]\bigg\}\bigg|_{p^2=m^2}\nn\\
 =\frac{k^2 X_a^4 c_{\alpha_3}^4s_{2\beta}^4s_{\alpha_3}^2}{2m^4}\bigg[&94+5\pi^2+4\sqrt{3}\pi+\left(95+2\sqrt{3}\pi\right)\ln\frac{\epsilon}{m^2}+21\ln^2\frac{\epsilon}{m^2}\bigg]\,.
\end{align}
Finally, the terms without $s_{\alpha_3}$ are given by
\begin{align}
 \hat\Sigma^{(2)\prime}_{h_3h_3}(m^2)\bigg|^{\ \mathcal{O}(s_{\alpha_3}^0)}=3 k^2 X_a^4 c_{\alpha_3}^4s_{2\beta}^4\bigg\{&\frac{\partial}{\partial p^2}T_{12345}(p^2,m^2,\epsilon,m^2,\epsilon,m^2)\nn\\
 &+\frac{\partial}{\partial p^2}T_{11234}(p^2,m^2,m^2,\epsilon,m^2,\epsilon)\nn\\
 &+\frac{\partial}{\partial p^2}T_{11234}(p^2,\epsilon,\epsilon,m^2,m^2,m^2)\nn\\
 &+\frac{\partial}{\partial p^2}C_0(0,p^2,p^2,m^2,m^2, \epsilon)B_0(m^2,\epsilon,m^2)\nn\\
 &+\frac{\partial}{\partial p^2}C_0(0,p^2,p^2,m^2,\epsilon, \epsilon)B_0(\epsilon,m^2,m^2)\nn\\
 &+6B_0^\prime(p^2,m^2,\epsilon)B_0^\prime(m^2,\epsilon,m^2)\bigg\}\bigg|_{p^2=m^2}\nn\\
 =\frac{k^2 X_a^4 c_{\alpha_3}^4s_{2\beta}^4}{4m^4}\bigg[&\frac{181}{3}+\frac{9}{2}\pi^2-12\pi^2\ln2+70\ln\frac{\epsilon}{m^2}\nn\\
 &\hspace{2.5cm}+\frac{39}{2}\ln^2\frac{\epsilon}{m^2}+18\zeta(3)\bigg]\,.
\end{align}
Finally, the loop-corrected decay width for the $h_3\to\tau^+\tau^-$ process, incorporating the external leg contributions, is found via the relation 
\begin{align}
 \hat\Gamma(h_3\to\tau^+\tau^-)=\Gamma^{(0)}(h_3\to\tau^+\tau^-)\bigg\{&1-\mathrm{Re}\hat\Sigma_{h_3h_3}^{(1)\prime}(m^2)-\mathrm{Re}\hat\Sigma_{h_3h_3}^{(2)\prime}(m^2)\nn\\
            &+\big(\mathrm{Re}\hat\Sigma_{h_3h_3}^{(1)\prime}(m^2)\big)^2-\frac12\big(\mathrm{Im}\hat\Sigma_{h_3h_3}^{(1)\prime}(m^2)\big)^2\nn\\
            &+\mathrm{Im}\hat\Sigma_{h_3h_3}^{(1)}(m^2)\cdot\mathrm{Im}\hat\Sigma_{h_3h_3}^{(1)\prime\prime}(m^2)+\mathcal{O}(k^3)\bigg\}  \,.
\end{align}
For the sake of brevity, we do not present the complete result, which can be straightforwardly obtained from the expressions in the previous equations.

%%%%%%%%%%%

\subsubsection{Numerical results}

%%%%%%%%%%% figure %%%%%%%%%%%
\begin{figure}
    \centering
    \includegraphics[width=.75\textwidth]{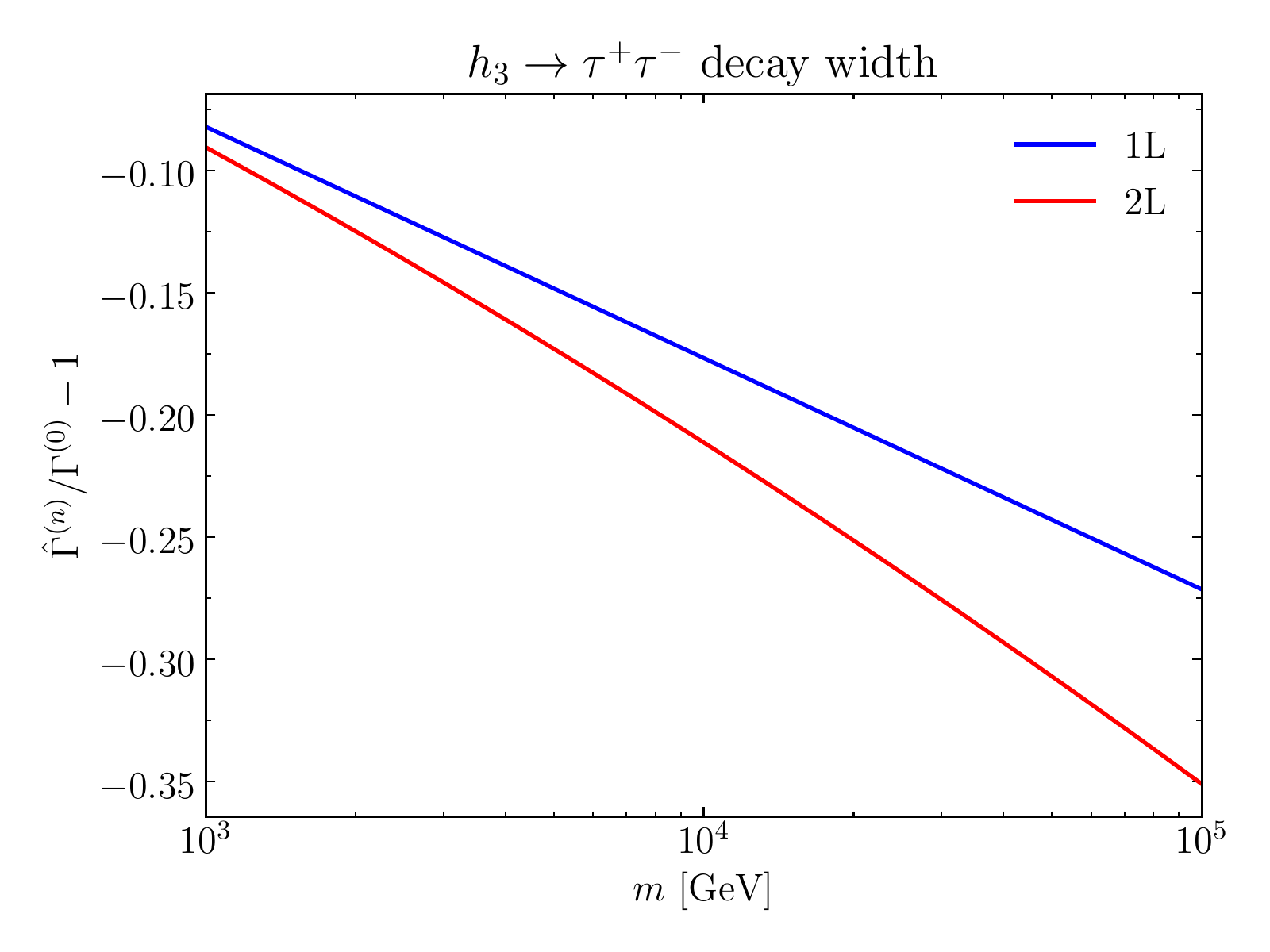}
    \caption{One-loop and two-loop external leg corrections to the $h_3\to\tau^+\tau^-$ decay width in the N2HDM relative to the tree-level result as a function of the heavy mass scale $m$ (see text).
    The input values in this plot are $\epsilon=(50\text{ GeV})^2$, $\tan\beta=1.26$, $s_{\alpha_3}=0.94$, and $X_a=3m$.}
    \label{fig:N2HDM_plot1}
\end{figure}
%%%%%%%%%%% figure %%%%%%%%%%%

%%%%%%%%%%% figure %%%%%%%%%%%
\begin{figure}
    \centering
    \includegraphics[width=.75\textwidth]{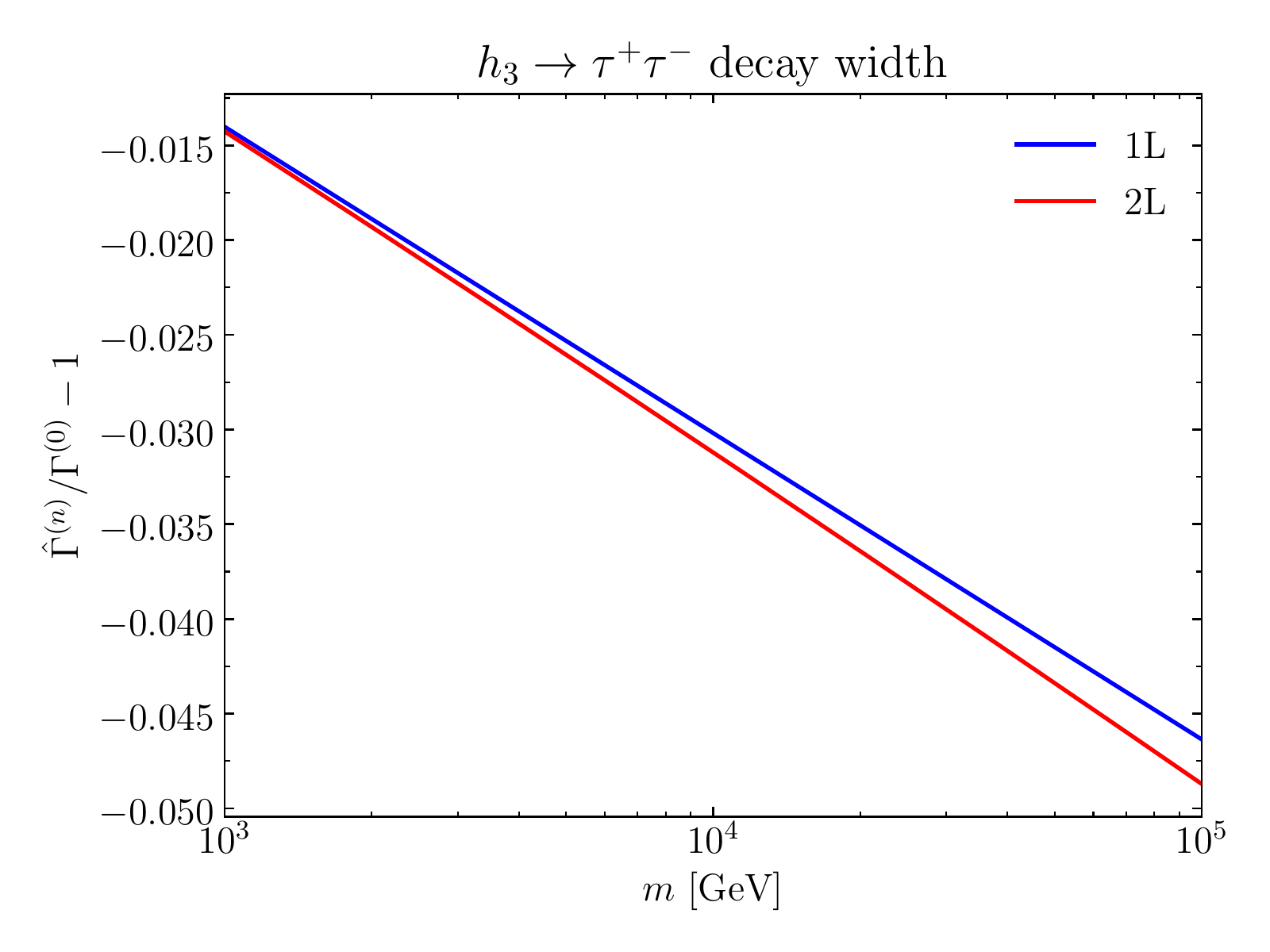}
    \caption{One-loop and two-loop external leg corrections to the $h_3\to\tau^+\tau^-$ decay width in the N2HDM relative to the tree-level result as a function of the heavy mass scale $m$ (see text). The input values in this plot are  $\epsilon=(50\text{ GeV})^2$, $\tan\beta=1.4$, $s_{\alpha_3}=0.99$, and $X_a=3m$.}
    \label{fig:N2HDM_plot2}
\end{figure}
%%%%%%%%%%% figure %%%%%%%%%%%

We now turn to some numerical examples in order to illustrate the possible size of the external leg corrections computed in the previous section. In \cref{fig:N2HDM_plot1,fig:N2HDM_plot2} we show the relative size of the one-loop (blue curve) and two-loop (red curve) external-leg corrections to the $h_3\to\tau^+\tau^-$ decay width with respect to the tree-level result as a function of the heavy mass scale $m$. For these plots a light mass scale of $\sqrt{\epsilon}=50\text{ GeV}$ and a rather large trilinear coupling $X_a=3m$ have been chosen. For $s_{\alpha_3}$ (and $\tan\beta$) we consider two different scenarios: for the first one --- inspired by one of the benchmark points in Ref.~\cite{Biekotter:2019kde} (see table 5 therein) --- we take $s_{\alpha_3}=0.94$. This choice ensures that the factor $c_{\alpha_3}$, entering with the same power as $X_a$, is not too small so that an overall suppression of this class of contributions is avoided. As can be observed in \cref{fig:N2HDM_plot1}, this scenario indeed gives rise to sizeable contributions at the one-loop and the two-loop level. For $m=1\text{ TeV}$, we find effects of $-8.2\%$ at the one-loop and $-0.84\%$ at the two-loop level. The corrections grow to $-27.1\%$ and $-8.0\%$ at one-loop and two-loop order, respectively, for $m=100\text{ TeV}$.

Values of $s_{\alpha_3}$ closer to unity are however more easily reconcilable with the experimental data on the Higgs signal at about $125\gev$, and we therefore consider a further scenario with $s_{\alpha_3}=0.99$, making use of the public code \texttt{ScannerS}~\cite{Muhlleitner:2020wwk}. As shown in \cref{fig:N2HDM_plot2}, in this case the size of the external leg corrections relative to the tree-level prediction for the $h_3\to\tau^+\tau^-$ decay width is (as expected) much smaller than for the first scenario. At one-loop order, the corrections increase from approximately $-1.4\%$ for $m=1\text{ TeV}$ to $-4.6\%$ for $m=100\text{ TeV}$, while at the two-loop level the corrections grow from $-0.025\%$ to $-0.24\%$ in the same range of $m$. The large relative increase of the two-loop corrections for increasing $m$ is due to the appearance of $\ln^2m^2/\epsilon$ terms at the two-loop level.

We briefly comment in this context on the constraints that have been applied for the above two scenarios. The analysis of \ccite{Biekotter:2019kde} was carried out for an N2HDM without trilinear couplings, and the same is also true for the implementation of the N2HDM in the code \texttt{ScannerS}. Concerning the theoretical and experimental constraints that the considered parameter points have to fulfill, we note that theoretical properties such as boundedness-from-below or perturbative unitarity (in the high-energy limit) are not significantly altered by the inclusion of trilinear couplings in the N2HDM scenario that we are considering here. On the other hand, the trilinear couplings can modify the production cross-sections and decay widths, such that a more thorough analysis would be needed in particular for the scenario with $s_{\alpha_3}=0.94$ in order to assess its compatibility with the latest experimental results.

In comparison to the MSSM scenarios discussed in the previous sections, we have found that the relative size of the two-loop corrections is larger in the considered N2HDM scenario. Nevertheless we have found also for the case of the N2HDM that the perturbative expansion remains well-behaved even if the mass ratio entering the logarithmic contributions in the external leg corrections becomes relatively large. In particular, the two-loop corrections to the $h_3\to\tau^+\tau^-$ decay width remain much smaller than their one-loop counterparts. Thus, also for the N2HDM case it appears safe to rely on a fixed-order analysis, and there does not seem to be a pressing need for developing a SCET-like resummation of the logarithmic contributions.

%% file: sec_conclusions.tex
If a new heavy BSM particle (or more than one) is discovered, the characterisation of its properties will be of foremost importance in order to unravel the underlying structure of the observed physics beyond the SM. A crucial ingredient in this context will be precise theoretical predictions to which the experimental measurements can be compared.

One re-occurring issue when calculating loop corrections for heavy BSM particles is the appearance of potentially large logarithmic corrections. In this work, we have pointed out the existence of large Sudakov-like logarithmic contributions related to external-leg corrections of heavy scalar particles in scenarios where in addition at least one light scalar particle is present, giving rise to the possibility of large trilinear couplings between the scalars and a significant hierarchy between the masses of the different scalars.

Working in a simple toy model, containing one light and two heavy scalars coupled to each other with a potentially large trilinear coupling, we discussed in detail how the occurrence of the large logarithmic corrections is related to a limit in which infrared singularities emerge. We showed how these singularities can be regulated by including the radiation of the light scalar particle or by resumming higher-order contributions involving the light scalar. On the other hand, if for the decay of a heavy scalar BSM particle the final states with and without the radiation of the light scalar particle can be experimentally distinguished from each other, potentially large logarithmic corrections involving the ratio of the two widely separated mass scales occur in the prediction for the decay width of the heavy particle.

In order to assess the size of these logarithmic corrections at higher orders, we derived the leading two-loop external-leg corrections in the considered toy model. In this context, we compared different choices of renormalisation schemes for the parameters entering our computation (masses and trilinear coupling). As a result, we found that the choice of an \msbar renormalisation for the involved masses leads to unphysically large corrections that are enhanced by powers of the BSM scale over the electroweak scale. We showed that those huge effects are absent if the involved masses are renormalised in the OS scheme. In the OS scheme, we found the numerical size of the two-loop corrections to be moderate. This implies that a resummation of the Sudakov-like logarithms, which should in principle be possible in SCET-like frameworks, is not expected to be mandatory for obtaining theoretical predictions of sufficient accuracy.

As a further illustration of the qualitative features that we had found in our analysis of the toy model we then considered specific examples of models that are popular for studying possible BSM phenomenology. In the MSSM, we investigated corrections related to external scalar top quarks as appearing e.g.\ for the decay of a gluino into a top quark and its scalar superpartner. As a second example, we discussed the decay of a heavy Higgs boson into two tau leptons in the N2HDM. For both models, we found large one-loop corrections but only moderate effects at the two-loop level. Thus, the analyses in the specific models confirm our results for the toy model indicating that the class of logarithmic contributions that we have investigated here is not expected to spoil the perturbative expansion. The corrections that we have obtained at the two-loop order turned out to be much smaller than their one-loop counterparts, suggesting that a fixed-order analysis at this level should be sufficient for obtaining reliable theoretical predictions.

%% file: app_loopfndef.tex
In this Section, we list the definitions of the one- and two-loop integrals that have been used
in this paper. We first introduce the shorthand notation
\begin{align}
 \mathcal{C}\equiv\frac{(2\pi\mu)^{2\epsilon_{\text{UV}}}}{i\pi^2}\,,
\end{align}
where $\epsilon_{\text{UV}}=(4-d)/2$, and $\mu$ is the regularisation scale (related to the renormalisation scale by the relation $Q^2=4\pi e^{-\gamma_E}\mu^2$). We will also use
\begin{align}
    \lln x\equiv \ln\frac{x}{Q^2}\,,
\end{align}
throughout the following discussion.

We would like to draw the reader's attention to our conventions for denoting loop integrals and loop functions: in the following, we refer to the (unrenormalised) complete loop integrals with bold capitals, such as $\mathbf{A}$, $\mathbf{B}$, $\mathbf{T}$, while we use normal script --- $A_0$, $B_0$, $T$ --- for the \textit{finite} parts of these integrals, defining the loop functions that appear in our computations.

First, the one-loop functions $A_0(x)$ and $B_0(p^2,x,y)$ are defined as the UV-finite parts of the integrals
\begin{align}
    \mathbf{A}(x)\equiv&\ \mathcal{C}\int\frac{d^dq}{q^2-x}\,,\nn\\
    \mathbf{B}(p^2,x,y)\equiv&\ \mathcal{C}\int\frac{d^dq}{(q^2-x)((q+p)^2-y)}\,,
\end{align}
as
\begin{align}
\label{EQ:defAB}
    A_0(x)\equiv&\ \lim_{\epsilon_{\text{UV}}\to0}\bigg[\mathbf{A}(x)-\frac{x}{\epsilon_{\text{UV}}}\bigg]\,,\nn\\
    B_0(p^2,x,y)\equiv&\ \lim_{\epsilon_{\text{UV}}\to0}\bigg[\mathbf{B}(p^2,x,y)-\frac{1}{\epsilon_{\text{UV}}}\bigg]\,,
\end{align}
while the function $C_0$ is equal to the finite integral
\begin{align}
\label{EQ:defC0}
    C_0(p_1^2,p_2^2,(p_1+p_2)^2,x,y,z)\equiv\ \mathcal{C}\int\frac{d^dq}{(q^2-x)((q+p_1)^2-y)((q+p_1+p_2)^2-z)}\,.
\end{align}

Next, the loop functions that enter the class of two-loop external-leg corrections (involving only trilinear couplings) that we are computing in this paper are the finite parts of the following two integrals
\begin{align}
\label{eq:defT11234_T12345}
    &\mathbf{T}_{11234}(p^2,x,y,z,u,v)\equiv\nn\\
    &\equiv \mathcal{C}^2\int\int  \frac{d^dq_1 d^dq_2}{(q_1^2-x)(q_1^2-y)((q_1+p)^2-z)((q_1-q_2)^2-u)(q_2^2-v)}\,,\nn\\
    &\mathbf{T}_{12345}(p^2,x,y,z,u,v)\equiv\nn\\
    &\equiv \mathcal{C}^2\int\int  \frac{d^dq_1 d^dq_2}{(q_1^2-x)((q_1+p)^2-y)((q_1-q_2)^2-z)(q_2^2-u)((q_2+p)^2-v)}\,,
\end{align}
which correspond to $-\mathbf{V}(p^2,z,x,v,u)$ (for degenerate mass arguments $x=y$) and $-\mathbf{M}(p^2,y,v,x,u,z)$ in the conventions of Refs.~\cite{Martin:2003qz, Martin:2003it,Martin:2005qm}, respectively. Note that we explicitly keep the momentum argument throughout this paper. 
We emphasise that we follow here the choice of Refs.~\cite{Martin:2003qz, Martin:2003it,Martin:2005qm} 
(which differs from the usual treatment of $T$-integrals) in that we consider only the finite part of the loop integral, removing also any piece of the form $\mathcal{O}(1/\epsilon_{\text{UV}})\times \mathcal{O}(\epsilon_{\text{UV}})$ in factors therein (in the present calculation, all such terms cancel with similar contributions arising from the subloop renormalisation). 
As a consequence of this choice, we do not need to include the $\mathcal{O}(\epsilon_{\text{UV}})$ part of counterterms entering subloop renormalisation.

When deriving expressions for the derivatives of $T_{11234}$ and $T_{12345}$ in the next section, we will also make use of the integrals
\begin{align}
    \mathbf{T}_{134}(x,y,z)\equiv&\ \mathcal{C}^2\int\int\frac{d^dq_1 d^dq_2}{(q_1^2-x)((q_1-q_2)^2-y)(q_2^2-z)}\,,\nn\\
    \mathbf{T}_{234}(x,y,z)\equiv&\ \mathcal{C}^2\int\int\frac{d^dq_1 d^dq_2}{((q_1+p)^2-x)((q_1-q_2)^2-y)(q_2^2-z)}\,,\nn\\
    \mathbf{T}_{2234}(p^2,x,y,z,u)\equiv&\ \mathcal{C}^2\int\int\frac{d^dq_1 d^dq_2}{((q_1+p)^2-x)((q_1+p)^2-y)((q_1-q_2)^2-z)(q_2^2-u)}\,,\nn\\
    \mathbf{T}_{1234}(p^2,x,y,z,u)\equiv&\ \mathcal{C}^2\int\int\frac{d^dq_1 d^dq_2}{(q_1^2-x)((q_1+p)^2-y)((q_1-q_2)^2-z)(q_2^2-u)}\,.
\end{align}
In terms of the notations of \ccite{Martin:2003qz, Martin:2003it,Martin:2005qm}, we have the following identifications
\begin{align}
   \mathbf{T}_{134}(x,y,z) =&-\mathbf{I}(x,y,z)\,,\nn\\
   \mathbf{T}_{234}(x,y,z) =&-\mathbf{S}(p^2,x,y,z)\,,\nn\\
   \mathbf{T}_{1234}(p^2,x,y,z,u) =&\ \mathbf{U}(p^2,y,x,u,z)\,,\nn\\
   \mathbf{T}_{2234}(p^2,x,x,y,z) =&\ \mathbf{T}(p^2,x,y,z)\,.
\end{align}
The minus signs differing between our conventions and those of \ccite{Martin:2003qz} arise from the fact that \ccite{Martin:2003qz} defines integrals in terms of Euclidean (rather than Lorentzian) momenta. Relations for the subtraction of UV-divergent pieces from these integrals can be found in Eqs.~(2.12)-(2.18) of \ccite{Martin:2003qz}.

%% file: app_derivativesloopfunctions.tex
\allowdisplaybreaks

We present here our derivation of expressions for the derivatives of the two-loop self-energy diagrams $T_{11234}$ and $T_{12345}$ --- shown in \cref{fig:phi3phi3_se_2L} --- which contribute to external leg corrections of heavy particles and contain potentially large terms as a consequence of the mass hierarchy between the involved particles.

%%%%%%%%%%%

\subsection{Setup of the calculation and intermediate results}

Our calculation makes use of the results of \ccite{Martin:2003qz, Martin:2003it,Martin:2005qm}, in which $T_{11234}$ and $T_{12345}$ correspond to $V$ and $M$, respectively. We start with the expressions of derivatives of the two-loop self-energy basis integrals with respect to the external momentum as well as mass arguments --- see in particular Eqs.~(3.22), (4.26), and (4.27) in \ccite{Martin:2003qz} --- and we extract the IR-divergent and (leading) finite terms of the derivatives of $T_{11234}$ and $T_{12345}$ for the two mass configurations
\begin{align*}
    \text{1:}&\quad m_1^2=\epsilon,\ m_2^2=m_3^2=m^2,\\
    \text{2:}&\quad m_1^2=0,\ m_2^2=m^2+\epsilon,\ m_3^2=m^2,
\end{align*}
used to regularise the IR divergence. We derive formulas for the related case 2': $m_1^2=0$, $m_2^2=m^2$, $m_3^2=m^2+\epsilon$ by applying the transformation $\epsilon\to-\epsilon$ and $m^2\to m^2+\epsilon$ to the results of case 2.

We require a number of limiting cases of the loop functions entering the relations for the derivatives: first, with one single mass scale we need
\begin{align}
    T_{234}(x,x,x,x)=&-S(x,x,x,x)=\frac{35}{8}x-\frac{11}{2}x\lln x+\frac{3}{2}x\lln^2x\,,\nn\\
    T_{2234}(x,x,x,x)=&\ T(x,x,x,x)=-\frac12+\frac12\lln^2 x-\lln x\,,\nn\\
    T_{12345}(x,x,0,x,0,x)=&-M(x,0,x,x,0,x)=-\frac1{x}\bigg[\pi^2\ln 2-\frac32\zeta(3)\bigg]\,,
\end{align}
where once again we have retained the external momentum argument explicitly (here $p^2=x$). Next, we also use limits with two distinct mass scales,
\begin{subequations}
\begin{align}
    B_0(x,0,y)=&\ 2-\lln y+\left(\frac{y}{x}-1\right)\ln\left(1-\frac{x}{y}\right)\,,\\
    T_{134}(0,x,y)=&-I(0,x,y)\nn\\
    =& -(x - y) \left[\mathrm{Li}_2 \frac{y}{x} - \lln(x - y)\ln\frac{x}{y} + \frac12 \lln^2 x -  \zeta(2)\right] \nn\\
     &- x \lln x (2 - \lln y) -2 y \lln y+ \frac52(x + y)\,,\nn\\
    T_{234}(s,0,0,x)=&-S(s,0,0,x)\nn\\
    =&\ x\mathrm{Li}_2\left(\frac{s}{x}\right)+\frac12x\lln^2 x-\left(2x-\frac12s\right)\lln x -\frac{x^2-s^2}{2s}\ln\left(1-\frac{s}{x}\right)\nn\\ &-\frac{13}{8}s+(2+\zeta(2))x\,,\\
    T_{234}(x,y,x,y)=&-S(x,y,y,x)\nn\\
    =&\ \frac{3}{8}x+4y-\left(\frac12x-y\right)\lln^2y \nn\\
    &-\left(x+\frac{y^2}{x}-2y\right)\left[\mathrm{Li}_2\left(1-\frac{x}{y}\right)-\zeta(2)\right]\nn\\
    &-\left(\frac32x-y\right)\lln x-5y \lln y+x\lln x\lln y\,,\\
    T_{2234}(s,x,x,0,0)=&\ T(s,x,0,0)\nn\\
    =&\ \mathrm{Li}_2\left(\frac{s}{x}\right)+\frac12\lln^2 x-\lln x+\left(1-\frac{x}{s}\right)\ln\left(1-\frac{s}{x}\right)-\frac12+\zeta(2)\,,\\
    T_{2234}(x,x,x,y,y)=&\ T(x,x,y,y)\nn\\
    =&-\frac12+\left(\frac{y}{x}-1\right)\bigg[\mathrm{Li}_2\left(1-\frac{x}{y}\right)-\zeta(2)\bigg]+\lln x(\lln y-1)-\frac12\lln^2 y\,,\\
    T_{2234}(x,y,y,x,y)=&\ T(x,y,y,x)\nn\\
    =&-\frac12+\left(1-\frac{y}{x}\right)\bigg[\mathrm{Li}_2\left(1-\frac{x}{y}\right)-\zeta(2)\bigg]+\lln x-2\lln y+\frac12\lln^2 y\,,\\
    T_{1234}(y,x,0,y,0)=&\ U(y,0,x,0,y)\nn\\
    =&\  3 - \frac{5}{2} \frac{x}{y}- \frac{\pi^2}{3}\frac{y}{x} + \frac{T_{134}(0, x, y)}{y} +
 y \lln y \left(\frac{1}{y} - \frac{1}{2 x} \lln y\right) \nn\\
 &+ \left(\frac{1}{x} - \frac{1}{y}\right) \bigg[(x - y) \frac{\pi^2}{6} \nn\\
 &\hspace{2cm}+ \lln (x - y) \bigg(\frac{1}{2} (x - y) \lln (x - y) + y \lln y - 2 x\bigg)\bigg]\,,\\
    T_{1234}(x,0,y,x,y)=&\ U(x,y,0,y,x)\nn\\
    =&\ \frac{11}{2} - 2 \lln x  - \lln y + \frac12 \lln^2 y+ \left(1+\frac{2y}{x}\right)\mathrm{Li}_2\left( 1 - \frac{x}{y}\right)\nn\\
    &- 2 \left(1 + \frac{y}{x}\right) \zeta(2)+ \ln\left(1-\frac{x}{y}\right)\bigg[\lln x - 1 + \frac{y}{x} (1 - \lln y)\bigg]\,.
\end{align}
\end{subequations}
Here $\mathrm{Li}_2$ denotes the dilogarithm, and we recall that $\zeta(2)=\mathrm{Li}_2(1)=\pi^2/6$.

In addition to these limits, we also had to derive several expansions of loop functions in powers of $\epsilon$, such as
\begin{subequations}
\begin{align}
\label{eq:loopfn_eps_exp}
    B_0(x,x,\epsilon)=&\ 2-\lln x-\pi\sqrt{\frac{\epsilon}{x}}+\frac{\epsilon}{2x}\bigg(\ln\frac{x}{\epsilon}+2\bigg)+\frac{\pi}{8}\left(\frac{\epsilon}{x}\right)^{3/2}-\frac{\epsilon^2}{12x^2}\nn\\
    &+\frac{\pi}{128}\left(\frac{\epsilon}{x}\right)^{5/2} - \frac{\epsilon^3}{120 x^3} + \frac{\pi}{1024}\left(\frac{\epsilon}{x}\right)^{7/2} - \frac{\epsilon^4}{840 x^4} + \frac{5 \pi}{32768} \left(\frac{\epsilon}{x}\right)^{9/2}\nn \\
    &- \frac{\epsilon^5}{5040 x^5} +\mathcal{O}\left(\left(\frac{\epsilon}{x}\right)^{11/2}\right)\,,\\
    T_{134}(x,x,\epsilon)=&-I(x,x,\epsilon)\nn\\
    =&\ x(\lln^2x-4\lln x+5)-\frac{1}{2}\epsilon\big[\lln^2x+2\lln x(2-\lln \epsilon)+3\big]\nn\\
    &+\frac{\epsilon^2}{6x}\left[\ln\frac{x}{\epsilon}+\frac{8}{3}\right]+\frac{\epsilon^3}{60x^2}\bigg[\ln\frac{x}{\epsilon}+\frac{31}{15}\bigg]+\frac{\epsilon^4}{420x^3}\bigg[\ln\frac{x}{\epsilon}+\frac{389}{210}\bigg]\nn\\
    &+\frac{\epsilon^5}{2520x^4}\bigg[\ln\frac{x}{\epsilon}+\frac{1097}{630}\bigg]+\mathcal{O}\left(\frac{\epsilon^6}{x^5}\right)\,,\\
    T_{1234}(x,\epsilon,x,x,x)=&\ U(x,x,\epsilon,x,x)\nn\\
    =&\ \frac{11}{2}-3\lln x+\frac12 \lln^2 x-\frac{2\pi^2}{3}+ \pi\sqrt{\frac{\epsilon}{x}}\ \lln x \nn\\
    &+ \frac{\epsilon}{x}\bigg[\frac12\lln x\ln\frac{\epsilon}{x}-\lln x-1+\frac{\pi^2}{6}\bigg]-\frac{\pi}{24}\left(\frac{\epsilon}{x}\right)^{3/2}[4+3\lln x]\nn\\
    &+\frac{\epsilon^2}{6x^2}\bigg[-\frac12\lln \epsilon+\lln x+\frac86\bigg]-\frac{\pi}{160}\left(\frac{\epsilon}{x}\right)^{5/2}\bigg[\lln x-\frac{2}{3}\bigg]\nn\\
    &+\frac{\epsilon^3}{144x^3}\big[1-\lln\epsilon+2\lln x\big]+\mathcal{O}\left(\left(\frac{\epsilon}{x}\right)^{5/2}\right)\,,\\
    T_{1234}(x,x,\epsilon,x,\epsilon)=&\ U(x,\epsilon,x,\epsilon,x)\nn\\
    =&\ \frac{11}{2}-3\lln x+\frac12 \lln^2 x-\frac{\pi^2}{3}+\pi \sqrt{\frac{\epsilon}{x}}(\lln x-2)\nn\\
    &+\frac{\epsilon}{x}\bigg[\frac12\lln x\ln \frac{\epsilon}{x}-\lln x-1+\frac{5\pi^2}{6}\bigg] \nn\\
    &+\frac{\pi}{2} \left(\frac{\epsilon}{x}\right)^{3/2}\bigg[\lln\epsilon -\frac54\lln x-\frac32\bigg]\nn\\
    &+\frac{\epsilon^2}{x^2}\bigg[-\frac18\ln^2\frac{\epsilon}{x}-\frac16\lln\epsilon+\frac14\lln x-\frac{3\pi^2}8+\frac{37}{36}\bigg]\nn\\
    &+\frac{\pi}{96}\left(\frac{\epsilon}{x}\right)^{5/2} \big[22 - 6 \lln\epsilon + 5 \lln x\big] + \frac{\epsilon^3}{288x^3}\big[95 - 66 \lln\epsilon + 69 \lln x\big]\nn\\
	&+\mathcal{O}\left(\left(\frac{\epsilon}{x}\right)^{7/2}\right)\,,\\
	T_{12345}(x,x,\epsilon,x,\epsilon,x)=&-M(x,\epsilon,x,x,\epsilon,x)\nn\\
	=&\ -\frac{1}{x}\bigg\{\pi^2\ln2-\frac32\zeta(3)+\pi\sqrt{\frac{\epsilon}{x}}\ln\frac{\epsilon}{x}- \frac{\epsilon}{2x}(1 + \pi^2) \nn\\
	&\hspace{1cm}+ \frac{\pi}{24}\left(\frac{\epsilon}{x}\right)^{3/2} \bigg[16+\ln\frac{\epsilon}{x}\bigg]-\frac{\epsilon^2}{144x^2}\bigg[5+18\pi^2+\ln\frac{\epsilon}{x}\bigg]\nn\\
	&\hspace{1cm}+\mathcal{O}\left(\left(\frac{\epsilon}{x}\right)^{5/2}\right)\bigg\}\,.
\end{align}
\end{subequations}
The expansions of the $B_0$ and $T_{134}$ integrals can be obtained directly from the known analytical expressions of these functions --- the former can be reproduced with \texttt{Package-X}~\cite{Patel:2015tea}, while the latter can be verified to correspond up to order $\epsilon$ to Eqs.~(3.15) and (3.21) in \ccite{Kumar:2016ltb}. For the $T_{1234}$ and $T_{12345}$ integrals, we can use Eqs.~(3.20), (3.21), (3.22), and (3.32) in \ccite{Martin:2003qz} to derive differential equations in the variable $\epsilon$.
The solutions of those differential equations yield the desired
integrals (as we know their expressions for $\epsilon=0$).

%%%%%%%%%%%% table %%%%%%%%%%%
\begin{table}[ht]
{\small
\centering
\begin{tabular}{| c | c c c c |}
\hline
 Integral &  \multicolumn{4}{|c|}{Numerical results (\texttt{TSIL} vs \cref{eq:loopfn_eps_exp})}\\
\hline\hline
  $T_{134}(x,x,\epsilon)/x$ & \texttt{TSIL} & Approx. $\mathcal{O}(\epsilon^3)$ & Approx. $\mathcal{O}(\epsilon^4)$ & Approx. $\mathcal{O}(\epsilon^5)$ \\
\hline
 $(a)$ & 2.706961 & 2.706961 & 2.706961 & 2.706961 \\
 $(b)$ & 2.649604 & 2.649604 & 2.649604 & 2.649604 \\
 $(c)$ & 1.984546 & 1.984533 & 1.984546 & 1.984546 \\
% $(b)$ & 1.290758 & 1.290346 & 1.290725 & 1.290755 \\
\hline
\hline
$T_{1234}(x,\epsilon,x,x,x)$ & \texttt{TSIL} & Approx. $\mathcal{O}(\epsilon)$ & Approx. $\mathcal{O}(\epsilon^2)$ & Approx. $\mathcal{O}(\epsilon^3)$ \\
\hline
 $(a)$ & -2.897500 & -2.897499 & -2.897500 & -2.897500 \\
 $(b)$ & -2.718365 & -2.717635 & -2.718365 & -2.718365 \\
 $(c)$ & -2.120744 & -2.066306 & -2.120920 & -2.120745 \\
\hline
\hline
$T_{1234}(x,x,\epsilon,x,\epsilon)$ & \texttt{TSIL} & Approx. $\mathcal{O}(\epsilon)$ & Approx. $\mathcal{O}(\epsilon^2)$ & Approx. $\mathcal{O}(\epsilon^3)$ \\
\hline
 $(a)$ &  0.330178 &  0.330195 &  0.330178 &  0.330178 \\
 $(b)$ & -0.000583 &  0.009712 & -0.000600 & -0.000583 \\
 $(c)$ & -0.814525 & -0.270416 & -0.838471 & -0.814698 \\
\hline
\hline
$x\cdot T_{12345}(x,x,\epsilon,x,\epsilon,x)$ & \texttt{TSIL} & Approx. $\mathcal{O}(\epsilon)$ & Approx. $\mathcal{O}(\epsilon^2)$ & $-$ \\
\hline
 $(a)$ & 4.748109 & 4.748108 & 4.748109 & $-$ \\
 $(b)$ & 3.538307 & 3.536898 & 3.538301 & $-$ \\
 $(c)$ & 1.823276 & 1.689842 & 1.812955 & $-$ \\
\hline
\end{tabular}
\caption{Sample values of the integrals $T_{134}(x,x,\epsilon)$, $T_{1234}(x,\epsilon,x,x,x)$, $T_{1234}(x,x,\epsilon,x,\epsilon)$, and $T_{12345}(x,x,\epsilon,x,\epsilon,x)$ --- multiplied by powers of $x$ to obtain a dimensionless quantity --- for the mass parameter assignments $(a)$, $(b)$, $(c)$ given in \cref{eq:massparam_abc}. We compare values obtained using the program \texttt{TSIL}~\cite{Martin:2005qm} with those from our approximate expansion, at different orders in $\epsilon$. Note that as the integral $T_{12345}$ is UV finite, there is no $Q$ dependence in the corresponding loop function.}
\label{tab:checktsil_T134_T1234_T12345}
}
\end{table}
%%%%%%%%%%%% table %%%%%%%%%%%

In \cref{tab:checktsil_T134_T1234_T12345} we provide some sample results of our checks of the expansions in $\epsilon$ of the integrals $T_{134}$, $T_{1234}$, and $T_{12345}$ for three sets of mass parameters:
\begin{align}
\label{eq:massparam_abc}
    (a):&\ x=1.0\cdot 10^4\text{ GeV}^2,\ \epsilon=1\text{ GeV}^2,\ Q^2=5.0\cdot 10^3\text{ GeV}^2;\nn\\
    (b):&\ x=1.0\cdot 10^4\text{ GeV}^2,\ \epsilon=1.0\cdot 10^2\text{ GeV}^2,\ Q^2=5.0\cdot 10^3\text{ GeV}^2;\nn\\
    (c):&\ x=1.0\cdot 10^4\text{ GeV}^2,\ \epsilon=2.0\cdot 10^3\text{ GeV}^2,\ Q^2=5.0\cdot 10^3\text{ GeV}^2.
\end{align}

%%%%%%%%%%%

\subsection{Example derivation}

In order to illustrate the procedure of our calculations, we present here how we derived the expansion of $T_{134}(x,x,\epsilon)$ in powers of $\epsilon$. Starting from Eq.~(5.3) in \ccite{Martin:2003qz}, which gives the first derivative\footnote{\ccite{Martin:2003qz} provides complete results for the derivatives of a basis of two-loop self-energy integrals with respect to external momentum and mass arguments. Solving this system of differential equations allows
a numerical evaluation of those basis integrals. This method is implemented in the public tool \texttt{TSIL}~\cite{Martin:2005qm}. } of the integral $I$
with respect to its first mass argument, and substituting the mass arguments with the two variables $x$ and $y$, we find
\begin{align}
    &\frac{\partial}{\partial y}T_{134}(x,x,y)=\nn\\
    &=\frac{4 x^2 -y^2 - 2 A_0(y) A_0(x) +  2 A_0(x)^2 - (y - 2 x) \big[A_0(y) + 2 A_0(x)- T_{134}(x, x,y)\big]}{y^2 + 2 x^2 - 2 (2 y x + x^2)}\,.
\end{align}
This is simply a differential equation in the function $T_{134}$, for which we also know the boundary condition at $y=0$
\begin{align}
    T_{134}(x,x,0)=x\big(\lln^2 x-4 \lln x+5\big)\,.
\end{align}
One can then straightforwardly solve this equation, and while the resulting function is fairly complicated, it can be expanded to arbitrary order in powers of $y=\epsilon\ll x$.

We should emphasise here that for the case of the $T_{134}$ integral one could in principle obtain the same result by directly expanding the analytical expression that is known for general mass assignments. However, we discuss the case of this integral as it provides a simple example of our setup to derive expansions of the more complicated integrals $T_{1234}$ and $T_{12345}$ --- for which analytical results are not known in general. Finally, for the derivatives of $T_{11234}$ and $T_{12345}$, we make use of the relations in Sec.~IV of \ccite{Martin:2003qz}.

%%%%%%%%%%%

\subsection{Mass configuration 1: \texorpdfstring{$m_1^2=\epsilon,\ m_2^2=m_3^2=m^2$}{m1\^{}2 = eps, m2\^{}2 = m3\^{}2 = m\^{}2}}

Setting $m_2^2=m_3^2=m^2$ and $m_1^2=\epsilon$, we find, up to order $\mathcal{O}(\epsilon^0)$
\begin{align}
\label{eq:loopfn_exp_caseI}
    \frac{d}{dp^2}&T_{11234}(p^2,m^2,m^2,\epsilon,m^2,\epsilon)\bigg|_{p^2=m^2}=\nn\\
    =& \frac{\pi (2-\lln m^2)}{4\sqrt{\epsilon}m^3} + \frac{-6 \lln\epsilon \lln m^2-3 \lln^2\epsilon+24 \lln\epsilon+9 \lln^2m^2-24 \lln m^2-\pi^2}{24 m^4}\,,\nn\\
    \frac{d}{dp^2}&T_{11234}(p^2,\epsilon,\epsilon,m^2,m^2,m^2)\bigg|_{p^2=m^2}=\nn\\
    =&-\frac{\lln m^2}{2 m^2 \epsilon} + \frac{3 \pi \lln m^2}{8 m^3 \sqrt{\epsilon}} + \frac{-50 + 6 \pi^2 + 3\lln \epsilon - 12 \lln m^2 +
  18 \lln\epsilon \lln m^2 - 18 \lln^2 m^2}{36 m^4}\,,\nn\\
  \frac{d}{dp^2}&T_{12345}(p^2,m^2,\epsilon,m^2,\epsilon,m^2)\bigg|_{p^2=m^2}=\nn\\
  &= \frac{1}{4m^4}\bigg[2+\ln\frac{m^2}{\epsilon}+\ln^2\frac{m^2}{\epsilon}\bigg]-\frac{\pi^2\ln2-3/2\zeta(3)}{m^4}\,.
\end{align}
We present in \cref{tab:checktsil_deriv_caseI} some example values of these three derivatives, multiplied by $m^4$ in order to obtain a dimensionless number, for the set of mass parameters $(a)$ in \cref{eq:massparam_abc}. We compare the results from \texttt{TSIL} (left column) to the values obtained with the approximate expressions of \cref{eq:loopfn_exp_caseI} (right column). We note that because \texttt{TSIL} provides results for the integrals $V$ and $M$ --- corresponding to $T_{11234}$ and $T_{12345}$ in our notation --- we need to take the derivatives with respect to $p^2$ numerically. We choose for this a step size of $\delta=10^{-3}\text{ GeV}^2$, although we have verified that the numerical derivatives are stable and depend only very slightly on $\delta$. Given that we derived approximate results only to order $\mathcal{O}(\epsilon^0)$, we choose a set of mass inputs with small $\epsilon$, for which we expect the higher-order terms to be moderate. Indeed, we find good agreement between the numerical results of \texttt{TSIL} and the approximate ones, with differences of only $0.06\%$, $3\cdot 10^{-4}\%$, and $2\%$, respectively.

%%%%%%%%%%%% table %%%%%%%%%%%
\begin{table}[ht]
\centering
\begin{tabular}{| c | c c |}
\hline
 Integral & \multicolumn{2}{|c|}{Numerical results}\\
  &  \texttt{TSIL} & Approx. $\mathcal{O}(\epsilon^0)$ \\
\hline
 $m^4\displaystyle{\frac{d}{dp^2}}T_{11234}(p^2,m^2,m^2,\epsilon,m^2,\epsilon)\bigg|_{p^2=m^2}$ & 85.552342
& 85.606671 \\
\hline
 $m^4\displaystyle{\frac{d}{dp^2}}T_{11234}(p^2,\epsilon,\epsilon,m^2,m^2,m^2)\bigg|_{p^2=m^2}$ & -3387.9644 & -3387.9533 \\
\hline
 $m^4\displaystyle{\frac{d}{dp^2}}T_{12345}(p^2,m^2,\epsilon,m^2,\epsilon,m^2)\bigg|_{p^2=m^2}$ & 21.636871 & 21.274760\\
\hline
\end{tabular}
\caption{Example values of the derivatives of $T_{11234}$ and $T_{12345}$, multiplied by $m^4$ to obtain dimensionless numbers, computed by \texttt{TSIL} and with the approximate expansions of \cref{eq:loopfn_exp_caseI}. The mass values are chosen according to $(a)$ in \cref{eq:massparam_abc}. The numerical derivatives that are necessary for the \texttt{TSIL} column are computed with a step size of $\delta=10^{-3}\text{ GeV}^2$.}
\label{tab:checktsil_deriv_caseI}
\end{table}
%%%%%%%%%%%% table %%%%%%%%%%%

%%%%%%%%%%%

\subsection{Mass configuration 2: \texorpdfstring{$m_1^2=0,\ m_2^2=m^2+\epsilon,\ m_3^2=m^2$}{m1\^{}2=0, m2\^{}2=m\^{}2 + eps}}
\label{app:deriv_case2}

Turning next to the mass configuration with $m_1^2=0$, $m_2^2=m^2+\epsilon$ ($\epsilon>0$), and $m_3^2=m^2$, we have
\begin{align}
\label{eq:loopfn_exp_caseII1}
    \frac{d}{dp^2}&T_{11234}(p^2,m^2+\epsilon,m^2+\epsilon,0,m^2,0)\bigg|_{p^2=m^2}=\nn\\
    =&\ \frac{2 - \lln m^2}{m^2\epsilon} + \frac{-\pi^2 + 6 \lln \epsilon - 3 \lln^2\epsilon - 6 \lln m^2 + 3 \lln^2m^2}{6 m^4}+\mathcal{O}(\epsilon)\,,\nn\\
    \frac{d}{dp^2}&T_{11234}(p^2,m^2,m^2,0,m^2+\epsilon,0)\bigg|_{p^2=m^2+\epsilon}=\nn\\
    =&\ \frac{\lln m^2 - 2}{m^2\epsilon} + \frac{2\pi^2 + 18 + 6 i \pi + (6 - 6 i \pi) \lln \epsilon - 3 \lln^2\epsilon - 12 \lln m^2 + 3 \lln^2m^2}{6 m^4} \nn\\
    \hphantom{=}&+\mathcal{O}(\epsilon)\,,
\end{align}
and
\begin{align}
\label{eq:loopfn_exp_caseII2}
   \frac{d}{dp^2}&T_{11234}(p^2,m_1^2,m_1^2,m^2+\epsilon,m^2+\epsilon,m^2)\bigg|_{p^2=m^2}\nn\\
   =&-\frac{\lln m^2}{2 m^2 m_1^2} + \frac{3 \pi \lln m^2}{8 m^3 m_1} \nn\\
    &+ \frac{-50 + 6 \pi^2 + 3\lln m_1^2 - 12 \lln m^2 +
  18 \lln m_1^2 \lln m^2 - 18 \lln^2 m^2}{36 m^4}\nn\\
    &+\frac{\epsilon}{m^2}\bigg[\frac{\pi \lln m^2}{8 m m_1^3} - \frac{1 + 2 \lln m^2}{4 m^2 m_1^2} + \frac{\pi (40 + 27 \lln m^2)}{192 m^3 m_1} -\frac{23 + 90 \lln m^2 - 42 \lln m_1^2}{144 m^4}\bigg] \nn\\
    &+\mathcal{O}(\epsilon^2)\,,\nn\\
   \frac{d}{dp^2}&T_{11234}(p^2,m_1^2,m_1^2,m^2,m^2,m^2+\epsilon)\bigg|_{p^2=m^2+\epsilon}\nn\\
   =&-\frac{\lln m^2}{2 m^2 m_1^2} + \frac{3 \pi \lln m^2}{8 m^3 m_1} \nn\\
    &+ \frac{-50 + 6 \pi^2 + 3\lln m_1^2 - 12 \lln m^2 +
  18 \lln m_1^2 \lln m^2 - 18 \lln^2 m^2}{36 m^4}\nn\\
    &+\frac{\epsilon}{m^2}\bigg[\frac{\pi \lln m^2}{8 m m_1^3} - \frac{3}{4 m^2 m_1^2} + \frac{\pi (-112 + 81 \lln m^2)}{192 m^3 m_1} \nn\\
    &\hspace{2cm}+\frac{329 - 48 \pi^2 - 138 \lln m^2 + 144 \lln^2 m^2 + 90 \lln m_1^2 -
 144 \lln m^2 \lln m_1^2}{144 m^4}\bigg] \nn\\
 \hphantom{=}&+\mathcal{O}(\epsilon^2)\,,
\end{align}
where we have retained a dependence on $m_1^2$ because the integral $T_{11234}(s,m_1^2,m_2^2,m_3^2,m_4^2) $ itself is IR-divergent in the limit $m_1\to0$, and finally
\begin{align}
\label{eq:loopfn_exp_caseII3}
    \frac{d}{dp^2}&T_{12345}(p^2,m^2+\epsilon,0,m^2,0,m^2+\epsilon)\bigg|_{p^2=m^2}=\nn\\
    =&\ \frac{1}{m^4}\bigg[\pi^2\bigg(\frac{1}{4}-\ln 2\bigg) +\frac{3}{2}\zeta(3)+\ln\frac{m^2}{\epsilon}+\ln^2\frac{m^2}{\epsilon}\bigg]+ \mathcal{O}(\epsilon)\,,\nn\\
    \frac{d}{dp^2}&T_{12345}(p^2,m^2,0,m^2+\epsilon,0,m^2)\bigg|_{p^2=m^2+\epsilon}=\nn\\
    =&\ \frac{1}{m^4}\bigg[-\pi^2\bigg(\frac{3}{4}+\ln 2\bigg) +\frac{3}{2}\zeta(3) + i \pi + (1 + 2i\pi)\ln\frac{m^2}{\epsilon}+\ln^2\frac{m^2}{\epsilon}\bigg]\nn\\
    \hphantom{=}&+\mathcal{O}(\epsilon)\,.
\end{align}

%%%%%%%%%%%% table %%%%%%%%%%%
\begin{table}[ht]
\centering
\begin{tabular}{| c | c c |}
\hline
 Integral & \multicolumn{2}{|c|}{Numerical results}\\
  &  \texttt{TSIL} & Expansion \\
\hline
 $m^4\displaystyle{\frac{d}{dp^2}}T_{11234}(p^2,m^2+\epsilon,m^2+\epsilon,0,m^2,0)\bigg|_{p^2=m^2}$ & -13022.295 & -13021.642 \\
\hline
 $m^4\displaystyle{\frac{d}{dp^2}}T_{11234}(p^2,m_1^2,m_1^2,m^2+\epsilon,m^2+\epsilon,m^2)\bigg|_{p^2=m^2}$ & -3361.5011 & -3361.3207 \\
\hline
 $m^4\displaystyle{\frac{d}{dp^2}}T_{12345}(p^2,m^2+\epsilon,0,m^2,0,m^2+\epsilon)\bigg|_{p^2=m^2}$ & 91.482800 & 91.470115 \\
\hline
\end{tabular}
\caption{Example values of the derivatives of $T_{11234}$ and $T_{12345}$, multiplied by $m^4$ to obtain dimensionless numbers, computed by \texttt{TSIL} and with the approximate expansions of \cref{eq:loopfn_exp_caseII1,eq:loopfn_exp_caseII2,eq:loopfn_exp_caseII3}. The mass values are chosen to be $m=100\text{ GeV}$, $\epsilon=1\text{ GeV}^2$, $m_1=1\text{ GeV}$, and the numerical derivatives are computed with a step size of  $\delta=10^{-3}\text{ GeV}^2$.}
\label{tab:checktsil_deriv_caseII}
\end{table}
%%%%%%%%%%%% table %%%%%%%%%%%

We present in \cref{tab:checktsil_deriv_caseII} some example values of the three derivatives of two-loop integrals for case 2. As in \cref{tab:checktsil_deriv_caseI}, we multiply the derivatives by $m^4$ in order to obtain dimensionless quantities, and we compare the results using \texttt{TSIL} (left column) with the approximate expansions of \cref{eq:loopfn_exp_caseII1,eq:loopfn_exp_caseII2,eq:loopfn_exp_caseII3} (right column). For the mass parameters, we take $m^2=(100\text{ GeV})^2$, $\epsilon=(1\text{ GeV})^2$, and $m_1=1\text{ GeV}$ (we recall that we need to take a non-zero value of $m_1$ to avoid a divergence in $T_{11234}(p^2,m_1^2,m_1^2,m^2+\epsilon,m^2+\epsilon,m^2)$ and its derivative). As desired, we find a very good agreement between the \texttt{TSIL} and approximate values, with discrepancies of only 0.05\%, 0.05\%, and 0.004\%, respectively, for the three derivatives.

As a final remark, we note that the light regulator mass $m_1$ that we needed to include for the integral $T_{11234}(p^2,m_1^2,m_1^2,m^2+\epsilon,m^2+\epsilon,m^2)$ could be generated by extending the resummation of the light-scalar contribution to two loops. We have explicitly verified that this resummation cures the IR divergence in the integral $T_{11234}(p^2,m_1^2,m_1^2,m^2+\epsilon,m^2+\epsilon,m^2)$ and its derivatives. We have however left to further work the derivation of expressions for the derivatives of the two other integrals ($T_{11234}(p^2,m^2+\epsilon,m^2+\epsilon,0,m^2,0)$ and $T_{12345}(p^2,m^2+\epsilon,0,m^2,0,m^2+\epsilon)$) with non-vanishing light-scalar masses.